\documentclass[showpacs,preprintnumbers,amsmath,amssymb,aps, prd]{revtex4}
\usepackage{graphicx}% Include figure files
\usepackage{grffile}
\usepackage{caption}
\usepackage{subcaption}
\usepackage{float}
\usepackage{xcolor}
\usepackage{dcolumn}% Align table columns on decimal point
\usepackage{bm}% bold math
\begin{document}
\title{Lorentz violation and noncommutative effect on superradiance scattering off
 Kerr-like black hole and on the shadow of it}
% Force line breaks with \
\author{Sohan Kumar Jha}
\affiliation{Chandernagore College, Chandernagore, Hooghly, West
Bengal, India}
\author{Anisur Rahaman}
\email{anisur.associates@iucaa.ac.in; manisurn@gmail.com
(Corresponding Author)} \affiliation{Durgapur Govt. College,
Durgapur, Burdwan - 713214, West Bengal, India}

\date{\today}% It is always \today, today, % but any date may be explicitly specified
\begin{abstract}
\begin{center}
Abstract
\end{center}
We consider a Lorentz violating non-commutating Kerr-like
spacetime and studied the superradiance effect and the shadow cast
by the back hole. We extensively study the different aspects of
the black hole associated with a generalized Kerr-like spacetime
metric endowed with the corrections licked with Lorentz violation
and non-commutativity effect jointly. We investigate the
superradiance effect, deviation of shape, and size of the
ergosphere, energy emission rate, and black hole shadow in this
generalized situation and study their variation taking different
admissible values of Lorentz violating parameter $l$ and
non-commutative parameter $b$. The admissible range has been
determined from the observation of the Event Horizon Telescope
(EHT) collaboration concerning $M87_8$ astronomical black hole. We
observe that the superradiance phenomena has a crucially depends
on the parameter $l$ and $b$ apart from its dependence on $a$
which is linked to the spin of the black hole. We also observe
that with the increase in Lorentz violating parameter $l$, the
size of the black hole shadow increases, and with the increase in
the non-commutative parameter $b$, the size of the black hole
decreases. We have made an attempt to constrain parameters $b$ of
a non-commutative Kerr-like black hole using the observation
available from the EHT collaboration, in the same way, we put
constrain on the Lorentz violating parameter $l$. This study shows
that black holes associated with non-commutative Kerr-like
spacetime may be a suitable candidate for an astrophysical black
hole.
\end{abstract}
\maketitle
\section{Introduction}
Different important optical phenomena have been encountered when
light approaches the vicinity of a black hole. Lansing (strong and
weak), superradiance, and formation of black hole shadow are
important in this context. From the time of Einstein, the study of
the optical effect in the vicinity of a black hole was started and
this field has been getting enriched through the investigation of
different scientists. In the lensing effect, the black hole
behaves as a natural celestial lens that makes a deviation in the
path of the light. Here bending takes place due to the strong
gravitational field of the black hole. Our focus is laid on the
study of the effect of quantum gravity study on the superradiance
phenomena and shadow of black holes.

In a gravitational system, the scattering of radiation off
absorbing rotating objects produce waves with amplitude larger
than incident one under certain conditions which is known as
rotational superradiance \cite{ZEL0, ZEL1}. In 1971, Zel'dovich
showed that scattering of radiation off rotating absorbing
surfaces result in waves with a larger amplitude as $ \omega <
m\Omega$ where $\omega$ is the frequency of the incident
monochromatic radiation with $m$, the azimuthal quantum number
with respect to the rotation axis and $\Omega$ is the angular
velocity of the rotating gravitational system. For review, we
would like to mention the lecture notes \cite{REVW}, and the
references therein. Rotational superradiance belongs to a wider
class of classical problems displaying stimulated or spontaneous
energy emission, such as the Vavilov-Cherenkov effect, the
anomalous Doppler effect. When quantum effects were incorporated,
it was argued that rotational superradiance would become a
spontaneous process and that rotating bodies including black holes
would slow down by the spontaneous emission of photons. From the
historic perspective, the discovery of black-hole evaporation was
well understood from the studies of black-hole superradiance
\cite{HAW}. Interest in the study of black-hole superradiance has
recently been revived in different areas, including astrophysics
high-energy physics via the gauge/gravity duality along with
fundamental issues in General Relativity. Superradiant
instabilities can be used to constrain the mass of ultralight
degrees of freedom \cite{INST00, INST0, INST1, INST2}, with
important applications to dark-matter searches. The black hole
superradiance is also associated with the existence of new
asymptotically flat hairy black-hole solutions \cite{HAIR} and
with phase transitions between spinning or charged black objects
and asymptotically anti-de Sitter (AdS) spacetime \cite{ADS0,
ADS1, ADS2} or in higher dimensions \cite{MSHO}. Finally, the
knowledge of superradiance is instrumental in describing the
stability of black holes and in determining the fate of the
gravitational collapse in confining geometries \cite{ADS0}

Shadow is a two-dimensional dark area in the celestial sphere
which is known as a black hole caused by the strong gravity of the
black hole. It was first examined by Synge in 1966 for a
Schwarzschild black hole \cite{SYNGE}, and the radius of the
shadow was given by Luminet \cite{JP}. The shadow of a
non-rotating black hole is a standard circle, while the shadow of
a rotating black hole elongates in the direction of the rotating
axis due to the dragging effects of spacetime \cite{BARDEEN,
CHANDRASEKHAR}. Hioki and Maeda \cite{KH} proposed two observables
based on the feature that points at the boundary of the Kerr
shadow to match the astronomical observations. One of which
roughly describes the size of the shadow and the other describes
the deformation of its shape from a reference circle. Also, using
the method given in \cite{TJDP} one can find the deviation from
circularity $\Delta C$ These various observables are very useful
in testing and constraining the parameters involved in the
modified theories of gravity.

The black hole itself was one of the important predictions of the
general theory of relativity. The announcement of capturing of the
shadow of the supermassive black hole $M87^*$ at the center of the
nearby galaxy Messier 87 by the Event Horizon Telescope (EHT) EHT
collaboration \cite{KA1, KA2, KA3, KA4, KA5, KA6} study concerning
the optical phenomena in the vicinity of black holes have got
renewed impetus. Various types of modified gravity along with
standard one were developed from time to time not only to resolve
ambiguities in different physical observable but also to have a
precise explanation of observations that are being available
through the use of the present times sophisticated instruments.
f(r) gravity, gravity modified with dark matter, dark energy,
monopole, etc. are a few examples. In this respect to make the
theory of gravity endowed with quantum correction is also a
fascinating direction of modification of the standard theory of
gravity. It is of great importance since quantization of gravity
is still not available in a mature form. Therefore the correction
that comes to incorporate the effect of the Planck scale is highly
appreciated in recent times. Two such concepts of taking into the
effect of the Planck scale are considered with important in the
recent time literature one is the LV effect and another is the
spacetime non-commutativity. Both the effect has been studied
earlier separably in the different physical systems however the
framework of having the combined treating these two on the same
footing effect is still lacking to the best of our knowledge. The
formulation of the Standard Model(SM) of particle physics and the
general theory of gravity (GTR) entirely depends on the principle
of Lorentz invariance. The GTR does not take into account the
quantum properties of particles, and SM, on the other hand,
neglects all gravitational effects of particles. At the Planck
scale, one cannot neglect gravitational interactions between
particles, and hence merger of SM with GTR in a single theory
become essential. It is indeed available from the quantum gravity
concept. At this scale, it is expected to face a violation of
Lorentz symmetry \cite{DM}. Several studies related to Lorentz
violation in different aspects have come in the literature
\cite{EMS1, EMS2, EMS3, EMS4, EMS5, EMS6, EMS7, EMS8, EMS9, EMS10,
EMS11, EMS12, EMS13, EMS14, EMS15, EMS16, EMS17, RB, GVL1, GVL2,
GVL3, GVL4, STRING1, STRING2, STRING3, STRING4}. The standard
model extension (SME) is an effective field theory that couples SM
to GTR \cite{DC, DC1, ESM1, ESM2} where Lorentz violation (LV) is
introduced. One of the theories that belong in this class is the
bumblebee model, where LV is introduced through an axial-vector
field $B_{\mu}$ which is known as the bumblebee field. Recently,
in \cite{DING, RC}, a Kerr-like solution was obtained from the
Einstein-bumblebee theory. In \cite{ARS}, it has been found that a
non-commutative Kerr-like solution is also possible from the
Einstein-bumblebee theory.

Non-commutative spacetime has been extensively studied in recent
years \cite{ZABONON, NONCOM, JINNON, ANINON, GITNON, CUINON}. The
correction due to noncommutative spacetime has been studied in
various fields. A fertile field for applying the idea of
noncommutative spacetime is black hole physics. Several ways are
available in the literature to implement a noncommutative
spacetime in theories of gravity \cite{PNICO, PAS, PAS1, SMEL,
HARI}. Through a modification of the matter source by a Gaussian
mass distribution noncommutativity has been brought in the black
hole physics in the article \cite{NICOLINI} and through Lorentzian
mass distribution noncommutativity has been introduced in the
article \cite{NOZARI}. An interesting extension concerning
thethermodynamic similarity between Reissner-Nordstr$\ddot{o}$m
black hole the noncommutative Schwarzschild black hole and the has
been made in the article \cite{KIM}. The thermodynamical aspects
of noncommutative black holes have been investigated by taking on
the tunneling formalism in the articles \cite{KNO, RABIN, MEHDI,
MEHDI1, MIAO, ISLAM, SUFI, GUPTA}. In \cite{LIANG}, by taking the
mass density to be a Lorentzian smeared mass distribution the
thermodynamic properties of noncommutative BTZ black holes have
been studied. So far we have found non-commutativity can be
implemented by modifying the point-like source of matter
designated by the Dirac delta function replacing it with a
distribution of matter. In the articles \cite{NOZARI, NICOLINI}
Gaussian and Lorentzian distributions are used to incorporate
non-commutativity. In this manuscript, we have introduced
non-commutativity into Kerr-like black hole \cite{DING} by
considering Lorentzian distribution as it has been used in
\cite{NOZARI}.

Correction due to quantum gravity is believed to be bestowed on
the theory of gravity introducing Lorentz violation,
non-commutativity of spacetime. In this article, we develop a
framework where we can study both the Lorentz violation and
non-commutativity of spacetime simultaneously on the same footing
and we carry out an investigation through this generalized
spacetime framework on the geometry, the photon orbit the energy
emission, and the black hole shadow. What light the obtained
information from the $\mathrm{M}87^{*}$ data can shade on this
modified framework that has been studied in detail. Constraining
of the LV parameter $(l)$ and the non-commutativity parameter
$(b)$ also has been executed from the data of the shadow of the
$\mathrm{M}87^{*}$.

We, therefore, make an attempt to have the combined effect of
these two and study their effect on the superradiance of the
shadow. Although a general study of the super-radiance phenomena
will be done here with the theories endowed with the quantum
correction we will be able to test this modified theory in the
light of the recently available data of the EHT collaboration. The
important information about the shadow is that it is found to have
an angular diameter $42\pm 3\mu$as with the deviation from
circularity $\Delta C= 0.1$ and axial ratio $\approx \frac{4}{3}$.
It helps us to constrain the free parameter entered into the
modified theory of gravity.

The manuscript is organized as follows. In Sec. II, we briefly
describe how the Kerr-like black hole metric is endowed with
Lorentz-violation and non-commutative of spacetime. In Sect.III,
we study the geometrical aspects concerning the horizon and
ergosphere of this modifies metric. Sec. IV is devoted to
superradiance scattering off the black hole corresponding to this
modified metric. In Sec. V we describe the photon orbit and shadow
corresponding to this black hole. Sec. VI contains the computation
of energy emission rate. In Sec. VII an attempt has been made to
constrain the parameters from the observation of the EHT
collaboration conserving the $\mathrm{M}87^{*}$ black hole
Sec.VIII contains a brief summary and conclusion of the work

\section{Modification of gravity containing
Lorentz-violation and non-commutativity of spacetime} Let us first
develop a framework where we can incorporate both the Lorentz
violation and the non-commutativity of spacetime on the same
footing. To this end, we give a brief description of how the
Lorentz violation effect was introduced.
\subsection{Lorentz violation effect}
Lorentz violation effect is believed to have the ability that it
may lead to significant effects on the properties of the black
holes anticipated beforehand. So it was attempted to introduce
through Einstein-bumblebee theory. It is an effective field theory
where the bumblebee field receives vacuum expectation through a
spontaneously breaking of symmetry the action of which is given by
\begin{eqnarray}
\mathcal{S}=\int d^{4} x \sqrt{-g}\left[\frac{1}{16 \pi
G_{N}}\left(\mathcal{R}+\varrho B^{\mu} B^{\nu} \mathcal{R}_{\mu
\nu}\right)-\frac{1}{4} B^{\mu \nu} B_{\mu
\nu}-V\left(B^{\mu}\right)\right]. \label{ACT}
\end{eqnarray}
where $\varrho^{2}$ represents the real coupling constant that
controls the non-minimal gravity interaction to bumblebee field
$B_{\mu}$. Like electromagnetic field the dynamics of the
bumblebee sector is described by the field strength tensor
corresponding to the bumblebee field which is defined by
\begin{eqnarray}
B_{\mu \nu}=\partial_{\mu} B_{\nu}-\partial_{\nu} B_{\mu}.
\end{eqnarray}
The field  $B_{\mu}$ receives a vacuum expectation value from a
suitable potential through the spontaneous breaking of the
symmetry of the theory.  The potential $V\left(B^{\mu}\right)$
that induces Lorentz symmetry breaking is given by
\begin{eqnarray}
V=V\left(B_{\mu} B^{\mu}\pm b^2)\right.
\end{eqnarray}
Here  $b^{2}$ is a real positive constant. It refers to a
non-vanishing vacuum expectation value for the field $B_{\mu}$. It
is assumed that the  has  a minimum described by the condition
\begin{eqnarray}
B_{\mu}B^{\mu}\pm b^2=0. \label{MINV}
\end{eqnarray}
The Eqn. (\ref{MINV}) ensures that a non-vanishing vacuum
expectation value for the field $B$
\begin{eqnarray}
<B^{\mu}>=b^{\mu}
\end{eqnarray}
will be received by the field $B_{\mu}$ from the potential $V$.
The field  $b^{\mu}$ indeed is a function of the spacetime
coordinates which have a constant magnitude $b_{\mu} b^{\mu}=\mp
b^{2}$  which may have a time-like as well as space-like nature
depending upon the choice of sign in front of  $b^{2}$. The
gravitational field equation in a vacuum that follows from the
action (\ref{ACT}) reads
\begin{eqnarray}
\mathcal{R}_{\mu \nu}-\frac{1}{2} g_{\mu \nu} \mathcal{R}=\kappa
T_{\mu \nu}^{B}.
\end{eqnarray}
where $\kappa=8 \pi G_{N}$ is the gravitational coupling and the
bumblebee energy-momentum tensor $T_{\mu \nu}^{B}$ has the
following expression
\begin{eqnarray}\nonumber
T_{\mu \nu}^{B}&=&B_{\mu \alpha} B_{\nu}^{\alpha}-\frac{1}{4}
g_{\mu \nu} B^{\alpha \beta} B_{\alpha \beta}-g_{\mu \nu} V+2
B_{\mu} B_{\nu} V^{\prime} \\\nonumber
&+&\frac{\varrho}{\kappa}[\frac{1}{2} g_{\mu \nu} B^{\alpha}
B^{\beta} R_{\alpha \beta}
-B_{\mu} B^{\alpha} R_{\alpha \nu}-B_{\nu} B^{\alpha} R_{\alpha \mu} \\
&+&\frac{1}{2} \nabla_{\alpha} \nabla_{\mu}\left(B^{\alpha}
B_{\nu}\right)+\frac{1}{2} \nabla_{\alpha}
\nabla_{\nu}\left(B^{\alpha} B_{\mu}\right)-\frac{1}{2}
\nabla^{2}\left(B^{\mu} B_{\nu}\right)-\frac{1}{2} g_{\mu \nu}
\nabla_{\alpha} \nabla_{\beta} \left(B^{\alpha} B^{\beta}\right)].
\end{eqnarray}
Here prime(') denotes differentiation with respect to the
argument,
\begin{eqnarray}
V^{\prime}=\left.\frac{\partial V(x)}{\partial
x}\right|_{x=B^{\mu} B_{\mu} \pm b^{2}}
\end{eqnarray}
Now following the road map of Casana et al. \cite{CASANA},  Ding
et al. obtained a Kerr-like solution \cite{DING} keeping in view
of the development of Koltz to reproduce the Kerr solution
\cite{KOLTZ}. According to the development of Koltz the
generalized form of radiating stationery axially symmetric black
hole metric can be written down as \cite{KOLTZ, DING}
\begin{eqnarray}
d s^{2}=-\gamma(\zeta, \theta) d
\tau^{2}+a[p(\zeta)-q(\theta)]\left(d \zeta^{2}+d
\theta^{2}+\frac{q}{a} d \phi^{2}\right)-2q(\theta)d\tau d\phi.
\label{METRIC}
\end{eqnarray}
where $a$ is a dimensional constant which is introduced matching
the dimension. The time $t$  and $\tau$ has the relation
\begin{eqnarray}
d \tau=d t-q d \phi.
\end{eqnarray}
In terms of $t$  Eqn. (\ref{METRIC}) turns into
\begin{eqnarray}\nonumber
&d s^{2}&=-\gamma(\zeta, \theta) d t^{2}+a[p(\zeta)
-q(\theta)]\left(d \zeta^{2}+d \theta^{2}\right) \\
&+&\left\{[1-\gamma(\zeta, \theta)] q^{2}(\theta)+p(\zeta)
q(\theta)\right\} d \phi^{2}-2 q(\theta)[1-\gamma(\zeta, \theta)]
d t d \phi.\label{METRIC1}
\end{eqnarray}
The metric ansatz (\ref{METRIC1}) is then used to compute the
gravitational field equations considering bumblebee space-like
nature of the field $b_{\mu}$:
\begin{eqnarray}
b_{\mu}=(0, b(\zeta, \theta), 0,0).
\end{eqnarray}
The space-like nature of the bumblebee was chosen ad in this
situation spacetime curvature showed up great radial variation
compared to its temporal variation. The condition
\begin{eqnarray}\nonumber
b_{\mu} b^{\mu}=b_{0}^{2}
\end{eqnarray}
where $b_{0}$ is a constant that led us to  find out $b_{\mu}$ in
the form
\begin{eqnarray}
b_{\mu}=\left(0, b_{0} \sqrt{a(p-q)}, 0,0\right). \label{FIELD}
\end{eqnarray}
in a straightforward manner. With this set up they established
that Einstein-bumblebee modified theory permitted to write the
bumblebee field as $b_{\mu}=\left(0, b_{0}\ell , 0,0\right)$ that
contained the LV parameter $\ell$ and landed onto a Kerr-like
metric which in the Boyer-Lindquist coordinate read
\begin{eqnarray}
d s^{2}=-\left(1-\frac{2 M r}{\rho^{2}}\right) d t^{2}-\frac{4 M r
a \sqrt{1+\ell} \sin ^{2} \theta}{\rho^{2}} d t d
\varphi+\frac{\rho^{2}}{\Delta} d r^{2}+\rho^{2} d
\theta^{2}+\frac{A \sin ^{2} \theta}{\rho^{2}} d \varphi^{2}.
\label{FINAL}
\end{eqnarray}
where
\begin{eqnarray}
\Delta=\frac{r^2-2 M r}{1+\ell}+a^{2}, A=\left[r^2+(1+\ell)
a^{2}\right]^{2}-\Delta(1+\ell)^{2} a^{2} \sin ^{2} \theta.
\end{eqnarray}
In the slow rotating \cite{KANZI} case, i.e. for $a^2\to 0$ the
metric has the form
\begin{eqnarray}
 d s^{2}\approx
-\left(1-\frac{2 M}{r}\right) d
t^{2}-\frac{Ma(1+l)sin^2\theta}{r}d\theta
d\varphi+\frac{1+\ell}{1-2 M / r} d r^{2}+r^{2}(d \theta^{2}+ \sin
^{2} \theta)d \varphi^{2},
\end{eqnarray}
If $\ell \rightarrow 0 $  it recovers the usual Kerr metric and
for $a \rightarrow 0$  it becomes
\begin{eqnarray}
d s^{2}=-\left(1-\frac{2 M}{r}\right) d t^{2}+\frac{1+\ell}{1-2 M
/ r} d r^{2}+r^{2} (d \theta^{2}+ \sin ^{2} \theta) d \varphi^{2},
\end{eqnarray}
\subsection{Non-commutative Kerr-like black hole}
Let us now describe in brief the incorporation of the effect of
non-commutativity. Like the Lorentz violation effect, the
non-commutativity is also believed to have the potential that
leads to significant effects on the properties of the black holes.
Therefore noncommutative spacetime in the theories gravity have
been the object of several interest \cite{NICO, ZABO}. Although an
ideal non-commutative extension of the standard theory has not yet
been available, it necessitates receiving the non-commutativity
effects in the frame of the commutative theory of standard general
relativity, in recent times, the authors in the articles \cite{
NICO, ZABO, SMA, SMA1, NICO1} made physically inspired, decent as
well as obedient non-commutativity amendments to Schwarzschild
black hole solutions. In particular, the study considering the
effects of noncommutativity on black hole physics has been an area
of huge interest. The generalization of quantum field theory by
non-commutativity based on coordinate coherent state formalism is
of interest in this respect which  cures the short distance
behavior of point-like structures \cite{SMA, SMA1,NICO1, KNOZ}. In
this approach, the particle mass $M$ is considered to be
distributed throughout a region of linear size instead of being
localized at a point. Therefore, noncommutativity is introduced by
modifying the mass density so that the Dirac delta function is
replaced by a Gaussian distribution \cite{NICO1} or alternatively
by a Lorentzian distribution \cite{KNOZ}.  The implementation of
these arguments leads to the replacement of the position
Dirac-delta function that describes a point-like structure, with
suitable function, e.g. Gaussian distribution function or
Lorentzian distribution describing smeared structures. In a real
sense, the description of matter would not be a Dirac-delta
distribution. It would be better described by a Gaussian
distribution or some other type of distribution such that it turns
into Dirac-delta function when the width of the distribution
approaches to zero.

To incorporate  the non-commutativity effect we consider that the
mass density of the black hole has a Lorentzian distribution as it
was found in article \cite{NOZARI, ANACLETO}
\begin{equation}
\rho_{b}=\frac{\sqrt{b} M}{\pi^{3 / 2}\left(\pi b+r^{2}\right)^{2}}.
\end{equation}
Here $b$ is the strength of non-commutativity of spacetime and $M$
is the total mass distributed throughout a region with a linear
size $\sqrt{b}$. For the smeared matter distribution, it can be
shown that \cite{ANACLETO}
\begin{equation}
\mathcal{M}_{b}=\int_{0}^{r} \rho_{b}(r) 4 \pi r^{2} d r =\frac{2
M}{\pi}\left(\tan ^{-1}\left(\frac{r}{\sqrt{\pi b}}\right)
-\frac{\sqrt{\pi b} r}{\pi b+r^{2}}\right) \approx -\frac{4
\sqrt{b} M}{\sqrt{\pi} r}+M+\mathcal{O}\left(b^{3 / 2}\right).
\end{equation}
It indicates that the Mass is point-like when the spread of the
distribution approaches towards a vanishing value since $\lim_{x
\to b }M_b = M $.

To combine these two effects we need a suitable framework where
tease two can fit suitably. Now, these two can be amalgamated if
we replace $M$ by $M_b$ in the expression of the spacetime metric
given in equation (\ref{FINAL}),  which is already endowed with
the LV effect. Therefore the generalized spacetime metric where
both the LV and non-commutating are bestowed simultaneously takes
the form
\begin{eqnarray}
d s^{2}=-\left(1-\frac{2 M_{b} r}{\rho^{2}}\right) d t^{2}-\frac{4
M_{b} r a \sqrt{1+l} \sin ^{2} \theta}{\rho^{2}} d t d
\varphi+\frac{\rho^{2}}{\Delta} d r^{2}+\rho^{2} d
\theta^{2}+\frac{A \sin ^{2} \theta}{\rho^{2}} d \varphi^{2},
\label{FINALL}
\end{eqnarray}
where
\begin{eqnarray}
\rho^{2}=r^{2}+(1+l) a^{2} \cos ^{2} \theta,\Delta=\frac{r^{2}
-2 M_{b} r}{1+l}+a^{2}, A=\left[r^{2}+(1+l)
a^{2}\right]^{2}-\Delta(1+l)^{2} a^{2} \sin ^{2} \theta.
\end{eqnarray}
The metric (\ref{FINALL}) carries the information of two
significant effects which have already been found needful to take
taken into account in the perspective of black physics. Although
these two are supposed to show a prominent role in the vicinity
Planck scale it is believed that these two have the ability to
lead to significant effects on the properties of the black holes
at an observable scale. The parameter $b$ is connected with the
noncommutativity of spacetime and the parameter $l$ is associated
with the Lorentz violation scenario. If the LV effect is switched
setting $l = 0$ it will render only the non-commutative effect and
the reverse will be the case if the non-commutating effect be
switched setting $b = 0$, and if both the effects are made
offsetting $l \rightarrow 0 $ and $b \rightarrow 0$, it recovers
the usual Kerr metric.
\section{Geometry concerning Horizon and ergo-sphere}
Let us  now focus to our  investigation with the metric developed
in equation (\ref{FINAL}). We get the expressions for Event
horizon and Cauchy horizon setting  $\Delta=0$,which are given by
\begin{eqnarray}
r_{\pm}=M\pm\frac{\sqrt{-\pi l a^2-\pi a^2+\pi M^2-8 \sqrt{\pi } M \sqrt{b}}}{\sqrt{\pi }},
\end{eqnarray}
where $\pm$ signs correspond to event horizon and Cauchy horizon
respectively. The event horizon and Cauchy horizon are labelled by
$r_{eh}$ and $r_{ch}$ respectively. What follows next is the
sketch of $\Delta$ for different values of $b$ and $l$

\begin{figure}[H]
\centering
\begin{subfigure}{.5\textwidth}
\centering
\includegraphics[width=.7\linewidth]{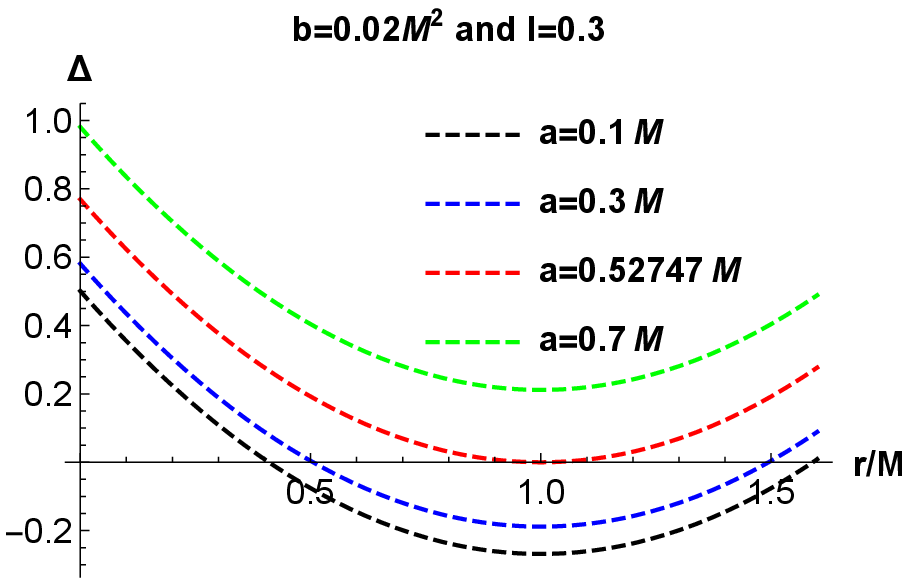}
\end{subfigure}%
\begin{subfigure}{.5\textwidth}
\centering
\includegraphics[width=.7\linewidth]{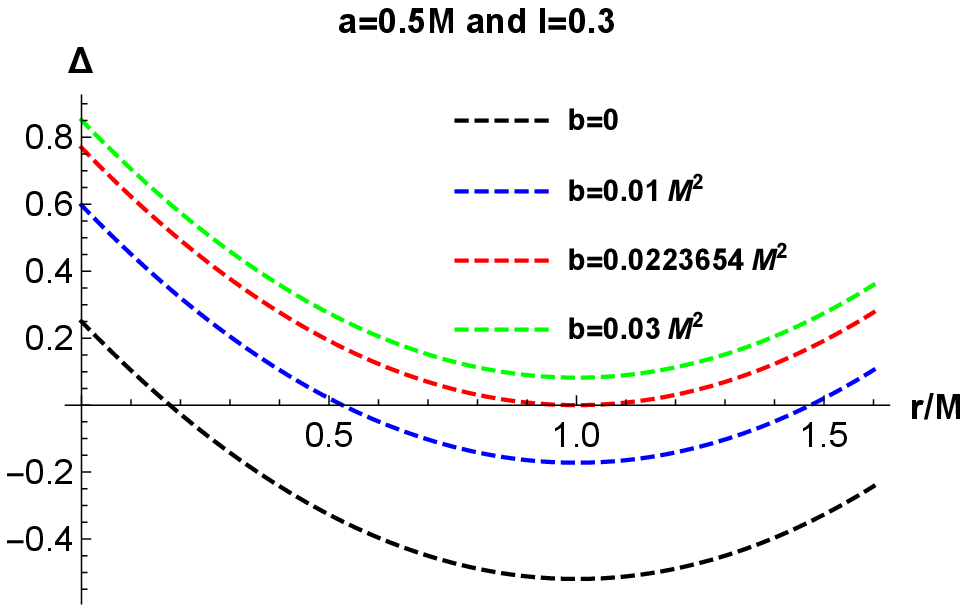}
\end{subfigure}
\caption{The left one gives variation of $\Delta$ for various
values of $a$ with $b=0.02M^{2}$ and $l=0.3$, and the right one
gives variation for various values of $b$ with $a=0.5M$ and
$l=0.3$.} \label{fig:test}
\end{figure}

\begin{figure}[H]
\centering
\begin{subfigure}{.5\textwidth}
\centering
\includegraphics[width=.7\linewidth]{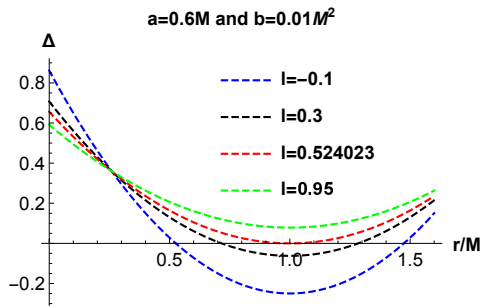}
\end{subfigure}
\caption{It gives variation of $\Delta$for various values of $l$ with $b=0.01M^{2}$ and $a=0.6$. }
\label{fig:test}
\end{figure}
From the above plots we see that there exists critical values of
$a$, for fixed values of $b$ and $l$, critical values of $b$ for
fixed values of $a,l$ and critical values of $l$ for fixed values
of $b,a$. The critical value of $a$, $b$ and $l$ are designated by
$a_c, b_c$ and $l_c$ respectively. In these cases $\Delta=0$ has
only one root. For $a < a_{c}$ we have black hole and for $a >
a_{c}$ we have naked singularity. Similarly for $b < b_{c}$ we
have black hole, but for $b > b_{c}$ we have naked singularity and
for $l < l_{c}$ signifies the black hole, however $l > l_{c}$
represents the naked singularity. Numerical computation shows that
we have $a_{c}=.52747M$ for $b=.02M^{2}$ and $l=.3$. Similarly for
$a=.5M$ and $l=.3$ we
have $b_{c}=.0223654M^{2}$, and for $a=.6M$ and $b=.01M^{2}$ we
find $l_{c}=.524023$.

There exists a black hole when the following inequality is
maintained
\begin{equation}
-\pi l a^2-\pi a^2+\pi M^2-8 \sqrt{\pi } M \sqrt{b} \geq 0,
\label{CON}
\end{equation}
However when in the equation (\ref{CON}) equality is maintained it
corresponds to extremal black holes, and when the equation
(\ref{CON}) is strictly greater than $0$ we have non-extremal
black holes which have both the Cauchy and Event horizons.
\begin{figure}[H]
\centering
\begin{subfigure}{.3\textwidth}
\centering
\includegraphics[width=.95\linewidth]{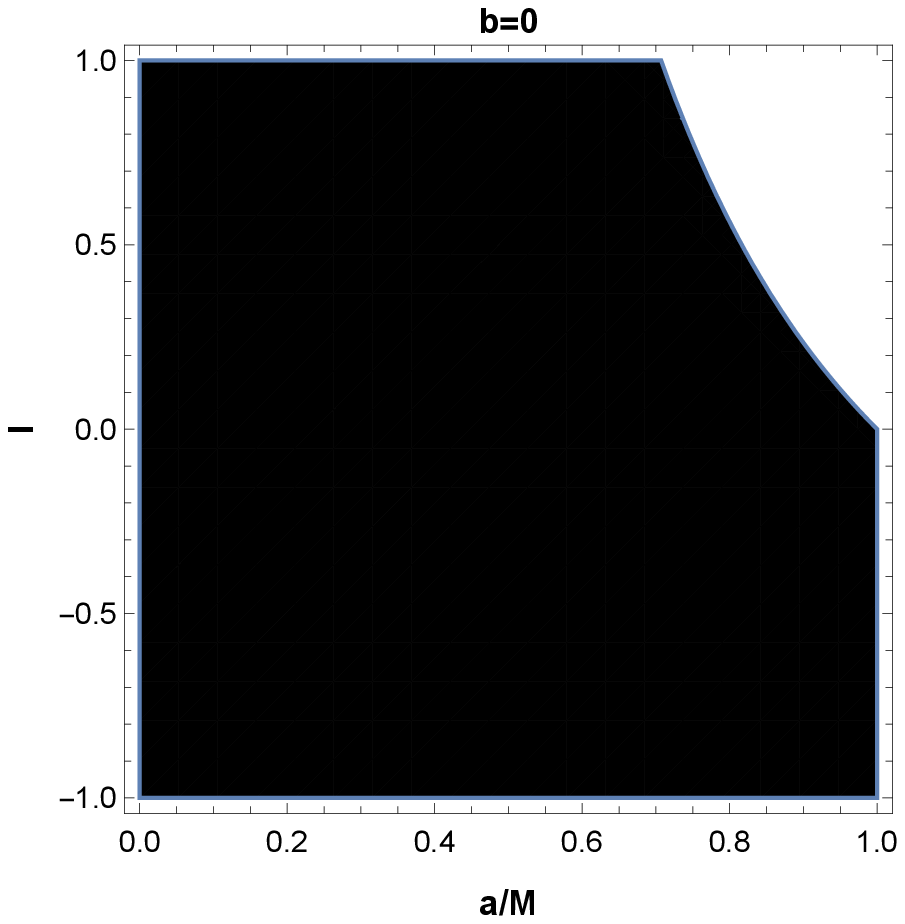}
\end{subfigure}%
\begin{subfigure}{.3\textwidth}
\centering
\includegraphics[width=.95\linewidth]{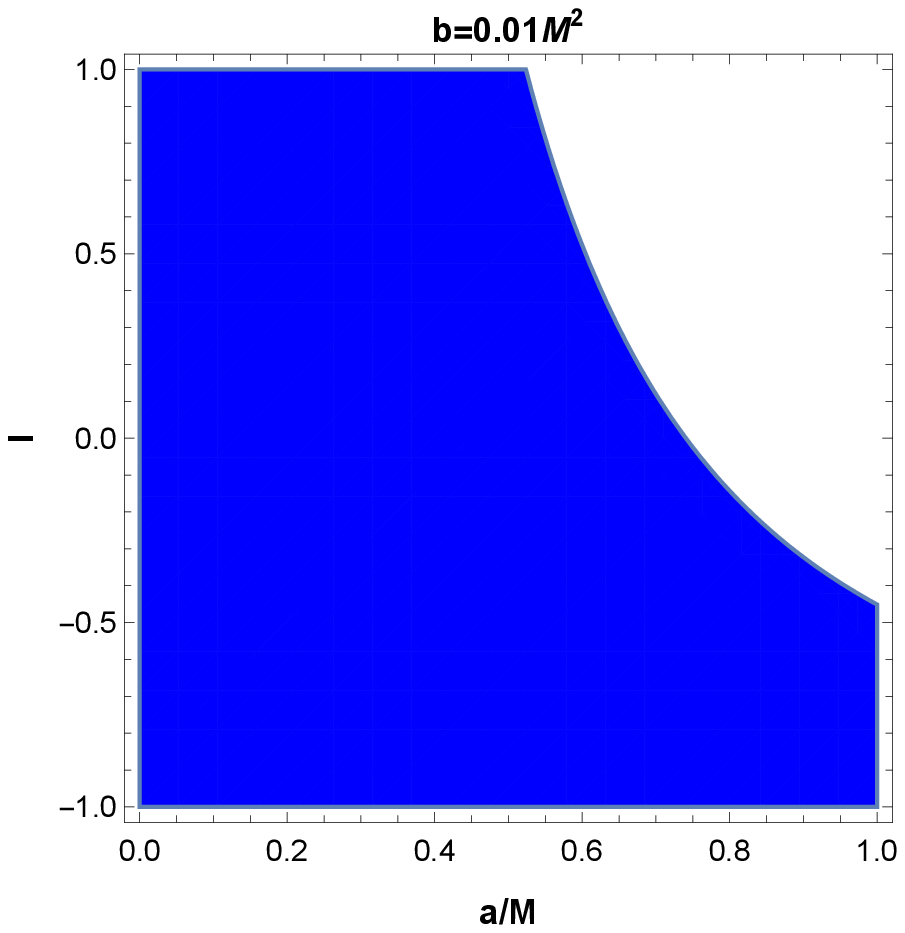}
\end{subfigure}%
\begin{subfigure}{.3\textwidth}
\centering
\includegraphics[width=.95\linewidth]{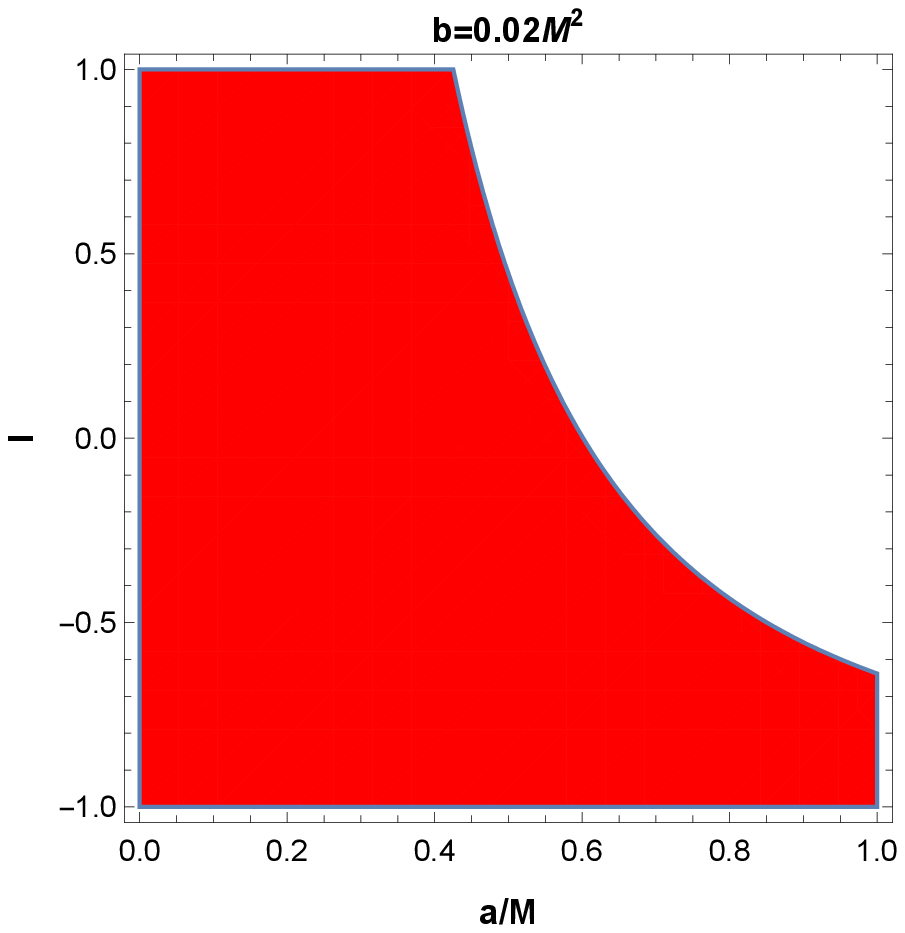}
\end{subfigure}
\par\smallskip
\begin{subfigure}{.3\textwidth}
\centering
\includegraphics[width=.95\linewidth]{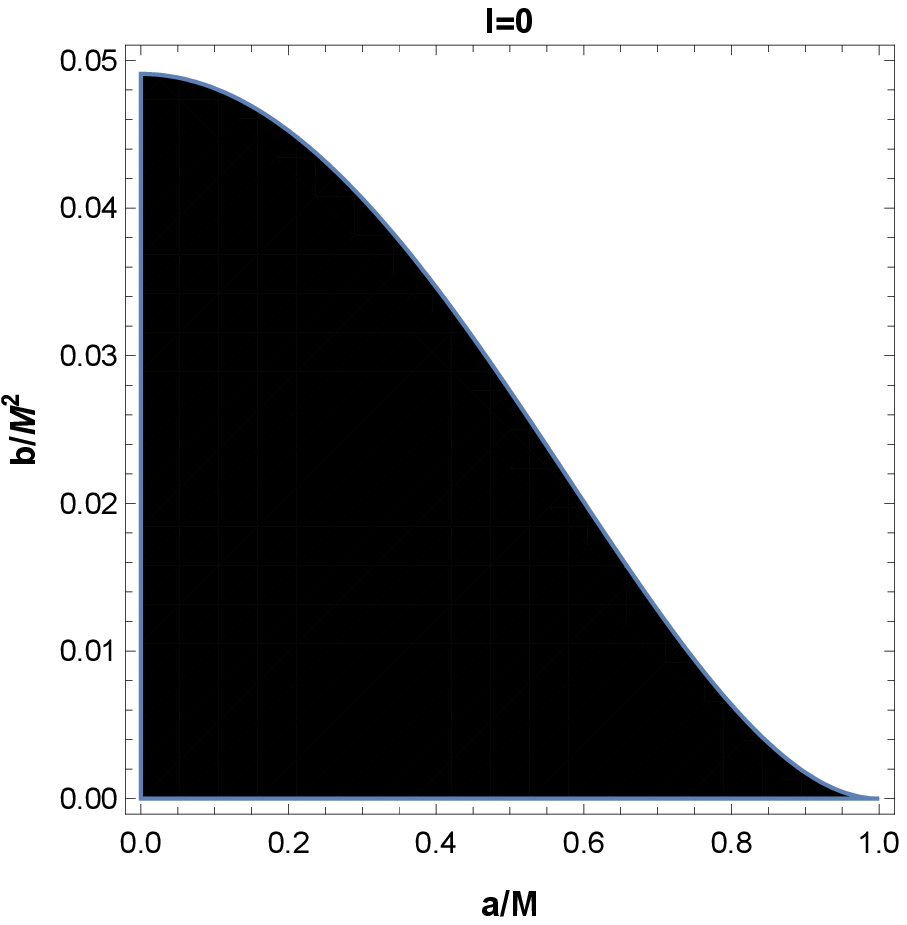}
\end{subfigure}%
\begin{subfigure}{.3\textwidth}
\centering
\includegraphics[width=.95\linewidth]{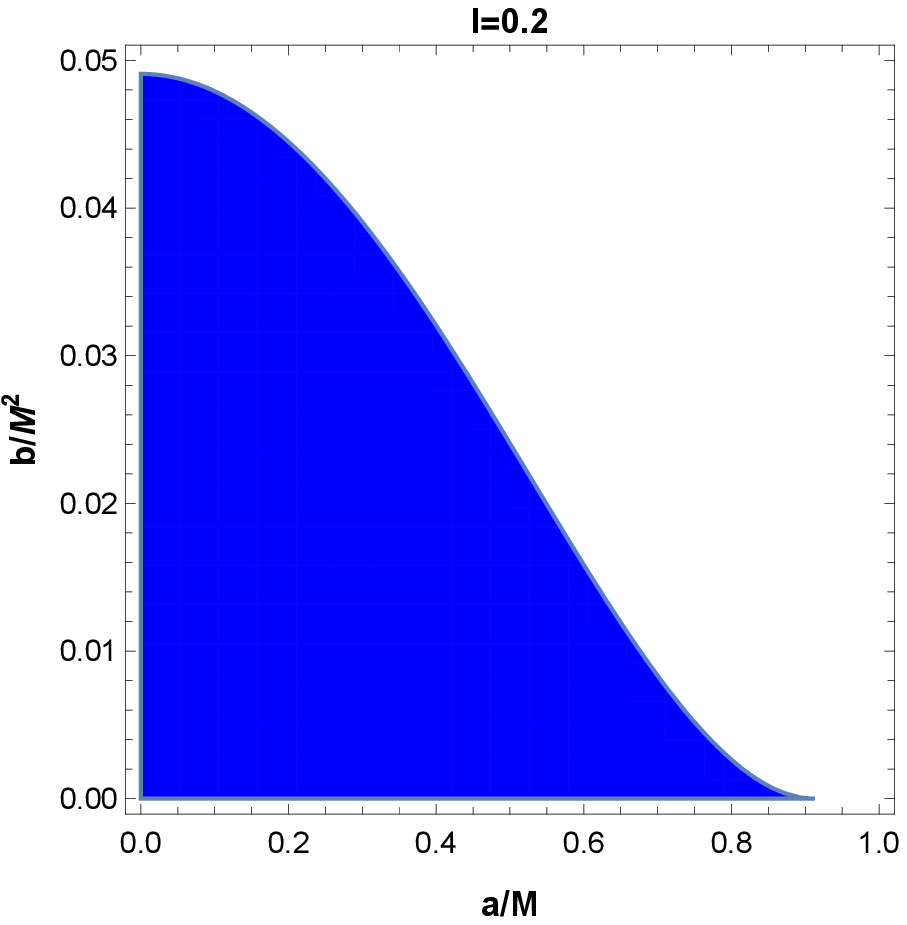}
\end{subfigure}%
\begin{subfigure}{.3\textwidth}
\centering
\includegraphics[width=.95\linewidth]{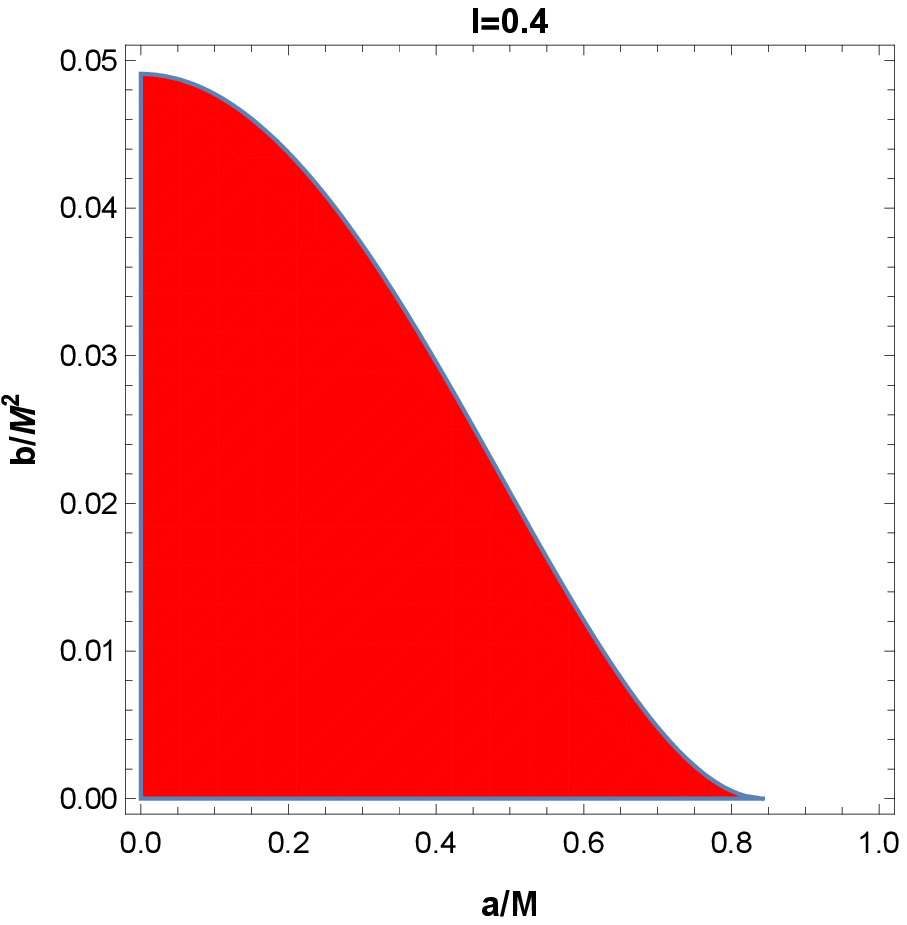}
\end{subfigure}
\caption{The upper one is the parameter space $(a/M-l)$ for
various values of $b$ and the lower one is the parameter space
$(a/M-b/M^{2})$ for various values of $l$. The colored regions
correspond to parameter space for which we have a black hole.}
\label{fig:region}
\end{figure}
From the above plots, we observe that as $l$ increases the
parameter space $(a/M - b/M^{2})$ for which we have black hole
gets shrunk and as $b$ increases the parameter space $(a/M - l)$
for which we have black hole also reduces.

Let us now focus on the static limit surface (SLS). At the SLS,
the asymptotic time-translational Killing vector becomes null
which is mathematically given by
\begin{equation}
g_{tt}=\rho^{2}-2M_{b}r=0.
\end{equation}
The real positive solutions of the above equation give radial
coordinates of ergosphere:
\begin{equation}
r^{ergo}_{\pm}=\frac{2 \sqrt{\pi } M\pm\sqrt{-4 \pi a^2 \cos ^2(\theta )
-4 \pi a^2 l \cos ^2(\theta )+4 \pi M^2-32 \sqrt{\pi } M \sqrt{b}}}{2 \sqrt{\pi }}.
\end{equation}

\begin{figure}[H]
\centering
\begin{subfigure}{.3\textwidth}
\centering
\includegraphics[width=.95\linewidth]{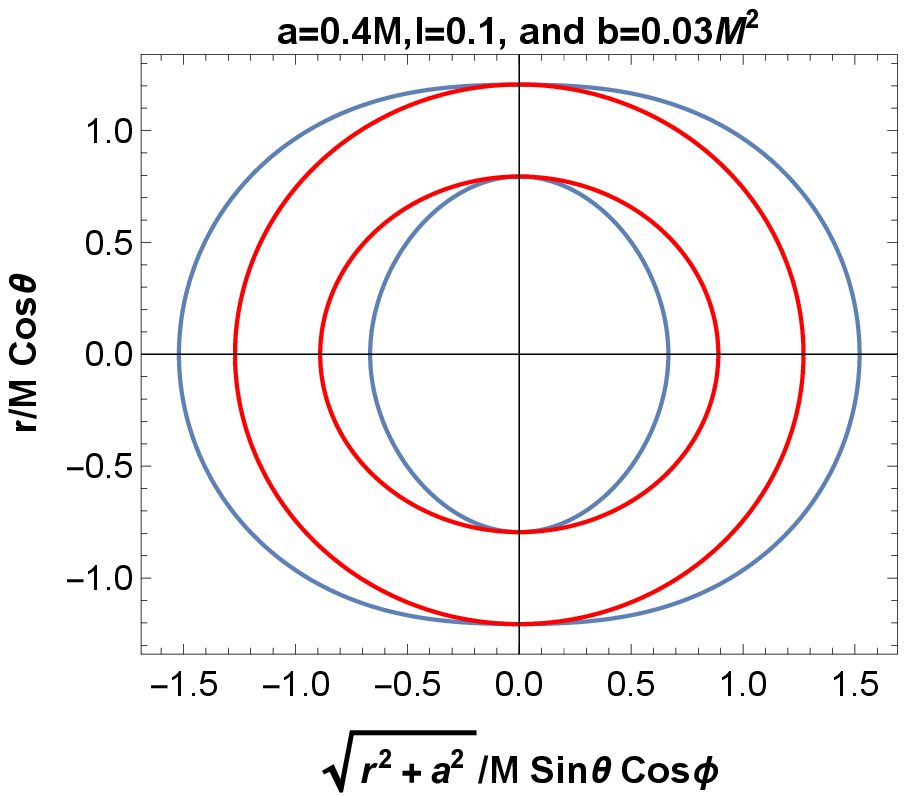}
\end{subfigure}%
\begin{subfigure}{.3\textwidth}
\centering
\includegraphics[width=.95\linewidth]{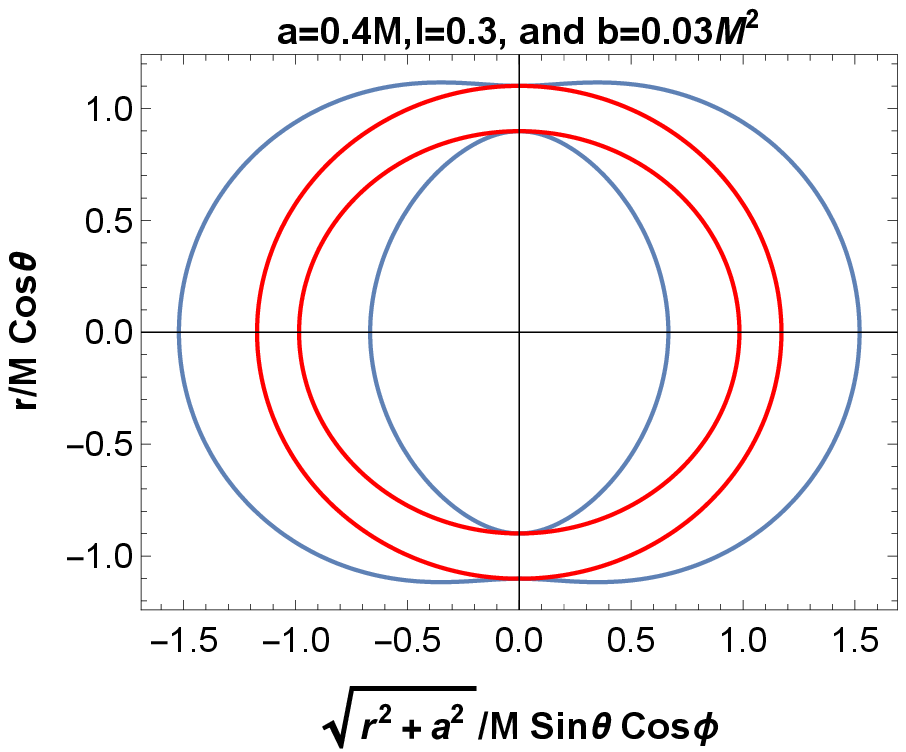}
\end{subfigure}%
\begin{subfigure}{.3\textwidth}
\centering
\includegraphics[width=.95\linewidth]{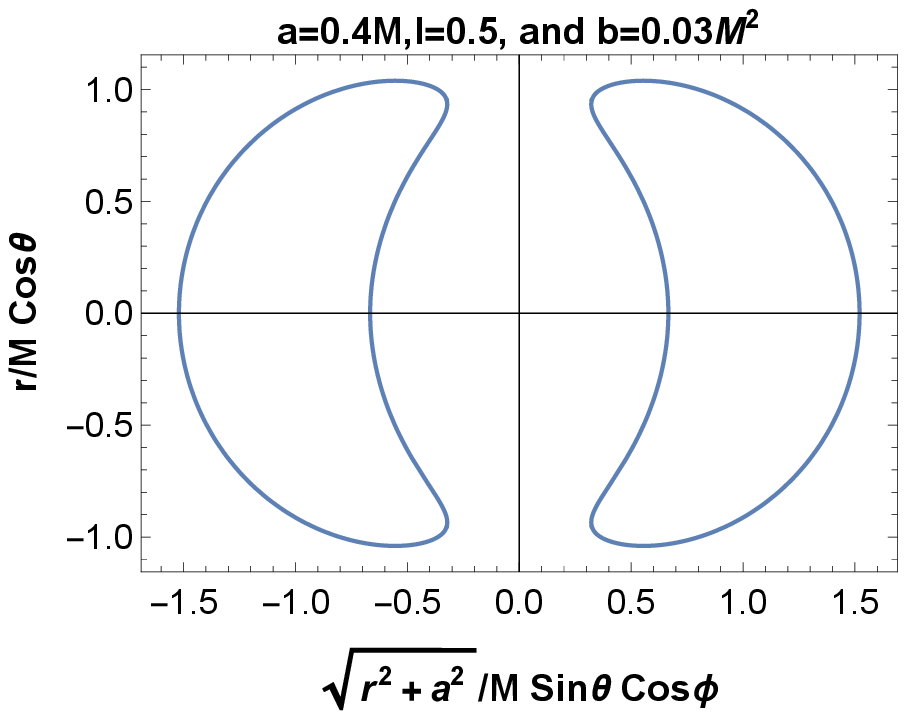}
\end{subfigure}
\par\smallskip
\begin{subfigure}{.3\textwidth}
\centering
\includegraphics[width=.95\linewidth]{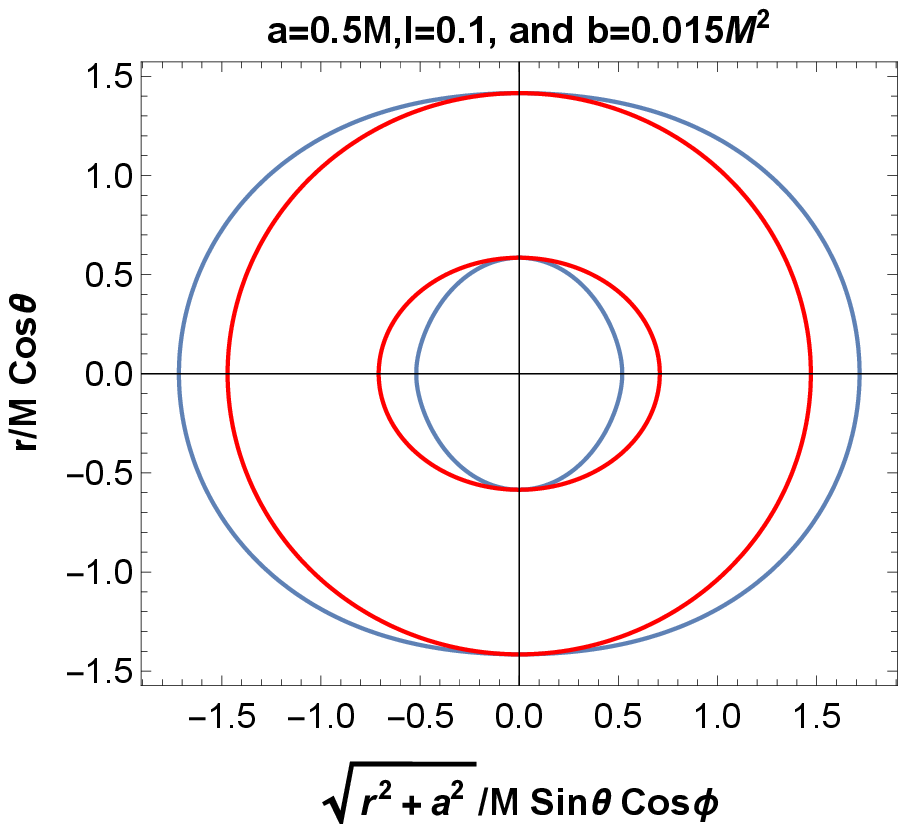}
\end{subfigure}%
\begin{subfigure}{.3\textwidth}
\centering
\includegraphics[width=.95\linewidth]{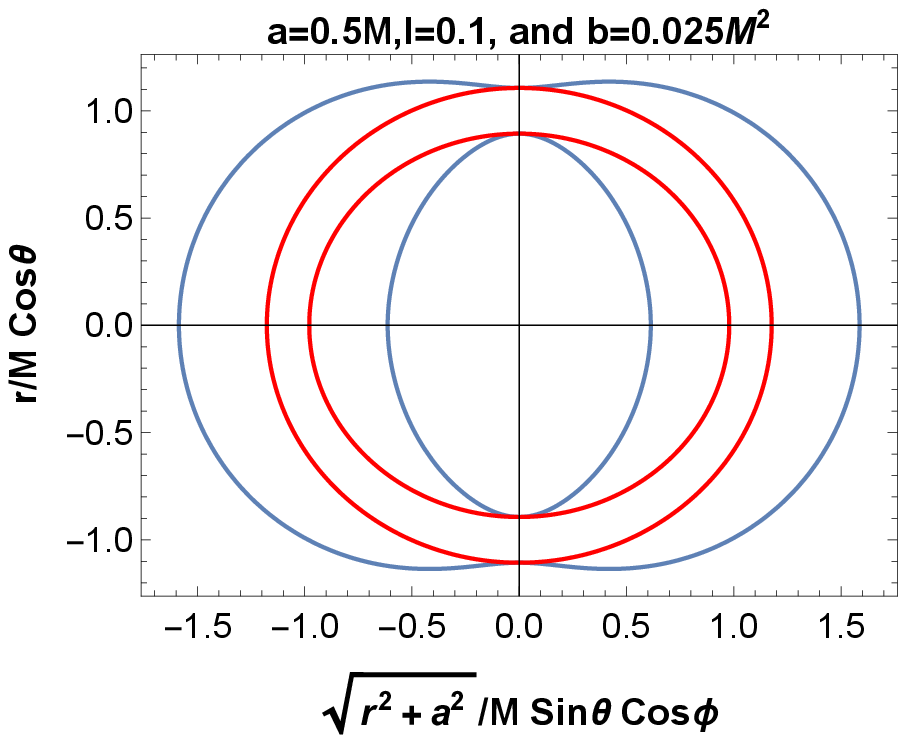}
\end{subfigure}%
\begin{subfigure}{.3\textwidth}
\centering
\includegraphics[width=.95\linewidth]{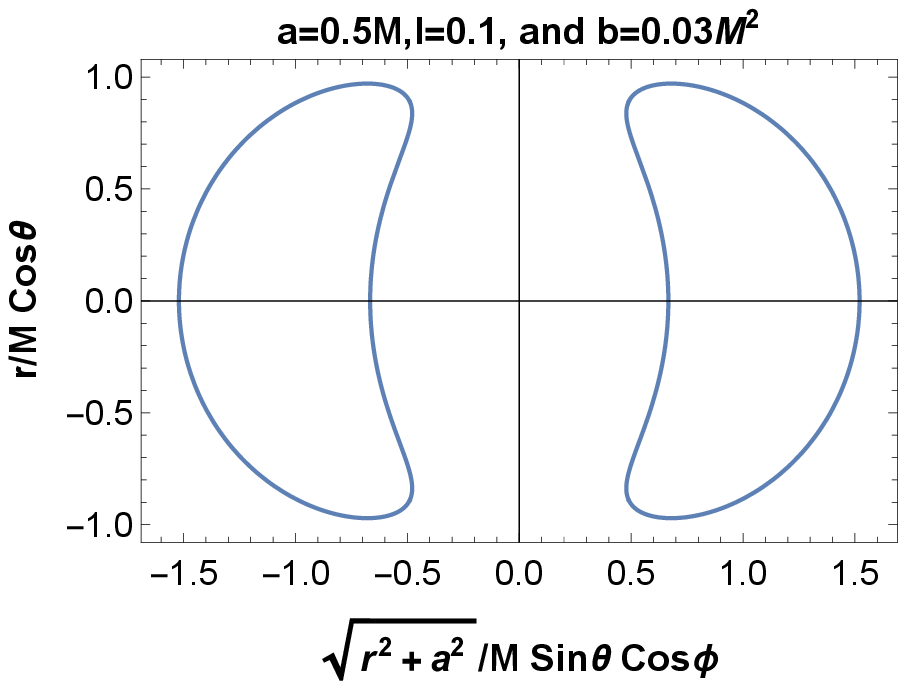}
\end{subfigure}
\caption{The cross-section of the event horizon (outer red line),
SLS (outer blue dotted line), and ergoregion of non-commutative
Kerr-like black holes.} \label{fig:ergo}
\end{figure}
The ergosphere, which lies between SLS and the event horizon, is
depicted above. Energy can be extracted from ergosphere
\cite{PENROSE}. From the above, we can conclude that the shape and
size of the ergosphere depend on rotational parameter $a$,
non-commutative parameter $b$, and LV parameter $l$. The size of
the ergosphere increases with the increase of $b$ and $l$. This
completes our discussion concerning the geometrical structure of
this spacetime. Let us now turn toward our main part of the
investigation which is connected to the optical properties in the
viscidity of this black hole. To this end let us first consider
the superradiance phenomena.

\section{Superradiance scattering of the scalar field off non-commutative
Kerr-like black hole} We bring the Klein-Gordon equation to curved
spacetime into action to study the superradiance scattering of a
scalar field $\Phi$.
\begin{eqnarray}
\left(\bigtriangledown_{\alpha}\bigtriangledown^{\alpha}+\mu^{2}\right)\Phi(t,r,\theta,\phi)
= \left[\frac{-1}{\sqrt{-g}}\partial_{\sigma}\left(g^{\sigma
\tau}\sqrt{-g}\partial_{\tau}\right)+\mu^{2}\right]\Phi(t,r,\theta,\phi)&=&0.
\label{KG}
\end{eqnarray}
Here $\mu$  represents the mass of the scalar field $\Phi$. We now
adopt the standard separation of variables method to the equation
Eqn.(\ref{KG}) in order  to separate it into radial and angular
part using the following ansatz. With the standard Boyer-Lindquist
coordinates $(t, r, \theta, \phi)$ we can write
\begin{eqnarray}
\Phi(t, r, \theta, \phi)=R_{\omega j m}(r) \Theta(\theta) e^{-i
\omega t} e^{i m \phi}, \quad j \geq 0, \quad-j \leq m \leq j,
\quad \omega>0, \label{PHI}
\end{eqnarray}
where $R_{\omega j m}(r)$ represents the radial function and
$\Theta(\theta)$ refers to the oblate spheroidal wave function.
The symbols $j$, $m$, and $\omega$ respectively stand for the
angular eigenfunction, angular quantum number, and the positive
frequency of the field under investigation as viewed by a far away
observer. Using the ansatz (\ref{PHI}), the differential equation
(\ref{KG}) is found to get separated into the followingtwo
ordinary differential equations. For radial part the equation
reads
\begin{eqnarray}
&&\frac{d}{d r}(\Delta \frac{d R_{\omega j m}(r)}{d
r})+(\frac{((r^2+a^{2}(1+\ell)) \omega-am\sqrt{1+\ell}
)^{2}}{\Delta(1+\ell)})R_{\omega l m}(r)
 \nonumber \\
&&-(\mu^{2} r^2+j(j+1)+a^{2}(1+\ell) \omega^{2}-2 m \omega
a\sqrt{1+\ell}) R_{\omega l m}(r)=0, \label{RE}
\end{eqnarray}
and the angular part of it is
\begin{eqnarray}
&&\sin \theta \frac{d}{d \theta}\left(\sin \theta \frac{d
\Theta_{\omega j m}(\theta)}{d \theta}\right)+\left(j(j+1) \sin
^{2} \theta-\left(\left(a\sqrt{1+\ell} \omega \sin ^{2}
\theta-m\right)^{2}\right)\right)\Theta_{\omega j m}(\theta)\nonumber \\
&& + a^{2}(1+\ell) \mu^{2} \sin ^{2} \theta \cos ^{2} \theta~
\Theta_{\omega j m}(\theta)=0.
\end{eqnarray}
Following the earlier investigation \cite{BEZERRA, KRANIOTIS} we
may have a general solution of the radial equation (\ref{RE}).
However, we are intended to study the scattering of the field
$\Phi$. So in this situation, we have used the asymptotic matching
procedure used in \cite{STRO1, STRO2, TEUK, PAGE, RAN}. The road
map of the important contributions \cite{STRO1, STRO2, TEUK, PAGE,
MK, RAN}  led us reach to the required result without using the
general solution.  First of all we consider  the radial part of
the equation (\ref{RE}) to find an asymptotic solution. Use of
Regge-Wheeler-like coordinate $r_{*}$ is helpful in this situation
in order to deal with the radial equation as per our requirement,
which is given by
\begin{eqnarray}
r_{*} \equiv \int d r \frac{r^2+a^{2}(1+\ell)}{\Delta},
\quad\left(r_{*} \rightarrow-\infty \quad \text{at event horizon},
\quad r_{*} \rightarrow \infty \quad \text{at infinity} \right).
\end{eqnarray}
To have the equation into the desired shape, we take on a new
radial function $\mathcal{R}_{\omega j
m}\left(r_{*}\right)=\sqrt{r^2+a^{2}(1+\ell)} R_{\omega j m}(r)$.
A few steps of  algebra, leads us to obtain the radial equation
with our desired form.
\begin{equation}
\frac{d^{2} \mathcal{R}_{\omega l m}\left(r_{*}\right)}{d
r_{*}^{2}}+V_{\omega j m}(r) \mathcal{R}_{\omega j
m}\left(r_{*}\right)=0. \label{RE1}
\end{equation}
An effective potential takes entry into the picture now and that
has the crucial role on the scattering that reads
\begin{eqnarray}
%\begin{split}
V_{\omega j m}(r)&=&\frac{1}{1+\ell}\left(\omega-\frac{m
\hat{a}}{r^2+\hat{a}^{2}}\right)^{2}-\frac{\Delta}{\left(r^2
+\hat{a}^{2}\right)^{2}}\left[\right. j(j+1)+\hat{a}^{2}
\omega^{2}-2 m \hat{a} \omega+\mu^{2} r^2
\\\nonumber
&&\left.+\sqrt{r^2+\hat{a}^{2}}\frac{d}{dr}\left(\frac{r\Delta
}{\left(r^2+\hat{a}^{2}\right)^{\frac{3}{2}}}\right)\right],
\label{POT}
%\end{split}
\end{eqnarray}
where $\hat{a}=a(1+\ell)^{\frac{1}{2}}$. So it appears that it is
equivalent to  the study of the scattering of the scalar field
$\Phi$ under this effective potential (\ref{POT}). It is
beneficial to study the asymptotic behavior of the scattering
potential at the event horizon and at spatial infinity in this
regard. The potential at the event horizon in the asymptotic limit
simplifies  into
\begin{eqnarray}
\lim _{r \rightarrow r_{eh}} V_{\omega j
m}(r)=\frac{1}{1+\ell}\left(\omega-m \hat{\Omega}_{h}\right)^{2}
\equiv k_{e h}^{2}.
\end{eqnarray}
and the same at spatial infinity turns into the following after a
few step of algebra
\begin{equation}
\lim _{r \rightarrow \infty} V_{\omega j m}(r)=\frac{\omega^{2}}{1+\ell}-\lim _{r
\rightarrow \infty} \frac{\mu^{2} r^2
\Delta}{\left(r^2+\tilde{a}^{2}\right)^{2}}=\frac{\omega^{2}}{1+\ell}
-\hat{\mu}^{2}\equiv k_{\infty}^{2}, ~~~
\hat{\mu}=\frac{\mu}{\sqrt{\ell+1}}.
\end{equation}
Note that the potential turns into a constant at the two extremal
points namely at event horizon and  at spatial infinity, however,
the numerical values of the constants are different indeed at the
two extremal points.

Since the behavior of the potential at the two extremal points are
known, we now move to observe the asymptotic behavior of the
radial solution. A little algebra shows that the radial equation
(\ref{RE1}) has the following
 solutions
\begin{equation}\label{AS}
R_{\omega j m}(r) \rightarrow\left\{\begin{array}{cl}
\frac{\mathcal{D}_{i n}^{eh} e^{-i k_{eh} r_{*}}}{\sqrt{r_{e h}^2
+\hat{a}^{2}}} & \text { for } r \rightarrow r_{e h} \\
\mathcal{D}_{i n}^{\infty} \frac{e^{-i k_{\infty}
r_{*}}}{r}+\mathcal{D}_{r e f}^{\infty} \frac{e^{i k_{\infty}
r_{*}}}{r} & \text { for } r \rightarrow \infty
\end{array}\right\}
\end{equation}
in the asymptotic limit.  Here $\mathcal{D}_{in}^{eh}$ be the
amplitude of the incoming scalar wave at event horizon($r_{eh}$),
and $\mathcal{D}_{in}^{\infty}$ is the corresponding quantity of
the incoming scalar wave at infinity $(\infty)$. The amplitude of
the reflected part of  scalar wave at infinity $(\infty)$ is
designated by $\mathcal{D}_{ref}^{\infty}$. So the stage is set to
compute  the Wronskian for the region adjacent to the event
horizon and at infinity. The Wronskian for the event horizon is
found out to be
\begin{equation}
W_{eh}=\left(R_{\omega j m}^{e h} \frac{d R_{\omega j m}^{* e
h}}{d r_{*}}-R_{\omega j m}^{* eh} \frac{d R_{\omega j m}^{eh}}{d
r_{*}}\right),
\end{equation}
and the Wronskian at infinity results out to
\begin{equation}
W_{\infty}=\left(R_{\omega j m}^{\infty} \frac{d R_{\omega j m}^{*
\infty}}{d r_{*}}-R_{\omega j m}^{* \infty} \frac{d R_{\omega j
m}^{\infty}}{d r_{*}}\right).
\end{equation}
The knowledge of standard theory of ordinary differential equation
provides  the information  that the Wronskian corresponding to the
solutions will be independent of $r^*$ since the solution are
linearly independent. Thus, the Wronskian evaluated at horizon is
compatible to equate with the Wronskian evaluated at infinity. In
fact, in the physical sense, it is  associated with  the flux
conservation of the process \cite{REVIEW}.  From the matching
condition an important relation between the amplitudes of incoming
and reflected waves at different regions of interest results.
\begin{equation}
\left|\mathcal{D}_{r e f}^{\infty}\right|^{2}=\left|\mathcal{D}_{i
n}^{\infty}\right|^{2}-\frac{k_{e
h}}{k_{\infty}}\left|\mathcal{D}_{i n}^{e h}\right|^{2}.
\label{AMP}
\end{equation}
A careful look reveals that if $\frac{k_{e h}}{k_{\infty}} <0$
i.e., $\omega<m \hat{\Omega}_{e h}$, the scalar wave will be
superradiantly amplified, since the relation $\left|\mathcal{D}_{r
e f}^{\infty}\right|^{2}>\left|\mathcal{D}_{i
n}^{\infty}\right|^{2}$  holds explicitly in this situation.
\subsection{Amplification factor $Z_{jm}$ for superradiance} 
\begin{figure}[H]
\centering
\begin{subfigure}{.58\textwidth}
%\centering
  \includegraphics[width=.9\linewidth]{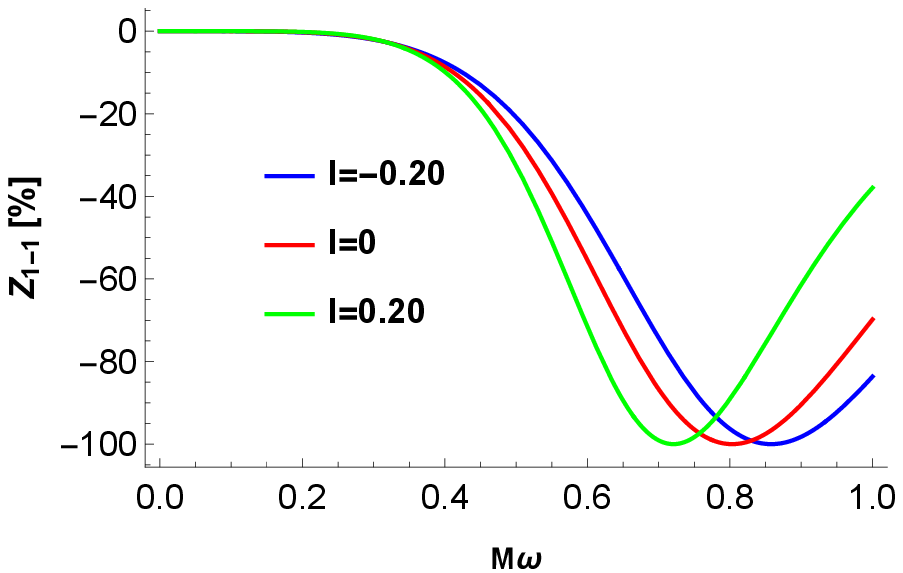}
%\caption{Critical radius for various values of $b$ with $ l=.1,k=.1$ and $\theta=\pi/2$}\hspace{1em}%
\end{subfigure}%
\begin{subfigure}{.58\textwidth}
%\centering
  \includegraphics[width=.8\linewidth]{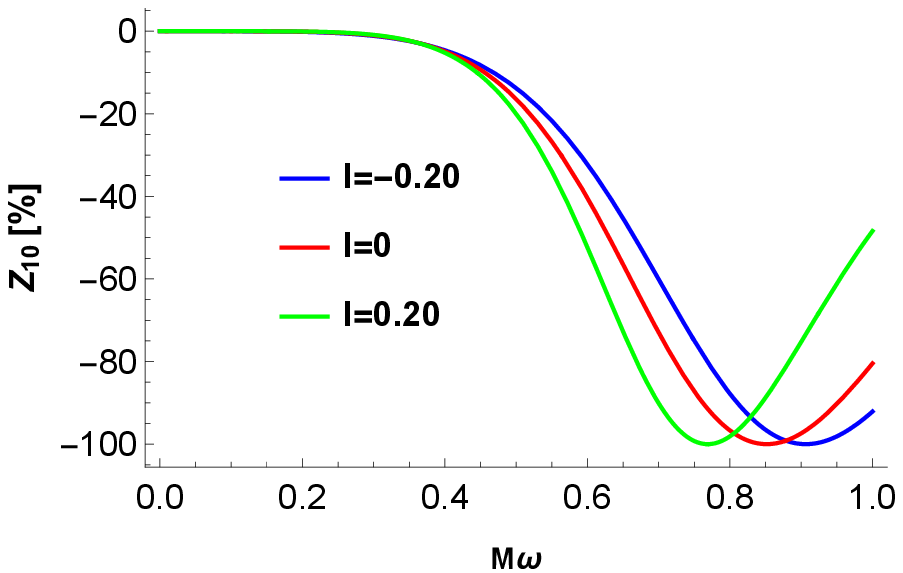}
%\caption{Critical radius for various values of $k$ with $ l=.1,b=.1$ and $\theta=\pi/2$}\hfill
\end{subfigure}
\caption{Variation of amplification factors with $\ell$ for
non-superradiant multipoles with $\hat{\mu}=0.1, b=0.01M^2$, and
$\hat{a}=0.3M$.} \label{non-superradiant}
\end{figure}

\begin{figure}[H]
\centering
\begin{subfigure}{.58\textwidth}
%\centering
  \includegraphics[width=.8\linewidth]{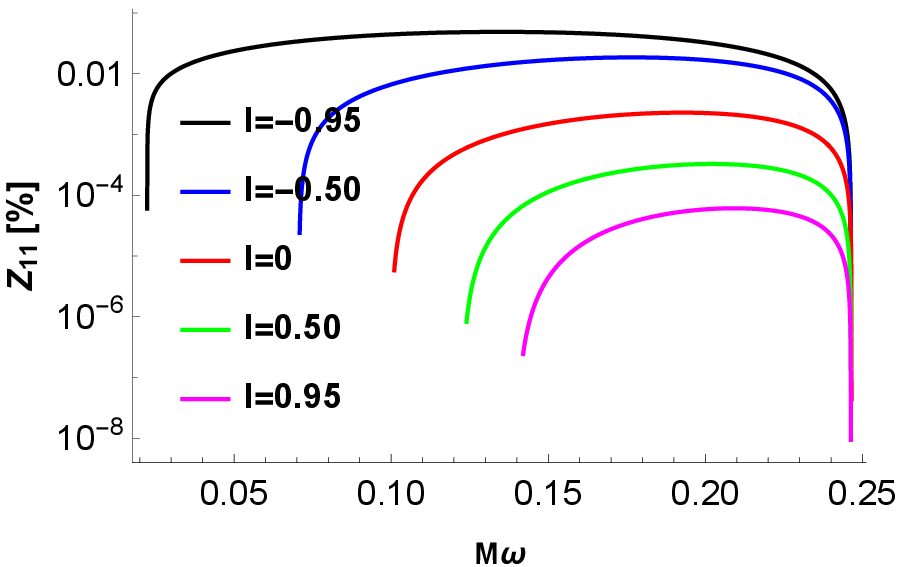}
%\caption{Critical radius for various values of $b$ with $ l=.1,k=.1$ and $\theta=\pi/2$}
\end{subfigure}%
\begin{subfigure}{.58\textwidth}
 %\centering
  \includegraphics[width=.8\linewidth]{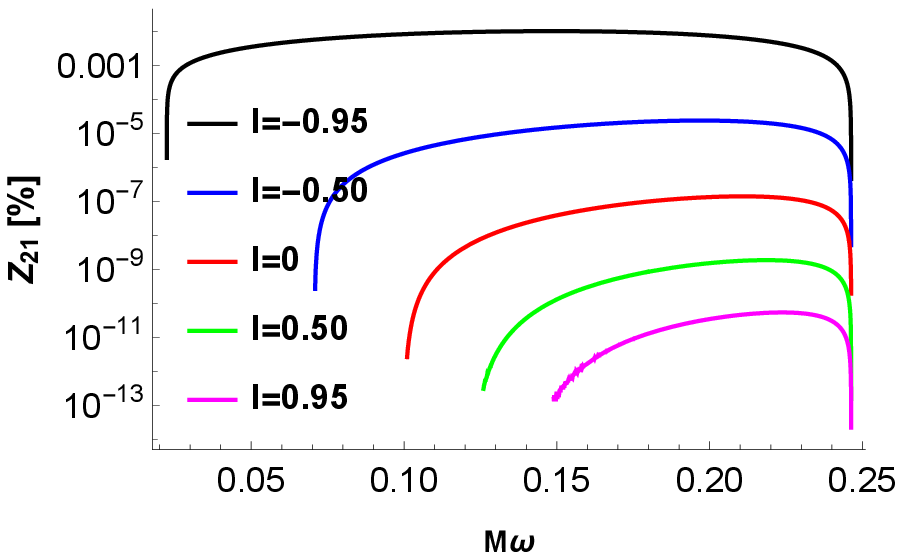}
%\caption{Critical radius for various values of $k$ with $ l=.1,b=.1$ and $\theta=\pi/2$}
\end{subfigure}
\begin{subfigure}{.58\textwidth}
\centering
  \includegraphics[width=.8\linewidth]{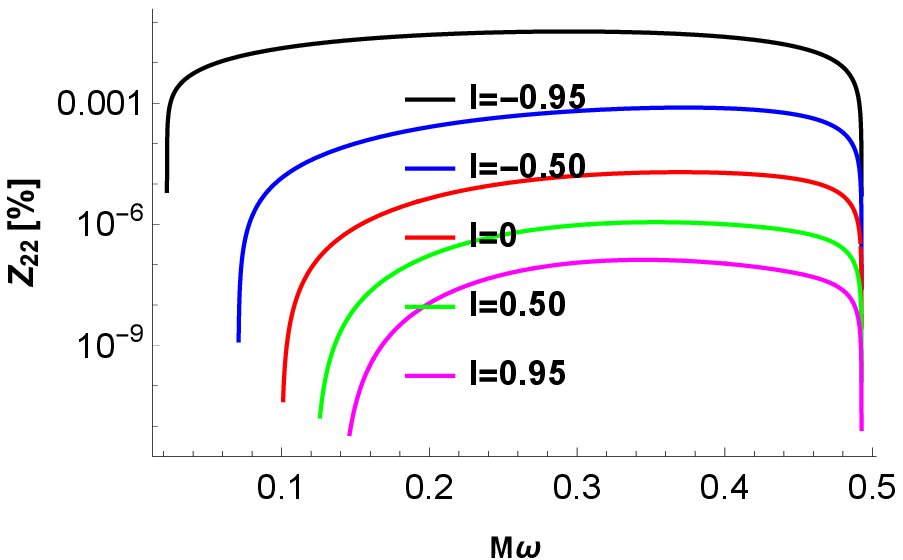}
%\caption{Critical radius for various values of $l$ with $k=.1,b=.1$ and $\theta=\pi/2$}
\end{subfigure}
\caption{Variation of amplification factors with $\ell$ for
various multipoles with $\hat{\mu}=0.1, b=0.02M^2$, and $\hat{a}=0.5M$.}
\label{zl}
\end{figure}

It is
straightforward to express the radial equation (\ref{RE}) in the
following form
\begin{eqnarray}\nonumber
&&\Delta^{2} \frac{d^{2} R_{\omega j m}(r)}{d r^{2}}+\Delta
\frac{d \Delta}{d r} \cdot \frac{d R_{\omega j m}(r)}{d r}\\
 &&+\left(\frac{\left(\left(r^2+\hat{a}^{2}\right)
\omega-\hat{a} m\right)^{2}}{1+\ell}-\Delta\left(\mu^{2}
r^2+j(j+1)+\hat{a}^{2} \omega^{2}-2 m \hat{a} \omega\right)\right)
R_{\omega j m}(r)=0. \label{RE2}
\end{eqnarray}
We now proceed to derive the solution for the near and the far
region and try to find out a single solution by matching the
solution for near-region  at infinitely with solution for the
far-region at its initial point such that this specific single
solution be useful in the vicinity of the cardinal region. it is
beneficial at this stage to introduce a  a new variable $y$ which
is defined by $y=\frac{r-r_{eh}}{r_{eh}-r_{ch}}$. In terms of $y$
the equation (\ref{RE2})  turns into
\begin{eqnarray}
&&\frac{y^{2}(y+1)^{2}}{(\ell+1)^{2}} \frac{\mathrm{d}^{2}
R_{\omega j m}(y)}{\mathrm{d} y^{2}}+\frac{y(y+1)(2
y+1)}{(\ell+1)^{2}} \frac{\mathrm{d} R_{\omega j m}(y)}{\mathrm{d}
y} \\\nonumber &&+\left(\frac{\mathcal{Q}^2
y^{4}}{1+\ell}+\frac{B^{2}}{1+\ell}-\frac{j(j+1)}{\ell+1}
y(y+1)-\frac{\hat{\mu}^{2} \mathcal{Q}^{2}}{\omega^{2}}
y^{3}(y+1)-\hat{\mu}^{2} r_{e h}^{2} y(y+1)-\frac{2 \hat{\mu}^{2}
r_{e h} \mathcal{Q}}{\omega} y^{2}(y+1) \right)R_{\omega j m}(y)=0,
\end{eqnarray}
Under under the approximation $\hat{a} \omega \ll 1$,  where
$\mathcal{Q}=\left(r_{e h}-r_{c h}\right) \omega$ and
$B=\frac{(\omega-m \hat{\Omega})}{r_{e h}-r_{c h}} r_{e h}^{2}$.
For the near-region, we have  $P y \ll 1$ and $\hat{\mu}^{2} r_{e
h}^{2} \ll 1$.  The above equation is simplified into
\begin{eqnarray}
y^{2}(y+1)^{2} \frac{\mathrm{d}^{2} R_{\omega j m}(y)}{\mathrm{d}
y^{2}}+y(y+1)(2 y+1) \frac{\mathrm{d} R_{\omega j
m}(y)}{\mathrm{d} y}+\left((\ell+1)B^{2}-j(j+1)(\ell+1)
y(y+1)\right) R_{\omega j m}(r)=0.
\end{eqnarray}
Since the Compton wavelength of the boson participating in the
scattering process is much smaller than the size of the black hole
the approximation $\left(\hat{\mu}^{2} r_{e h}^{2} \ll 1\right)$
holds good.  The general solution of the above equation in terms
of associated Legendre function of the first kind
$P_{\lambda}^{\nu}(y)$ can be written down as
\begin{eqnarray}
R_{\omega j m}(y)=d
P^{2i\sqrt{1+\ell}B}_{\frac{\sqrt{1+4j(j+1)
(l+1)}-1}{2}}(1+2y).
\end{eqnarray}
If we now  use the relation
\begin{equation}
P_{\lambda}^{\nu}(z)=\frac{1}{\Gamma(1-\nu)}\left(\frac{1+z}{1-z}\right)^{\nu
/ 2}{ }_{2} F_{1}\left(-\lambda, \lambda+1 ; 1-\nu ;
\frac{1-z}{2}\right).
\end{equation}
it enables us to  express $R_{\omega j m}(y)$ in terms of the
ordinary hypergeometric functions ${ }_{2} F_{1}(a, b ; c ; z)$ :
\begin{equation}
R_{\omega j m}(y)=d\left(\frac{y}{y+1}\right)^{-i\sqrt{\ell+1} B}{
}_{2} F_{1}\left(\frac{1-\sqrt{1+4(\ell+1) j(j+1)}}{2},
\frac{1+\sqrt{1+4(\ell+1) j(j+1)}}{2} ; 1-2 i\sqrt{\ell+1} B
;-y\right).\label{NEAR}
\end{equation}
We are intended to find out a single solution using the matching
condition at the desired position where the two solutions mingle
with each other. In this respect, we need to observe the large $y$
behavior of the above expression. The Eqn. (\ref{NEAR}) for large
y, i.e.,$(y \to\infty$) reduces to
\begin{eqnarray}
R_{\text {near-large } y} \sim d &&\left(\frac{\Gamma(\sqrt{1+4(\ell+1) j(j+1)})
 \Gamma(1-2 i\sqrt{\ell+1} B)}{\Gamma\left(\frac{1+\sqrt{1+4(\ell+1) j(j+1)}}{2}
 -2 i\sqrt{\ell+1} B\right) \Gamma\left(\frac{1+\sqrt{1+4(\ell+1) j(j+1)}}{2}\right)}
 y^{\frac{\sqrt{1+4(\ell+1) j(j+1)}-1}{2}}+\right.\\
&& \frac{\Gamma(-\sqrt{1+4(\ell+1) j(j+1)}) \Gamma(1-2
i\sqrt{\ell+1} B)}{\Gamma\left(\frac{1-\sqrt{1+4(\ell+1)
j(j+1)}}{2}\right) \Gamma\left(\frac{1-\sqrt{1+4(\ell+1)
j(j+1)}}{2}-2 i\sqrt{\ell+1} B\right)}
y^{\left.-\frac{\sqrt{1+4(\ell+1) j(j+1)}+1}{2}\right)} .
\label{NF}
\end{eqnarray}
For the far-region, we can use the  approximations $y+1 \approx y$
and $\hat{\mu}^{2} r_{e h}^{2} \ll 1$. We may drop all the terms
except those which describe the free motion with momentum $j$ and
that reduces equation (\ref{RE2}) to
\begin{equation}
\frac{\mathrm{d}^{2} R_{\omega j m}(y)}{\mathrm{d}
y^{2}}+\frac{2}{y} \frac{\mathrm{d} R_{\omega j m}(y)}{\mathrm{d}
y}+\left(k_{l}^{2}-\frac{j(j+1)(\ell+1)}{y^{2}}\right) R_{\omega j
m}(y)=0, \label{FAR}
\end{equation}
where $k_{l} \equiv \frac{P\sqrt{1+\ell}}{\omega}
\sqrt{\omega^{2}-\mu^{2}}$. Equation (\ref{FAR}) has the
general solution
\begin{eqnarray}
R_{\omega j m, \text { far }}=e^{-i k y}(c_{1}
y^{\frac{\sqrt{1+4(\ell+1)
 j(j+1)}-1}{2}} M(\frac{1+\sqrt{1+4(\ell+1) j(j+1)}}{2},
 1+\sqrt{1+4(\ell+1) l(l+1)}, 2 i k_{l} y)+ \\
c_{2} y^{-\frac{\sqrt{1+4(\alpha+1) j(j+1)}+1}{2}}
M(\frac{1-\sqrt{1+4(\ell+1) j(j+1)}}{2}, 1-\sqrt{1+4(\ell+1)
j(j+1)}, 2 i k_{l} y)), \nonumber \label{FARR}
\end{eqnarray}
Here $M(a; b; y)$ stands for the confluent hypergeometric Kummer
function of the first kind.
 In order to match the solution with (\ref{NF}), we look for the small
$y$ behavior of the solution (\ref{FARR}). The solution (\ref{NF})
and (\ref{FN}) are susceptible for matching, since these two have
common region of interest. The matching of the asymptotic
solutions (\ref{NF}) and (\ref{FN}) enables us to compute the
scalar wave flux at infinity that resulting in
\begin{eqnarray}
c_{1}=& d \frac{\Gamma(\sqrt{1+4(\ell+1) j(j+1)}) \Gamma(1-2
i\sqrt{\ell+1} B)}{\Gamma\left(\frac{1+\sqrt{1+4(\ell+1)
j(j+1)}}{2} -2 i\sqrt{\ell+1} B\right)
\Gamma\left(\frac{1+\sqrt{1+4(\ell+1) j(j+1)}}{2}\right)},
\\\nonumber c_{2}=& d \frac{\Gamma(-\sqrt{1+4(\ell+1) j(j+1)})
\Gamma(1-2 i\sqrt{\ell+1}
B)}{\Gamma\left(\frac{1-\sqrt{1+4(\ell+1) j(j+1)}}{2}-2
i\sqrt{\ell+1} B\right) \Gamma\left(\frac{1-\sqrt{1+4(\ell+1)
j(j+1)}}{2}\right)}. \label{DD}
\end{eqnarray}
For small $y (y \to 0)$, the equation (\ref{FARR}) turns into
\begin{equation}
 R_{\omega j m, \text { far-small } \mathrm{y}} \sim y^{-\frac{1+\sqrt{1+4(\ell+1) j(j+1)}}{2}}. \label{FN}
\end{equation}
Since these two solutions (\ref{NF}), and (\ref{FN}) have a common
region of interest, the solutions are susceptible for matching. We
therefore compute the scalar wave flux at infinity resulting by
matching the asymptotic solutions (\ref{NF}) and
\begin{eqnarray}
c_{1}=& d \frac{\Gamma(\sqrt{1+4(\ell+1) j(j+1)}) \Gamma(1-2
i\sqrt{\ell+1} B)}{\Gamma\left(\frac{1+\sqrt{1+4(\ell+1)
j(j+1)}}{2} -2 i\sqrt{\ell+1} B\right)
\Gamma\left(\frac{1+\sqrt{1+4(\ell+1) j(j+1)}}{2}\right)},
\\\nonumber c_{2}=& d \frac{\Gamma(-\sqrt{1+4(\ell+1) j(j+1)})
\Gamma(1-2 i\sqrt{\ell+1}
B)}{\Gamma\left(\frac{1-\sqrt{1+4(\ell+1) j(j+1)}}{2}-2
i\sqrt{\ell+1} B\right) \Gamma\left(\frac{1-\sqrt{1+4(\ell+1)
j(j+1)}}{2}\right)}. \label{DD}
\end{eqnarray}
We expand equation (\ref{FARR}) around infinity which after
expansion results
\begin{eqnarray}
c_{1} \frac{\Gamma(1+\sqrt{1+4(\ell+1) j(j+1)})}{\Gamma\left(\frac{1+\sqrt{1+4(\ell+1)
j(j+1)}}{2}\right)} k_{l}^{-\frac{1+\sqrt{1+4(\ell+1)
j(j+1)}}{2}}\left((-2 i)^{-\frac{1+\sqrt{1+4(\ell+1)
 j(j+1)}}{2}} \frac{e^{-i k_{l} y}}{y}+(2 i)^{-\frac{1+\sqrt{1+4(\ell+1) j(j+1)}}{2}}
  \frac{e^{i k_{l} y}}{y}\right)+ \\\nonumber
c_{2} \frac{\Gamma(1-\sqrt{1+4(\ell+1)
j(j+1)})}{\Gamma\left(\frac{1-\sqrt{1+4(\ell+1) j(j+1)}}{2}\right)}
k_{l}^{\frac{\sqrt{1+4(\ell+1) j(j+1)}-1}{2}}\left((-2
i)^{\frac{\sqrt{1+4(\ell+1) j(j+1)}-1}{2}} \frac{e^{-i k_{l}
y}}{y}+(2 i)^{\frac{\sqrt{1+4(\ell+1) j(j+1)}-1}{2}} \frac{e^{i
k_{l} y}}{y}\right).
\end{eqnarray}
With the approximations $\frac{1}{y} \sim \frac{\mathcal{Q}}{\omega} \cdot
\frac{1}{r}, \quad e^{\pm i k_{l} y} \sim e^{\pm i
\sqrt{(1+\ell)(\omega^{2}-\mu^{2})} r}$, if we match the above
solution with the radial solution \eqref{AS}
$$
R_{\infty}(r) \sim \mathcal{D}_{i n}^{\infty} \frac{e^{-i
\sqrt{\frac{\omega^{2}}{1+\ell}-\hat{\mu}^{2}}
r^{*}}}{r}+\mathcal{D}_{r e f}^{\infty} \frac{e^{i
\sqrt{\frac{\omega^{2}}{1+\ell}-\hat{\mu}^{2}} r^{*}}}{r}, \quad
\text { for } \quad r \rightarrow \infty
$$
we get
$$
\begin{array}{c}
\mathcal{D}_{i n}^{\infty}=\frac{\mathcal{Q}}{\omega}\left(c_{1}(-2 i)^{-\frac{1+\sqrt{1+4(\ell+1)j(j+1)}}{2}}
 \frac{\Gamma(1+\sqrt{1+4(\ell+1)j(j+1)})}{\Gamma\left(\frac{1+\sqrt{1+4(\ell+1)j(j+1)}}{2}\right)}
 k_{l}^{-\frac{1+\sqrt{1+4(\ell+1)j(j+1)}}{2}}+\right. \\
\left.c_{2}(-2 i)^{\frac{\sqrt{1+4(\ell+1)j(j+1)}-1}{2}}
\frac{\Gamma(1-\sqrt{1+4(\ell+1)j(j+1)})}{\Gamma\left(\frac{1-\sqrt{1+4(\ell+1)j(j+1)}}{2}\right)}
k_{l}^{\frac{\sqrt{1+4(\ell+1)j(j+1)}-1}{2}}\right),
\end{array}
$$
and
$$
\begin{array}{l}
\mathcal{D}_{r e f}^{\infty}=\frac{\mathcal{Q}}{\omega}\left(c_{1}(2 i)^{-\frac{1+\sqrt{1+4(\ell+1)j(j+1)}}{2}}
 \frac{\Gamma(1+\sqrt{1+4(\ell+1)j(j+1)})}{\Gamma\left(\frac{1+\sqrt{1+4(\ell+1)j(j+1)}}{2}\right)}
  k_{l}^{-\frac{1+\sqrt{1+4(\ell+1)j(j+1)}}{2}}+\right. \\
\left.c_{2}(2 i) \frac{\sqrt{1+4(\ell+1)j(j+1)}-1}{2}
\frac{\Gamma(1-\sqrt{1+4(\ell+1)j(j+1)})}{\Gamma\left(\frac{1-\sqrt{1+4(\ell+1)j(j+1)}}{2}\right)}
k_{l}^{\frac{\sqrt{1+4(\ell+1)j(j+1)}-1}{2}}\right) .
\end{array}
$$
Substituting  the expressions of $c_{1}$ and $c_{2}$ from Eqn.
(\ref{DD}) into the above expressions we have
\begin{eqnarray}
\mathcal{D}_{in}^{\infty}&=&\frac{d(-2
i)^{-\frac{1+\sqrt{1+4(\ell+1)
j(j+1)}}{2}}}{\sqrt{(1+\ell)(\omega^{2}-\mu^{2})}} \cdot
\frac{\Gamma(\sqrt{1+4(\ell+1) j(j+1)}) \Gamma(1+\sqrt{1+4(\ell+1)
j(j+1)})}{\Gamma\left(\frac{1+\sqrt{1+4(\ell+1)j(j+1)}}{2}-2
i\sqrt{\ell+1}
B\right)\left(\Gamma\left(\frac{1+\sqrt{1+4(\ell+1)j(j+1)}}{2}\right)\right)^{2}}\times
\\\nonumber &&\Gamma(1-2 i\sqrt{\ell+1} B)
k_{l}^{\frac{1-\sqrt{1+4(\ell+1)j(j+1)}}{2}}+\frac{d(-2
i)^{\frac{\sqrt{1+4(\ell+1)j(j+1)}-1}{2}}}{\sqrt{(1+\ell)(\omega^{2}-\hat{\mu}^{2})}}
\times \\\nonumber &&\frac{\Gamma(1-\sqrt{1+4(\ell+1)j(j+1)})
\Gamma(-\sqrt{1+4(\ell+1)j(j+1)})}{\left(\Gamma\left(\frac{1-\sqrt{1+4(\ell+1)j(j+1)}}{2}\right)\right)^{2}
\Gamma\left(\frac{1-\sqrt{1+4(\ell+1)j(j+1)}}{2}-2 i\sqrt{\ell+1}
B\right)} \Gamma(1-2 i\sqrt{\ell+1} B)
k_{l}^{\frac{1+\sqrt{1+4(\ell+1)j(j+1)}}{2}},
\end{eqnarray}
and
\begin{eqnarray}
\mathcal{D}_{ref}^{\infty}&=&\frac{d(2
i)^{-\frac{1+\sqrt{1+4(\ell+1)
j(j+1)}}{2}}}{\sqrt{(1+\ell)(\omega^{2}-\mu^{2})}} \cdot
\frac{\Gamma(\sqrt{1+4(\ell+1) j(j+1)}) \Gamma(1+\sqrt{1+4(\ell+1)
j(j+1)})}{\Gamma\left(\frac{1+\sqrt{1+4(\ell+1)j(j+1)}}{2}-2
i\sqrt{\ell+1}
B\right)\left(\Gamma\left(\frac{1+\sqrt{1+4(\ell+1)j(j+1)}}{2}\right)\right)^{2}}\times
\\\nonumber &&\Gamma(1-2 i\sqrt{\ell+1} B)
k_{l}^{\frac{1-\sqrt{1+4(\ell+1)j(j+1)}}{2}}+\frac{d(2
i)^{\frac{\sqrt{1+4(\ell+1)j(j+1)}-1}{2}}}{\sqrt{(1+\ell)(\omega^{2}-\hat{\mu}^{2})}}
\times \\\nonumber &&\frac{\Gamma(1-\sqrt{1+4(\ell+1)j(j+1)})
\Gamma(-\sqrt{1+4(\ell+1)j(j+1)})}{\left(\Gamma\left(\frac{1-\sqrt{1+4(\ell+1)j(j+1)}}{2}\right)\right)^{2}
\Gamma\left(\frac{1-\sqrt{1+4(\ell+1)j(j+1)}}{2}-2 i\sqrt{\ell+1}
B\right)} \Gamma(1-2 i\sqrt{\ell+1} B)
k_{l}^{\frac{1+\sqrt{1+4(\ell+1)j(j+1)}}{2}}.
\end{eqnarray}
The amplification factor ultimately results out to be
\begin{equation}
Z_{j m} \equiv \frac{\left|\mathcal{D}_{r e
f}^{\infty}\right|^{2}}{\left|\mathcal{D}_{i
n}^{\infty}\right|^{2}}-1. \label{AMPZ}
\end{equation}
Equation (\ref{AMPZ}) is a general expression of the amplification
factor obtained by making use of the asymptotic matching method.
When
$\frac{\left|\mathcal{D}_{ref}^{\infty}\right|^{2}}{\left|\mathcal{D}_{i
n}^{\infty}\right|^{2}}$ acquires a value greater than unity there
will be a gain in amplification factor that corresponds to
superradiance phenomena. However, a negative value of the
amplification factor indicates a loss that corresponds to the
non-appearance of superradiance. To study the effect of Lorentz
violation on the superradiance phenomena, it will be useful to
plot $Z_{j m}$ versus $M\omega$ for different LV parameters. In
Fig. (6), we present the variation $Z_{j m}$ versus $M\omega$ for
the leading multipoles $j= 1$, and $2$ taking different values
(both negative and positive) of LV Parameter. From the fig.
(\ref{non-superradiant}) along with fig. (\ref{zl}), it is evident
that superradiance for a particular $j$ occurs when the allowed
values of $m$ are restricted to $m > 0$.

For negative $m$, amplification factor takes negative value which
refers to the nonoccurrence of superradiance. The plots also show
transparently that with the decrease in the value of the LV
parameters the superradiance process enhances and the reverse is
the case when the value of the LV parameter decreases. In Fig.
(\ref{zb}) we have also studied the effect of the parameter $b$ on
the superradiance scenario. It shows that the superradiance
scenario gets diminished with the increase in the value of the
parameter $b$. In \cite{ARS} we have noticed that the size of the
shadow decreases with the increase in the value of both the
parameters $l$ and $b$. The only difference is that $l$ can take
both positive values, however, $b$ as per definition can not be
negative. Therefore, an indirect relation of superradiance with
the size of the shadow is being revealed through this analysis. A
decrease in the value of $b$ and $l$ indicate the increase in the
size of the shadow.

\begin{figure}[H]
\centering
\begin{subfigure}{.58\textwidth}
%\centering
  \includegraphics[width=.8\linewidth]{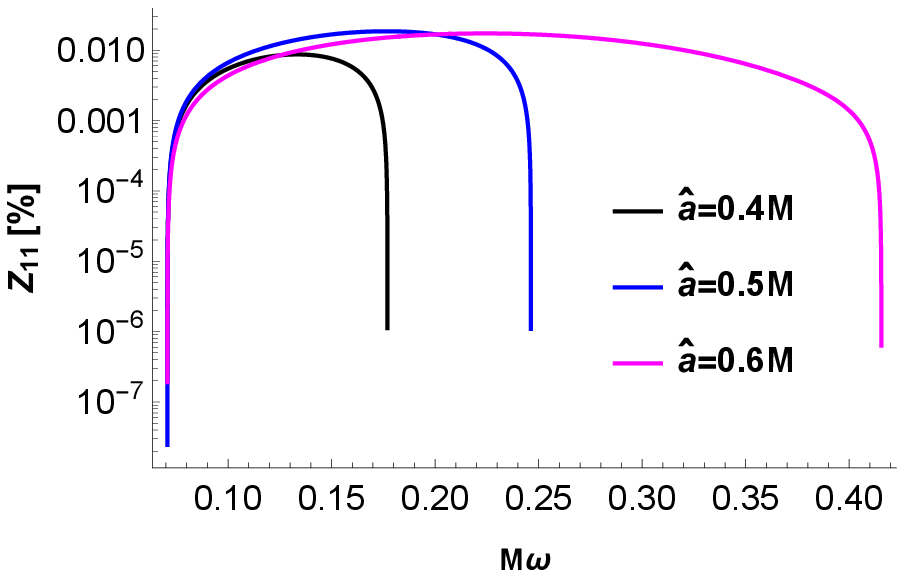}
%\caption{Critical radius for various values of $b$ with $ l=.1,k=.1$ and $\theta=\pi/2$}
\end{subfigure}%
\begin{subfigure}{.58\textwidth}
%\centering
  \includegraphics[width=.8\linewidth]{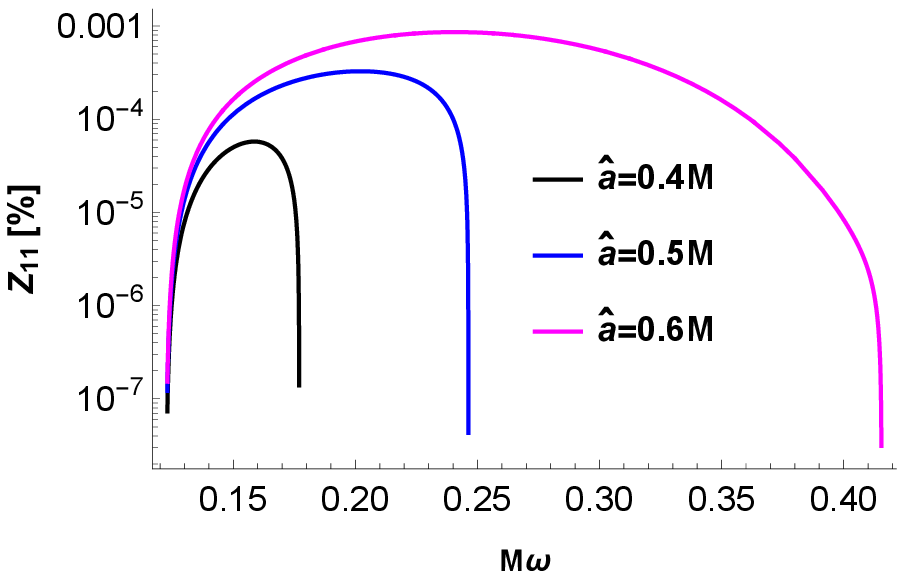}
%\caption{Critical radius for various values of $k$ with $ l=.1,b=.1$ and $\theta=\pi/2$}
\end{subfigure}
\begin{subfigure}{.58\textwidth}
%\centering
  \includegraphics[width=.8\linewidth]{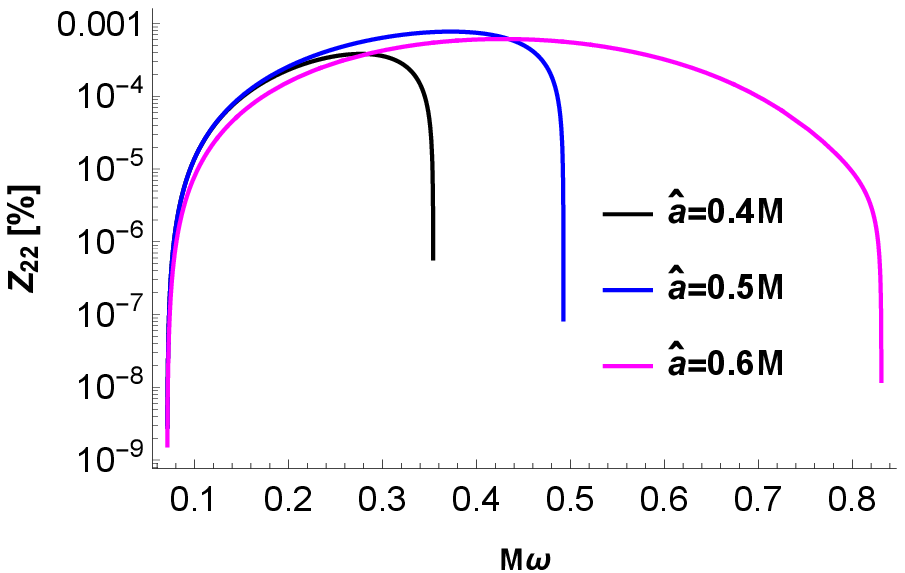}
%\caption{Critical radius for various values of $l$ with $k=.1,b=.1$ and $\theta=\pi/2$}
\end{subfigure}%
\begin{subfigure}{.58\textwidth}
%\centering
  \includegraphics[width=.8\linewidth]{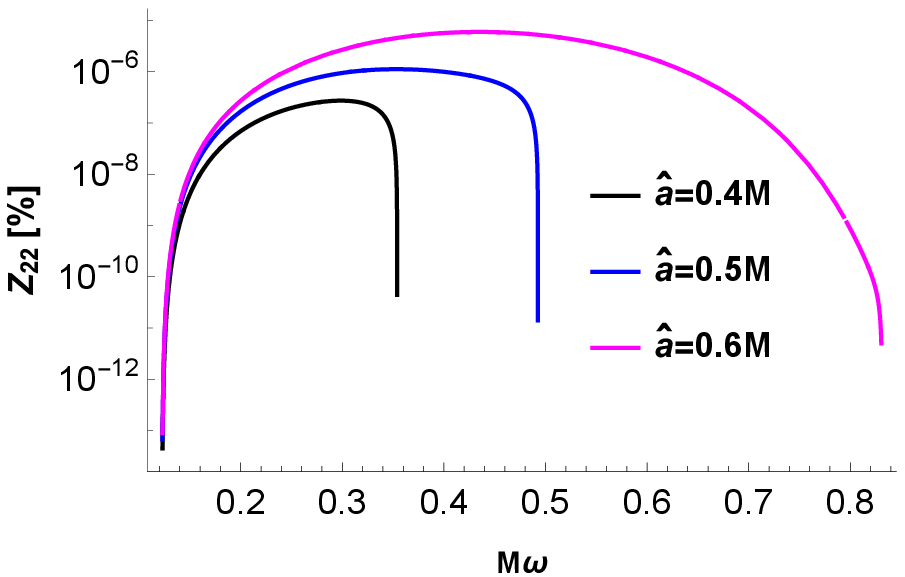}
%\caption{Critical radius for various values of $l$ with $k=.1,b=.1$ and $\theta=\pi/2$}
\end{subfigure}
\caption{Variation of amplification factors with $\hat{a}$ for
various multipoles with $\hat{\mu}=0.1$ and $b=0.02M^2$. For left ones
$\ell=-0.5$ and for right ones $\ell=0.5$.} \label{za}
\end{figure}

\begin{figure}[H]
\centering
\begin{subfigure}{.58\textwidth}
%\centering
  \includegraphics[width=.8\linewidth]{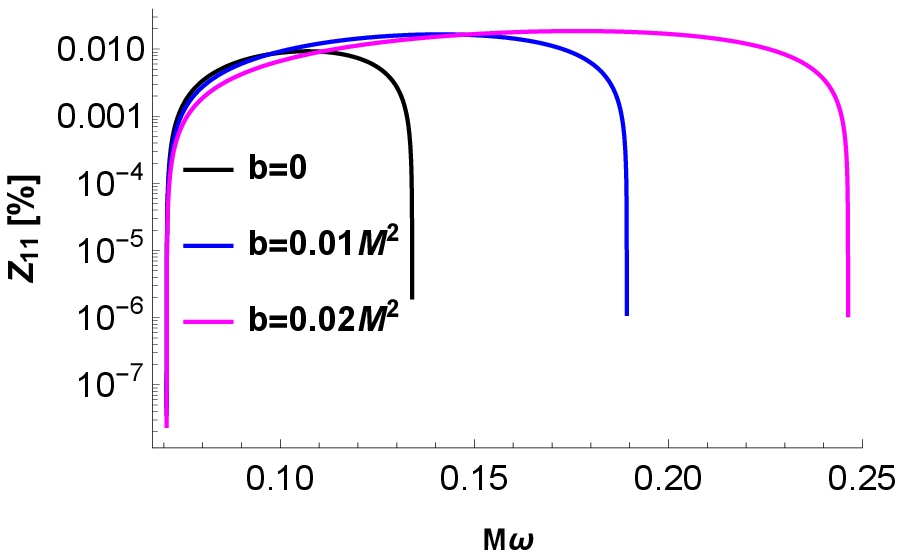}
%\caption{Critical radius for various values of $b$ with $ l=.1,k=.1$ and $\theta=\pi/2$}
\end{subfigure}%
\begin{subfigure}{.58\textwidth}
%\centering
  \includegraphics[width=.8\linewidth]{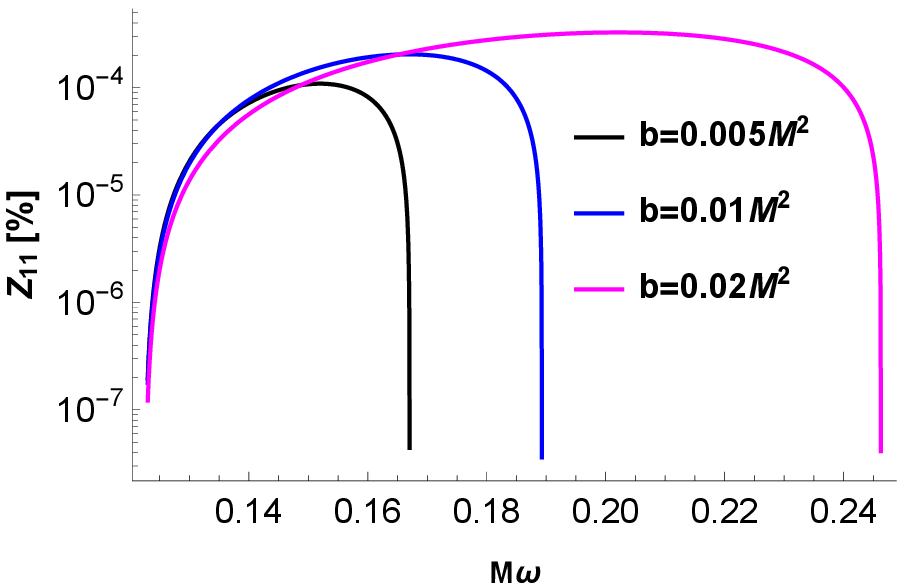}
%\caption{Critical radius for various values of $k$ with $ l=.1,b=.1$ and $\theta=\pi/2$}
\end{subfigure}
\begin{subfigure}{.58\textwidth}
%\centering
  \includegraphics[width=.8\linewidth]{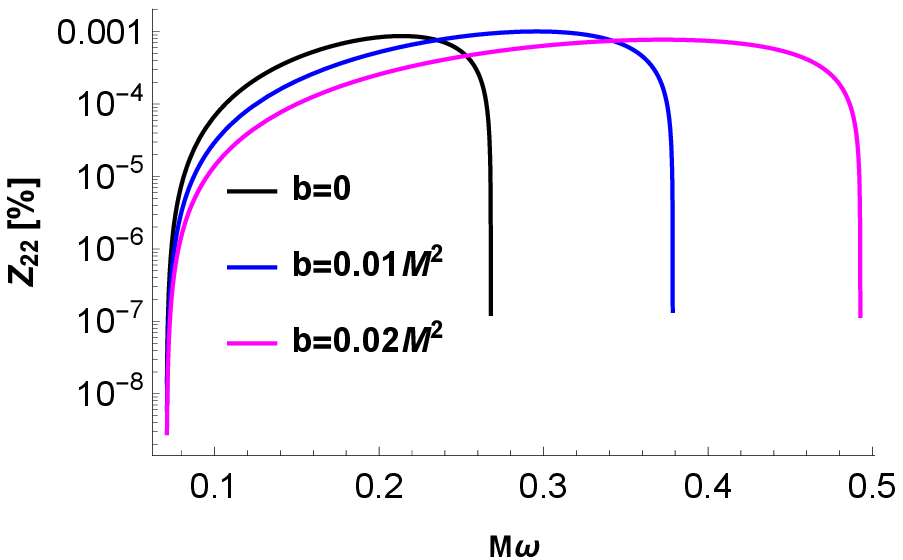}
%\caption{Critical radius for various values of $l$ with $k=.1,b=.1$ and $\theta=\pi/2$}
\end{subfigure}%
\begin{subfigure}{.58\textwidth}
%\centering
  \includegraphics[width=.8\linewidth]{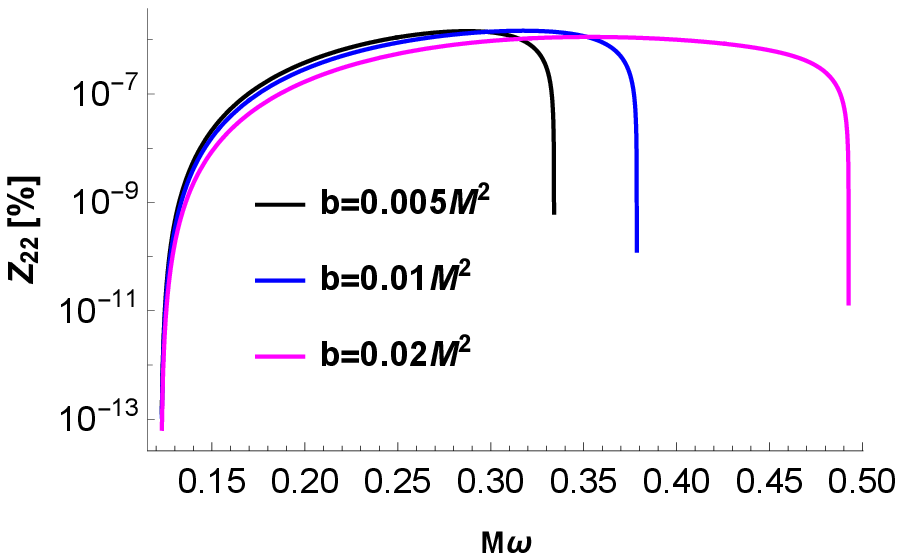}
%\caption{Critical radius for various values of $l$ with $k=.1,b=.1$ and $\theta=\pi/2$}
\end{subfigure}
\caption{Variation of amplification factors with $b$ for various
multipoles with $\hat{\mu}=0.1$ and $\hat{a}=0.5M$. For left ones
$\ell=-0.5$ and for right ones $\ell=1$.} \label{zb}
\end{figure}

We observe that for the negative value of $m$, the amplification
factor takes negative value. Therefore, superradiance does not
occur. The plots also transpires that with the superradiance
process enhances with the decrease in the value of the LV
parameters, and it diminishes with the increase in value of the LV
parameter. In Fig. (\ref{zb}), we have also studied the effect of
the parameter $b$ related to the non-commutativity of the
sapcetime on the superradiance process. It shows that with the
increase in the value of the parameter $b$ the superradiance
process gets diminished. However, with the increase in $a$ the
superradiance effect increases as is found from Fig.(\ref{za}).

\subsection{Superradiant instability for Lorentz violating and non-commutative
Kerr-like black hole}
From equation (\ref{RE}) we have
\begin{eqnarray}
\Delta \frac{d}{d r}\left(\Delta \frac{d R_{\omega j m}}{d
r}\right)+\mathcal{F} R_{\omega j m}=0, \label{MRE}
\end{eqnarray}
where for a slowly rotating black hole $(\hat{a} \omega \ll 1)$
$$
\mathcal{F} \equiv\frac{\left(\left(r^2+\hat{a}^{2}\right) \omega-m
\hat{a}\right)^{2}}{1+\ell}+\Delta\left(2 m \hat{a}
\omega-l(l+1)-\mu^{2}r^2\right).
$$
If we now look for the black hole bomb mechanism, we would have
have the following solutions for the radial equation (\ref{MRE})
$$
R_{\omega j m} \sim\left\{\begin{array}{ll}
e^{-i(\omega-m \hat{\Omega}) r_{*}} & \text { as } r \rightarrow r_{e h}
\left(r_{*} \rightarrow-\infty\right) \\
\frac{e^{-\sqrt{\mu^{2}-\omega^{2} r_{*}}}}{r} & \text { as } r
\rightarrow \infty  \left(r_{*} \rightarrow \infty\right)
\end{array}\right.
$$
The above solution represents the physical boundary conditions
that the scalar wave at the black hole horizon is purely ingoing
while at spatial infinity it is decaying exponentially (bounded)
solution, provided that $\omega^{2}<\mu^{2}$. With the new radial
function
$$
\psi_{\omega j m} \equiv \sqrt{\Delta} R_{\omega j m},
$$
the radial equation (\ref{MRE}) turns into
$$
\left(\frac{d^{2}}{d r^{2}}+\omega^{2}-V\right)
\psi_{\omega j m}=0.
$$
with
\begin{figure}[H]
\centering
\begin{subfigure}{.34\textwidth}
%\centering
  \includegraphics[width=.9\linewidth]{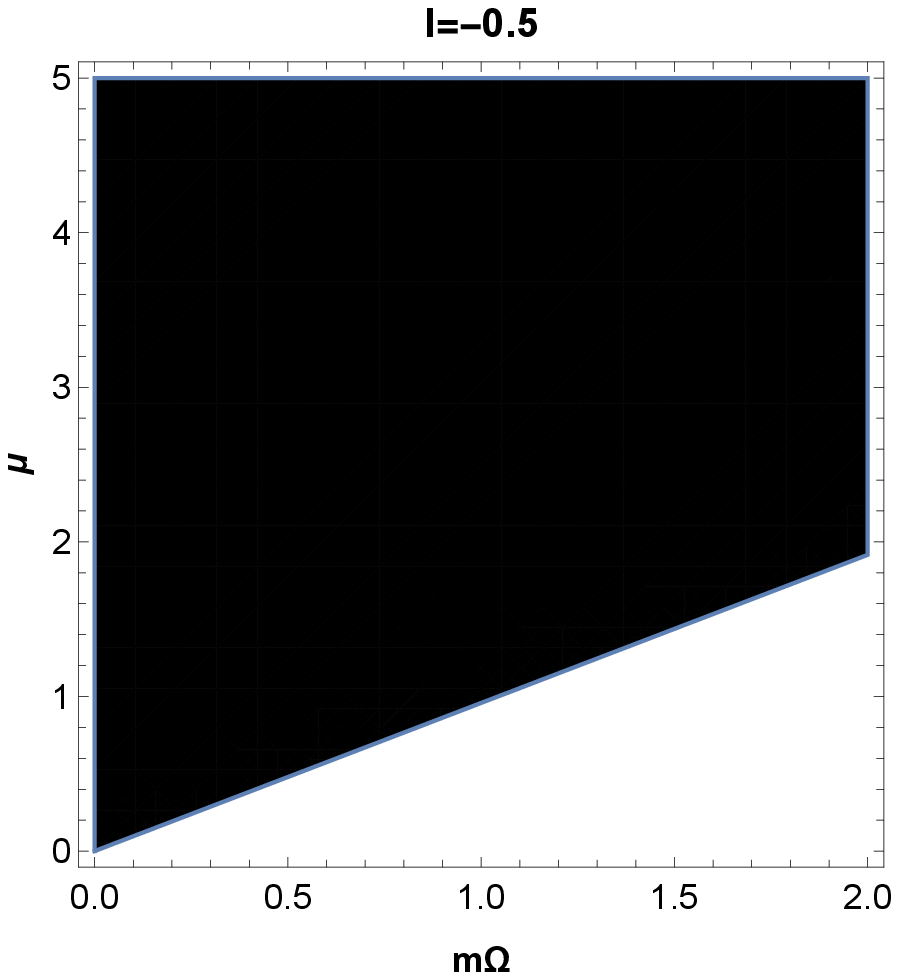}
%\caption{Critical radius for various values of $b$ with $ l=.1,k=.1$ and $\theta=\pi/2$}
\end{subfigure}%
\begin{subfigure}{.34\textwidth}
%\centering
  \includegraphics[width=.9\linewidth]{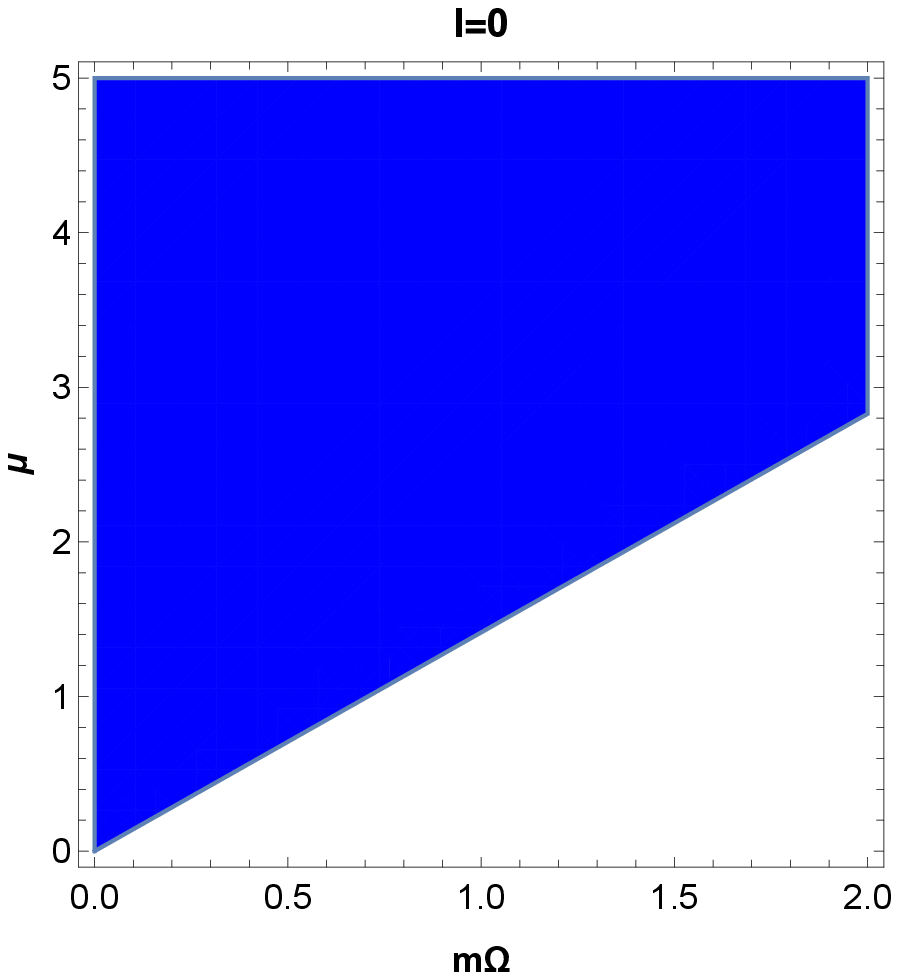}
%\caption{Critical radius for various values of $k$ with $ l=.1,b=.1$ and $\theta=\pi/2$}
\end{subfigure}%
\begin{subfigure}{.34\textwidth}
%\centering
  \includegraphics[width=.9\linewidth]{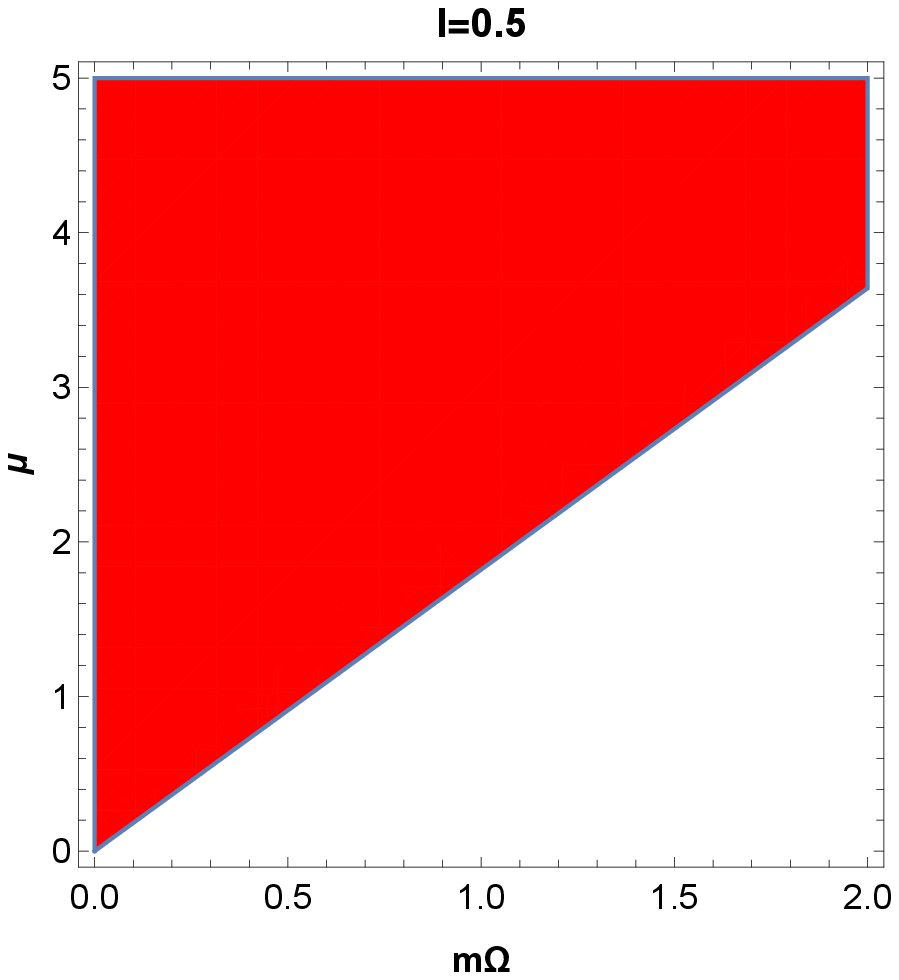}
%\caption{Critical radius for various values of $l$ with $k=.1,b=.1$ and $\theta=\pi/2$}
\end{subfigure}%
\caption{Parameter space($m\Omega$-$\mu$) for massive scalar field
where colored area represents region with stable dynamics and
non-colored area represents region with unstable dynamics. Here $a=0.4M \quad \text{and}\quad b=0.01M^2$}
\label{fig:test}
\end{figure}

\begin{figure}[H]
\centering
\begin{subfigure}{.34\textwidth}
%\centering
  \includegraphics[width=.9\linewidth]{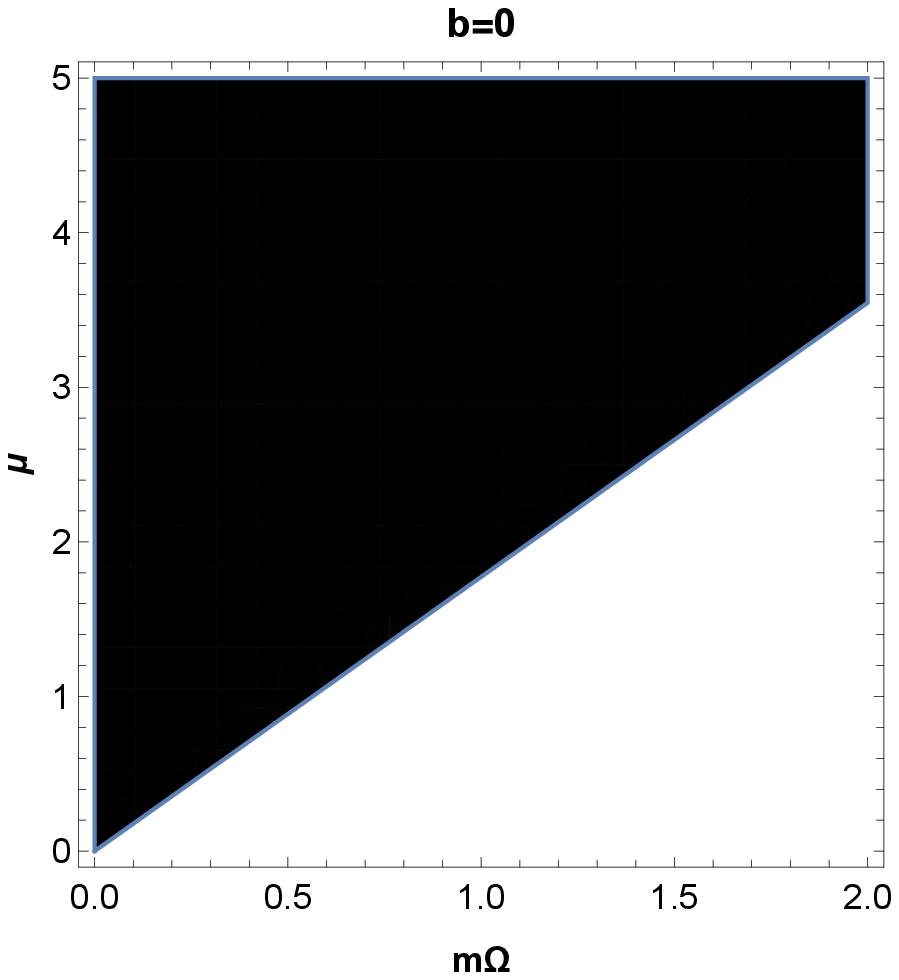}
%\caption{Critical radius for various values of $b$ with $ l=.1,k=.1$ and $\theta=\pi/2$}
\end{subfigure}%
\begin{subfigure}{.34\textwidth}
%\centering
  \includegraphics[width=.9\linewidth]{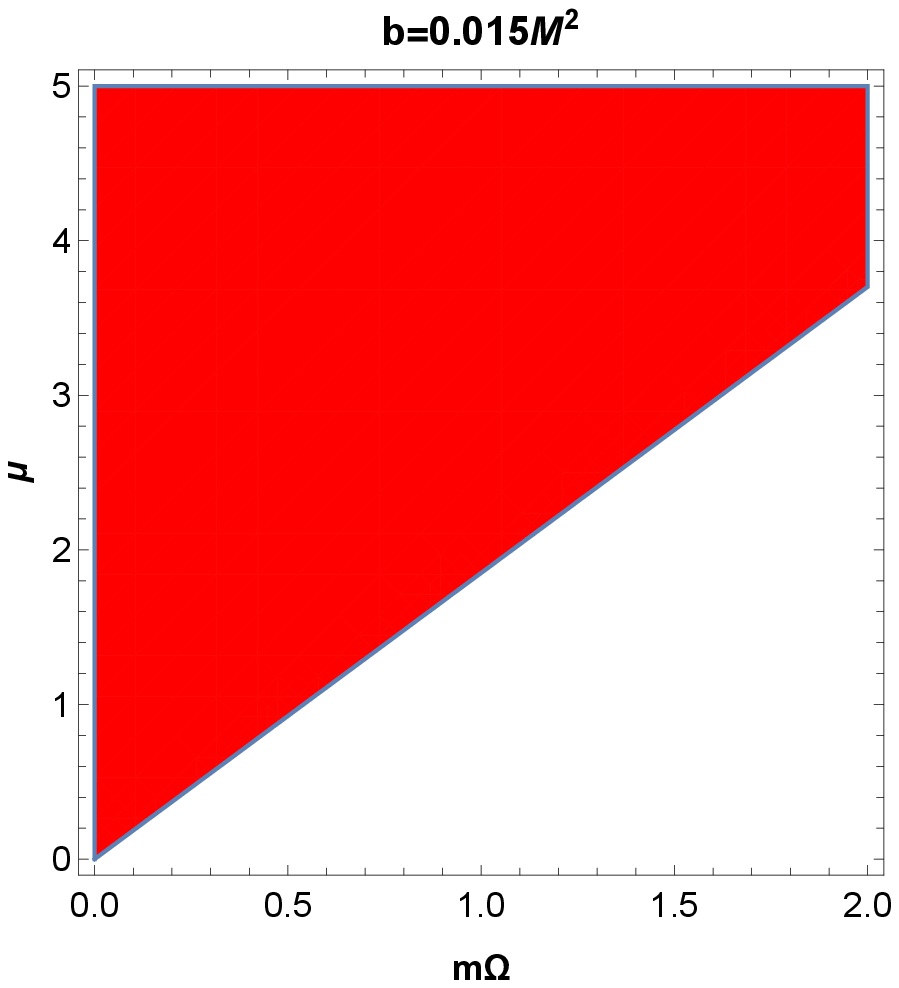}
%\caption{Critical radius for various values of $k$ with $ l=.1,b=.1$ and $\theta=\pi/2$}
\end{subfigure}%
\begin{subfigure}{.34\textwidth}
%\centering
  \includegraphics[width=.9\linewidth]{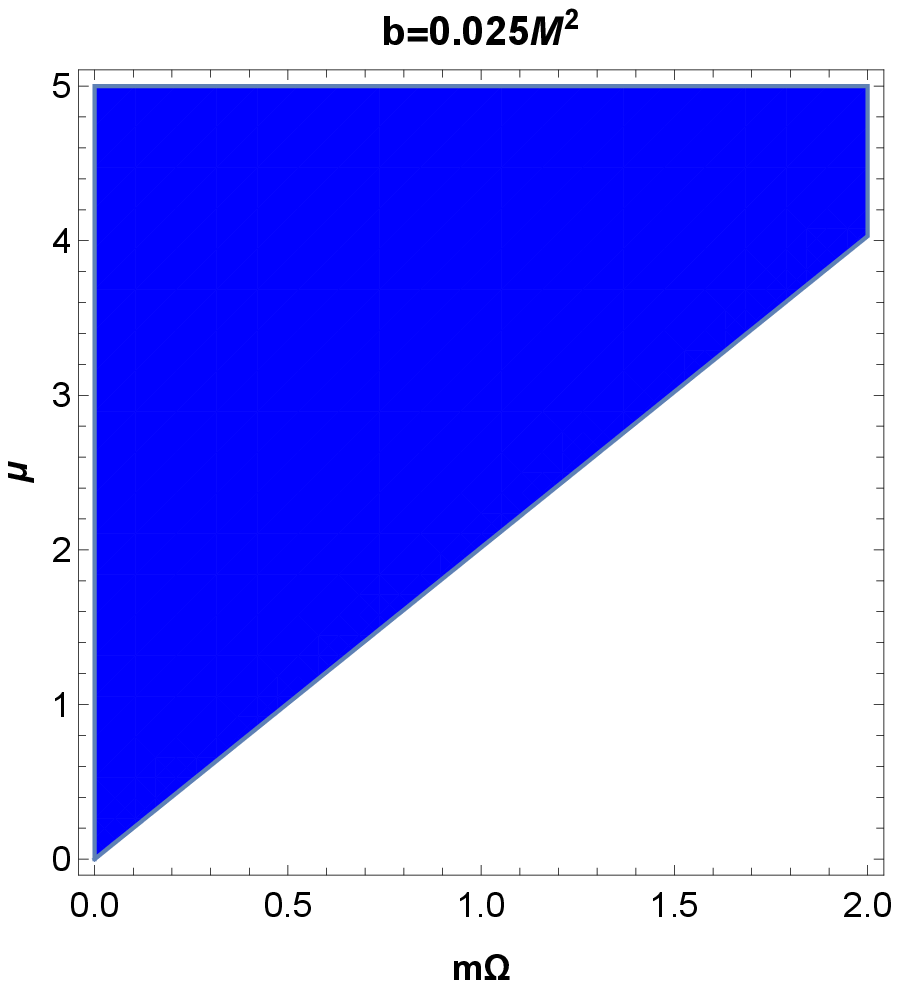}
%\caption{Critical radius for various values of $l$ with $k=.1,b=.1$ and $\theta=\pi/2$}
\end{subfigure}%
\caption{Parameter space($m\Omega$-$\mu$) for massive scalar field
where colored area represents region with stable dynamics and
non-colored area represents region with unstable dynamics.
Here $a=0.4M\quad \text{and}\quad l=0.5$}
\label{fig:test}
\end{figure}

$$
\omega^{2}-V=\frac{\mathcal{F}+\frac{M \left(M-\frac{8 \sqrt{t}}{\sqrt{\pi }}\right)-a^2 (l+1)}{(l+1)^2}}{\Delta^{2}},
$$
which is the Regge-Wheel equation. By discarding the terms
$\mathcal{O}\left(1 / r^{2}\right)$ the asymptotic form of the
effective potential $V(r)$ looks
$$
V(r)=\mu^{2}\left(1+\ell\right)-\left(1+\ell\right)\frac{4 M
\omega^{2}}{r}+(\ell+1) \frac{2 M \mu^{2}}{r}.
$$
To realize the trapping meaningfully by the above effective
potential it is necessary that its asymptotic derivative be
positive i.e. $V^{\prime} \rightarrow 0^{+}$ as $r \rightarrow
\infty$ \cite{HOD}.  This along with the fact that superradiance
amplification of scattered waves occurs when $\omega<m
\hat{\Omega}$ we get the regime
$$
\frac{\mu}{\sqrt{2}}<\omega^{2}<m \hat{\Omega},
$$
in which the integrated system of non-commutative Kerr bumblebee
black hole and massive scalar field may experience a superradiant
instability, known as the black hole bomb. The dynamics of the
massive scalar field in non-commutative Kerr like
black hole will
remain stable when $\mu\geq\sqrt{2}m\hat{\Omega}$.

\section{Photon orbit and black hole shadow}
In this section, we study the black hole shadow related to this
modified theory. There are several studies related to the black
hole shadow from which we will get the necessary inputs for the
study \cite{SDO1, SDO2, SDO3, SDO4}. In order to study the shadow,
we introduce two conserved parameters $\xi$ and $\eta$ as usual
which are defined by
\begin{eqnarray}
\xi=\frac{L_{z}}{E} \quad \textrm{and} \quad \eta=\frac{\mathcal{Q}}{E^{2}},
\end{eqnarray}
where $E, L_{z}$, and $\mathcal{Q}$ are the energy, the axial
component of the angular momentum, and the Carter constant
respectively. Then the null geodesics in the bumblebee rotating
black hole spacetime in terms of $\xi$ are given by
\begin{eqnarray}\nonumber
\rho^{2} \frac{d r}{d \lambda}=\pm \sqrt{R}, \quad \rho^{2} \frac{d
\theta}{d \lambda}=\pm \sqrt{\Theta}, \\\nonumber
(1+\ell) \Delta \rho^{2} \frac{d t}{d \lambda}=A-2 \sqrt{1+\ell} \operatorname{Mra\xi},\\
(1+\ell) \Delta \rho^{2} \frac{d \phi}{d \lambda}=2 \sqrt{1+\ell}
M r a+\frac{\xi}{\sin ^{2} \theta}\left(\rho^{2}-2 M r\right),
\end{eqnarray}
where $\lambda$ is the affine parameter and
\begin{eqnarray}
R(r)=\left[\frac{r^{2}+(1+l) a^{2}}{\sqrt{1+l}}-a
\xi\right]^{2}-\Delta\left[\eta+(\xi-a\sqrt{1+l})^{2}\right],\quad
\Theta(\theta)=\eta+(1+l) a^{2} \cos ^{2} \theta-\xi^{2} \cot
^{2} \theta.
\end{eqnarray}
The radial equation of motion can be written down in the familiar
form
\begin{eqnarray}
\left(\rho^{2} \frac{d r}{d \lambda}\right)^{2}+V_{e f f}=0.
\end{eqnarray}
The effective potential $V_{e f f}$ then reads
\begin{eqnarray}
V_{e f f}=-\left[\frac{r^{2}+(1+l) a^{2}}{\sqrt{1+l}}-a
\xi\right]^{2}+\Delta\left[\eta+(\xi-a\sqrt{1+l})^{2}\right].
\end{eqnarray}
The following equations describe the unstable spherical orbit on
the equatorial plane, $\theta=\frac{\pi}{2}$.
\begin{eqnarray}
\theta=\frac{\pi}{2},\quad R(r)=0,\quad \frac{d R}{d r}=0,\quad
\frac{d^{2} R}{d r^{2}}<0,\quad \textrm{and} \quad \eta=0.
\label{CONDITION1}
\end{eqnarray}
We plot the potential $V_{e f f}$  versus $r/M$ with
$\xi=\xi_{c}+0.2$,  where $\xi_{c}$ is the value of $\xi$ for
equatorial spherical unstable orbit.
\begin{figure}[H]
\centering
\begin{subfigure}{.5\textwidth}
\centering
\includegraphics[width=.65\linewidth]{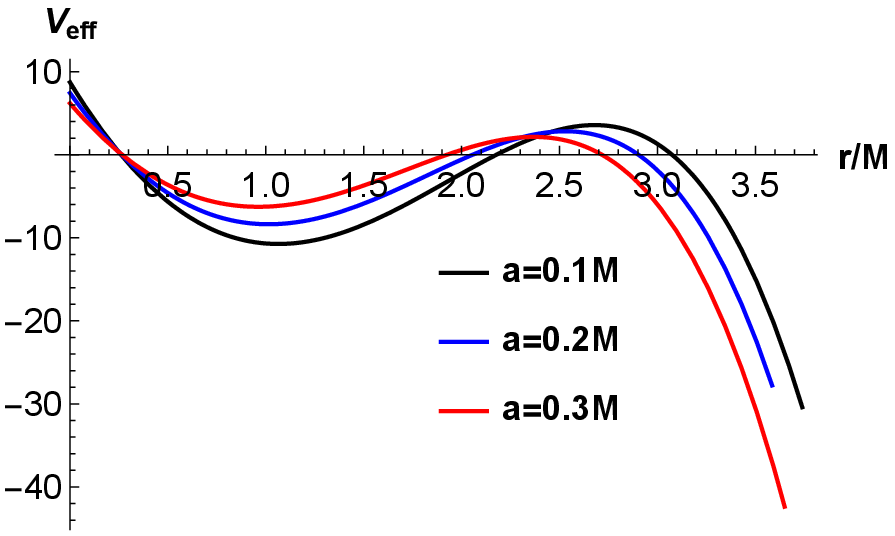}
\end{subfigure}%
\begin{subfigure}{.5\textwidth}
\centering
\includegraphics[width=.65\linewidth]{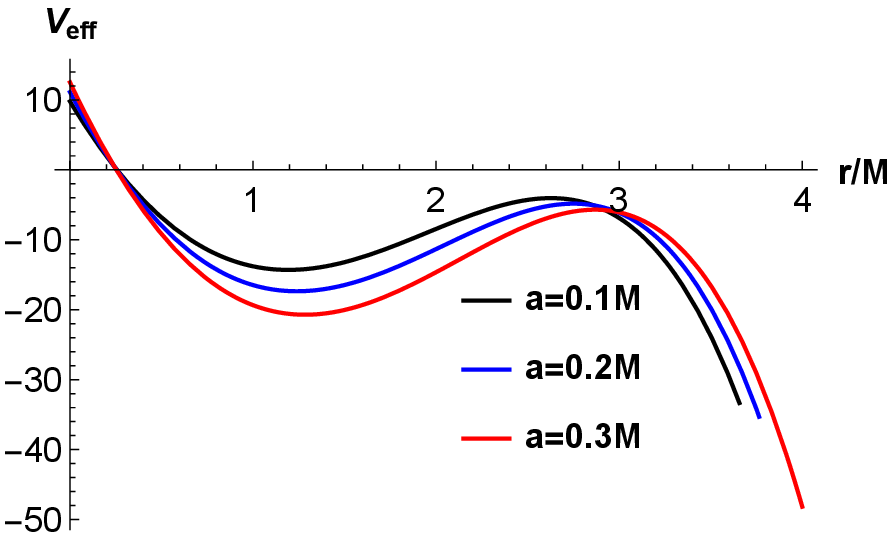}
\end{subfigure}
\caption{The left and the right panels describe the effective
potential for prograde orbits and the retrograde orbits
respectively for various values of $a$ with $b=0.01M^{2}$ and
$l=0.1$.} \label{fig:test}
\end{figure}

\begin{figure}[H]
\centering
\begin{subfigure}{.5\textwidth}
\centering
\includegraphics[width=.65\linewidth]{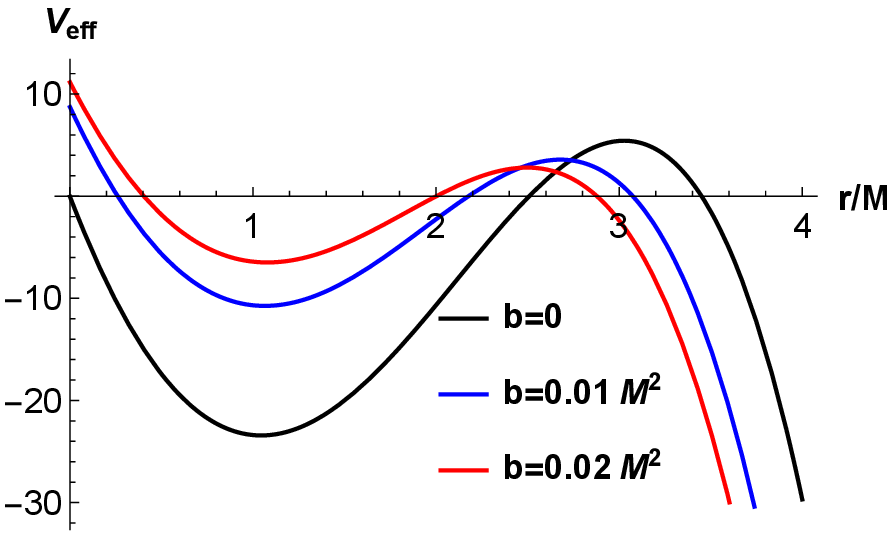}
\end{subfigure}%
\begin{subfigure}{.5\textwidth}
\centering
\includegraphics[width=.65\linewidth]{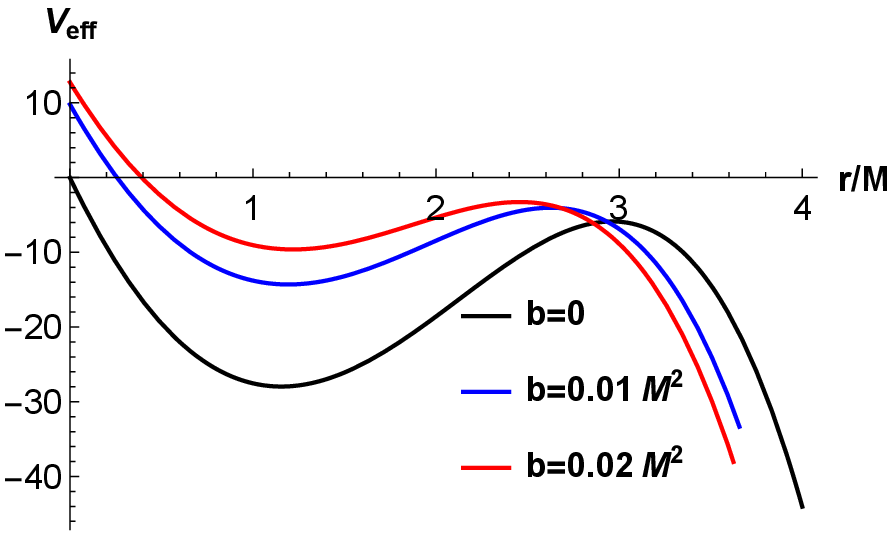}
\end{subfigure}
\caption{The left and the right panels describe the effective
potential for prograde orbits and the retrograde orbits
respectively for various values of $b$ with $a=0.1M$ and $l=0.1$.}
\label{fig:test}
\end{figure}

\begin{figure}[H]
\centering
\begin{subfigure}{.5\textwidth}
\centering
\includegraphics[width=.65\linewidth]{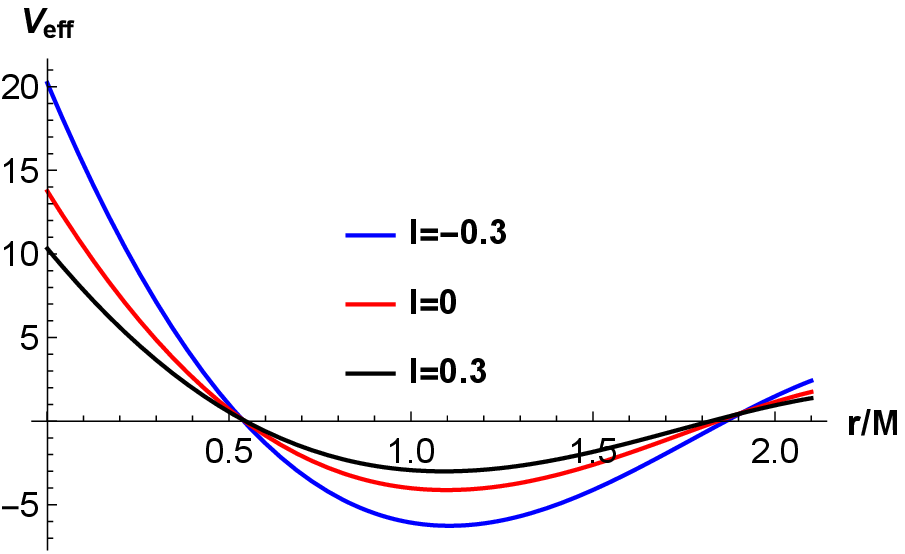}
\end{subfigure}%
\begin{subfigure}{.5\textwidth}
\centering
\includegraphics[width=.65\linewidth]{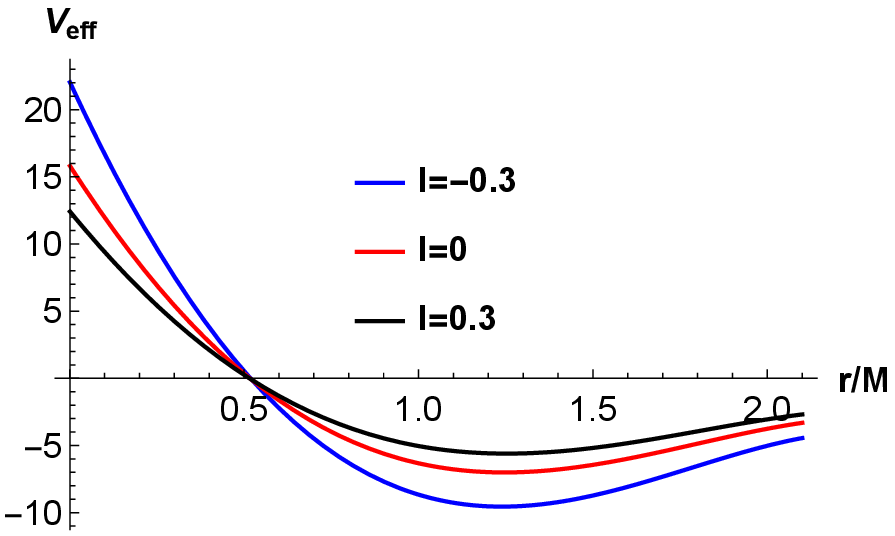}
\end{subfigure}
\caption{The left panel describes the effective potential for
prograde orbits and right panel describes the retrograde orbits
for various values of $l$ with $a=0.1M$ and $b=.03M^{2}$.}
\label{fig:test}
\end{figure}
The plots depicted above show that the turning points for prograde
orbits shift towards the left as $a$ or $b$ increases. We also
plot-critical radii of prograde and retrograde orbits for the
different scenarios in the Fig. furnished below.
\begin{figure}[H]
\centering
\begin{subfigure}{.5\textwidth}
\centering
\includegraphics[width=.65\linewidth]{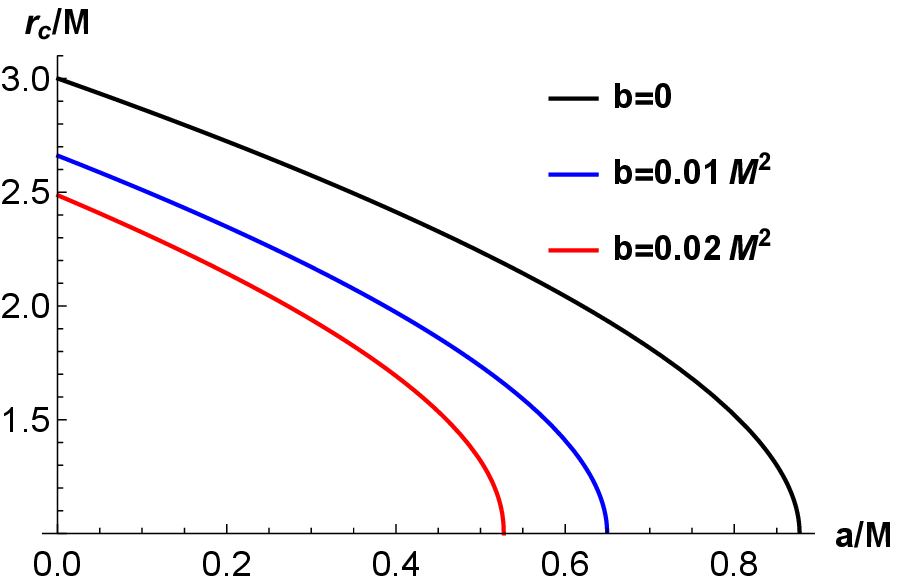}
\end{subfigure}%
\begin{subfigure}{.5\textwidth}
\centering
\includegraphics[width=.65\linewidth]{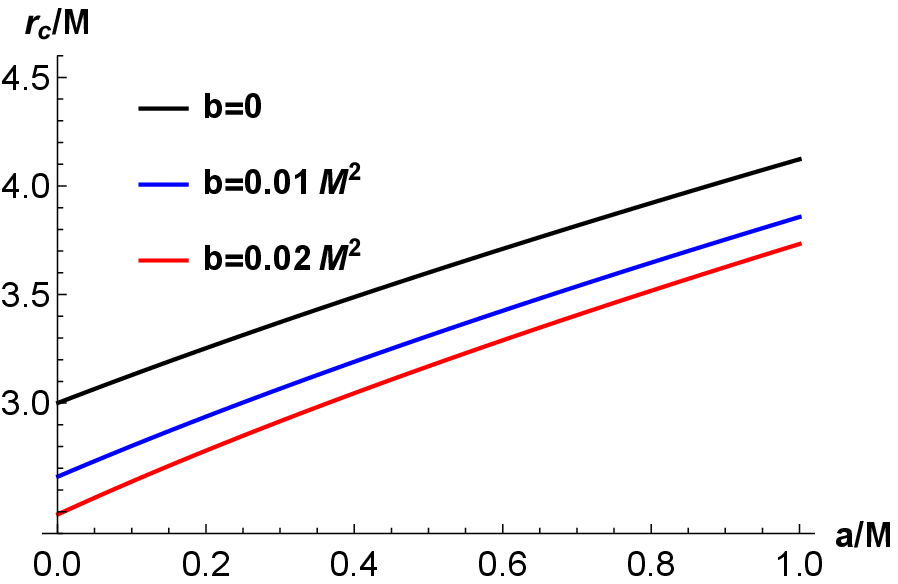}
\end{subfigure}
\caption{The left panel shows the variation of critical radius for
prograde orbits and the right panel shows the variation of
critical radius for retrograde orbits for various values of $b$
with $l=0.3$.} \label{fig:test}
\end{figure}

\begin{figure}[H]
\centering
\begin{subfigure}{.5\textwidth}
\centering
\includegraphics[width=.65\linewidth]{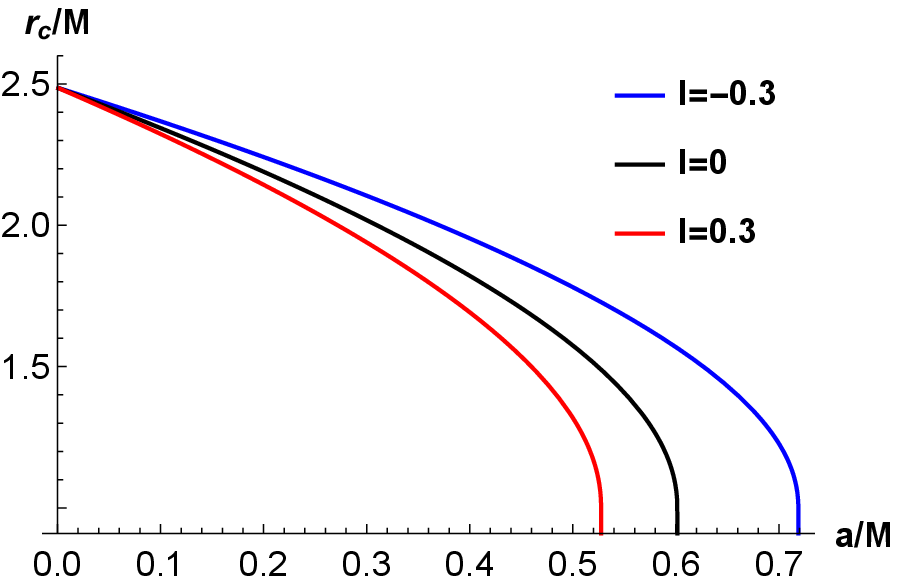}
\end{subfigure}%
\begin{subfigure}{.5\textwidth}
\centering
\includegraphics[width=.65\linewidth]{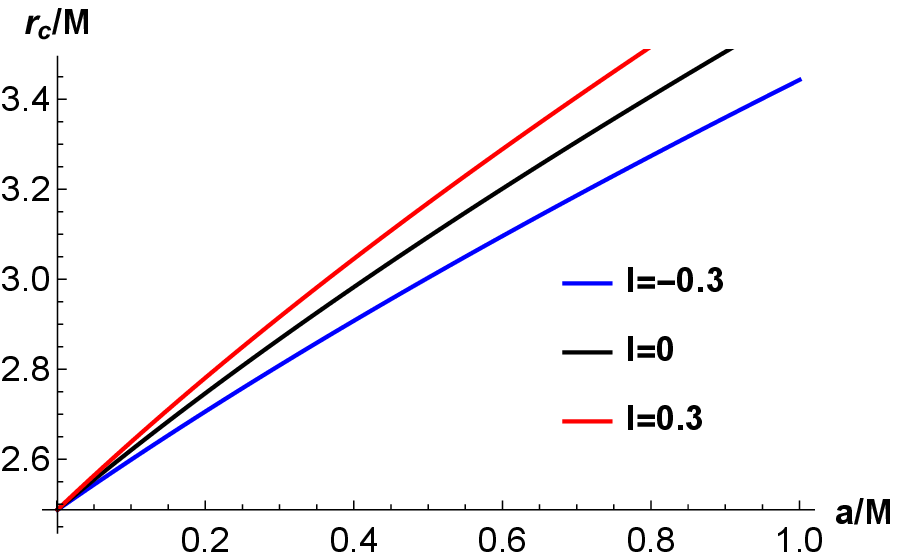}
\end{subfigure}
\caption{The left panel shows the variation of critical radius for
prograde orbits and the right one shows the variation of the
critical radius for retrograde orbits for various values of $l$
with $b=0.02M^{2}$. } \label{fig:test}
\end{figure}
It can be concluded from the above plots that critical radii, both
for the prograde and retrograde orbits, decrease with an increase
in the non-commutative parameter $b$. On the other hand, the
critical radius for prograde orbit decreases with the increase in
$l$ but for retrograde orbit, it increases with the increase in
$l$. For more generic orbits $\theta \neq \pi / 2$ and $\eta \neq
0,$ the solution of Eqn. (\ref{CONDITION1})$ r=r_{s}$, gives the
$r-$ constant orbit, which is also called spherical orbit and the
conserved parameters of the spherical orbits are given by
\begin{eqnarray}
\xi_{s}&=&\frac{\left(a^{2}\left(1+l\right)+r^{2}\right)\Delta^{'}(r)
-4r\Delta(r)}{a\sqrt{1+l}\Delta^{'}(r)},\\\nonumber
\eta_{s}&=&\frac{r^{2}\left(8\Delta(r)\left(2a^{2}\left(1+l\right)
+r\Delta^{'}(r)\right)-r^{2}\Delta^{'}(r)^{2}
-16\Delta^{'}(r)^{2}\right)}{a^{2}\left(1+l\right)\Delta^{'}(r)^{2}},
\end{eqnarray}
where $'$ stands for differentiation with respect to radial
coordinate. The above expressions in the limit $l \rightarrow 0$
and $b \rightarrow 0$ reduce to those for Kerr black hole. It
would be useful at this point to introduce two celestial
coordinates for a better study of the shadow. The two celestial
coordinates, which are used to describe the shape of the shadow
that an observer sees in the sky, can be given by
\begin{eqnarray}\nonumber
\alpha(\xi, \eta ; \theta)&=&\lim _{r \rightarrow \infty} \frac{-r
p^{(\varphi)}}{p^{(t)}} = -\xi_{s} \csc \theta,\\\nonumber
\beta(\xi, \eta ; \theta)&=&\lim _{r \rightarrow \infty} \frac{r
p^{(\theta)}}{p^{(t)}}
=\sqrt{\left(\eta_{s}+a^{2} \cos ^{2} \theta-\xi_{s}^{2} \cot ^{2} \theta\right)},\\
\end{eqnarray}
where $\left(p^{(t)}, p^{(r)}, p^{(\theta)}, p^{(\phi)}\right)$
are the tetrad components of the photon momentum with respect to
locally non-rotating reference frames \cite{BARDEEN}.

With these inputs, we now plot black hole shadows for various cases
which are depicted in the figures below.

\begin{figure}[H]
\centering
%\begin{subfigure}{.5\textwidth}
%\centering
%\includegraphics[width=.7\linewidth]{shadow a=0.5 l=0.png}
%\caption{Sapes of the shadow for various values of $b$ with $a/M=.5, l=0$ and $\theta=\pi/2$}
%\end{subfigure}%
\begin{subfigure}{.5\textwidth}
\centering
\includegraphics[width=.7\linewidth]{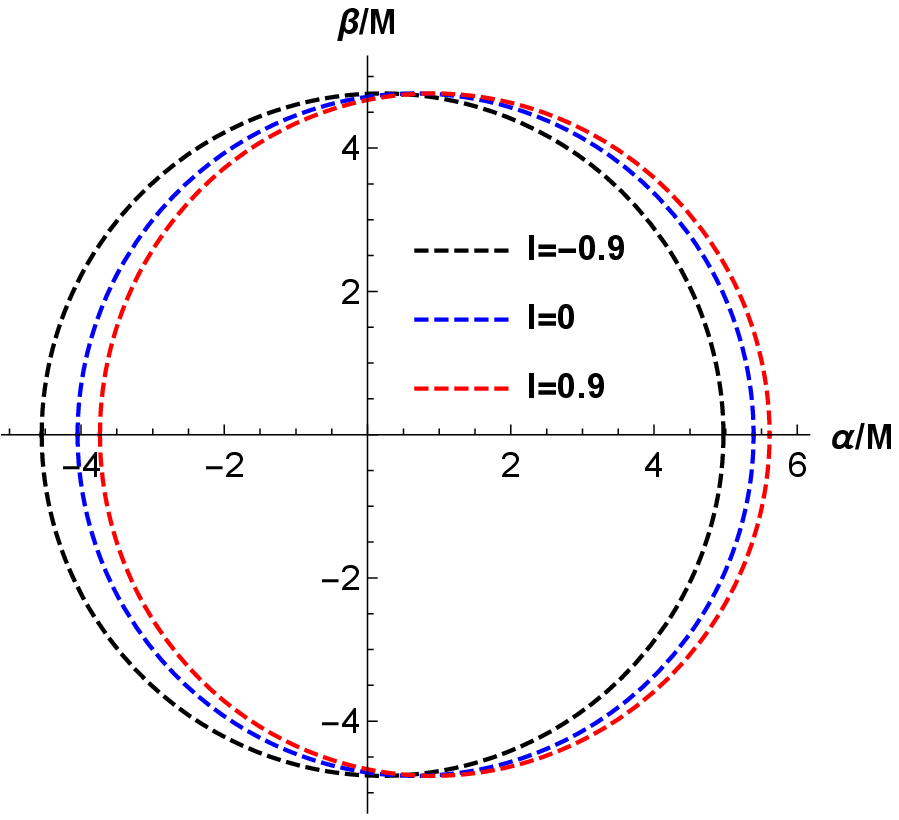}
\end{subfigure}%
\begin{subfigure}{.5\textwidth}
\centering
\includegraphics[width=.7\linewidth]{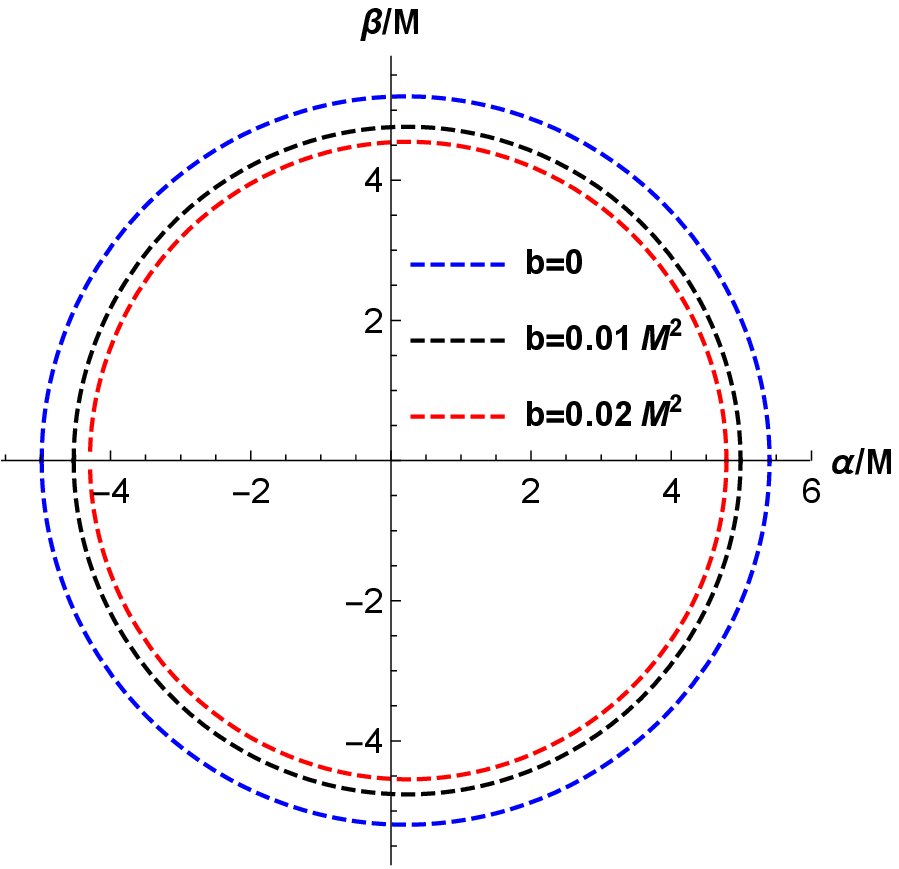}
\end{subfigure}%
\caption{The left panel gives shapes of the shadow for various
values of $l$ with $a=0.1M$, $b=0.01M^{2}$, and $\theta=\pi/2$.
The right panel gives shapes of the shadow for various values of
$b$ with $a=0.1M$ , $l=0.1$, and $\theta=\pi/2$.} \label{fig:test}
\end{figure}

\begin{figure}[H]
\centering
%\centering
\begin{subfigure}{.5\textwidth}
\centering
\includegraphics[width=.7\linewidth]{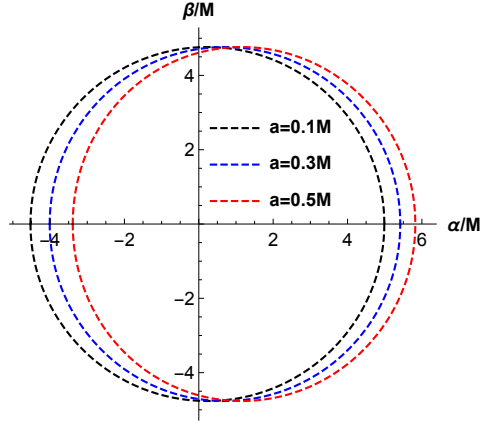}
\end{subfigure}
\caption{ The shapes of the shadow for various values of $a$ with
$b=0.01M^{2}$, $l=0.1$, and $\theta=\pi/2$. } \label{fig:test}
\end{figure}
From the above plots, we observe that the size of the shadow
increases with an increase in $l$, whereas it decreases with an
increase in $b$. Besides, if we increase $a$ then the shadow
shifts toward the right.

Using the parameters which are introduced by Hioki and Maeda
\cite{KH}, we analyze the deviation from the circularity form
$\left(\delta_{s}\right)$ and the size $\left(R_{s}\right)$ of the
shadow cast by the black hole.
\begin{figure}[H]
\centering
\begin{subfigure}{.5\textwidth}
\centering
\includegraphics[width=.7\linewidth]{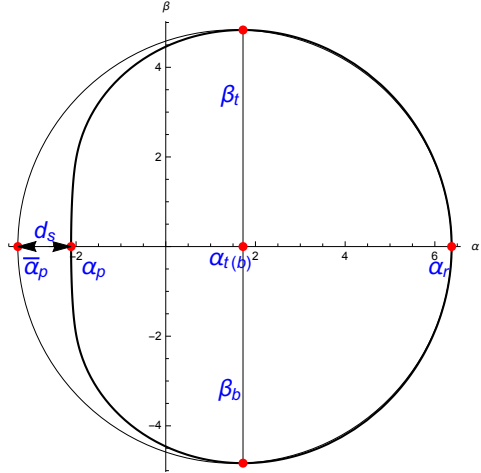}
\end{subfigure}
\caption{The black hole shadow and reference circle. $ds$ is the
distance between the left points of the shadow and the reference
circle.} \label{fig:test}
\end{figure}
For calculating these parameters, we consider five points
$\left(\alpha_{t}, \beta_{t}\right),\left(\alpha_{b}, \beta_{b}\right),\left(\alpha_{r}, 0\right)$
$\left(\alpha_{p}, 0\right)$ and $\left(\bar{\alpha}_{p},
0\right)$ which are top, bottom, rightmost, leftmost of the shadow
and leftmost of the reference circle respectively. So, we have
$$
R_{s}=\frac{\left(\alpha_{t}-\alpha_{r}\right)^{2}+\beta_{t}^{2}}{2\left|\alpha_{t}-\alpha_{r}\right|}
$$
and
$$
\delta_{s}=\frac{\left|\bar{\alpha}_{p}-\alpha_{p}\right|}{R_{s}}.
$$
In the following Fig. We plot $R_{s}$ and $\delta_{s}$ for various
scenarios to study how $R_{s}$ and $\delta_{s}$ varies with
parameters of the modified theory of gravity.
\begin{figure}[H]
\centering
\begin{subfigure}{.5\textwidth}
\centering
\includegraphics[width=.7\linewidth]{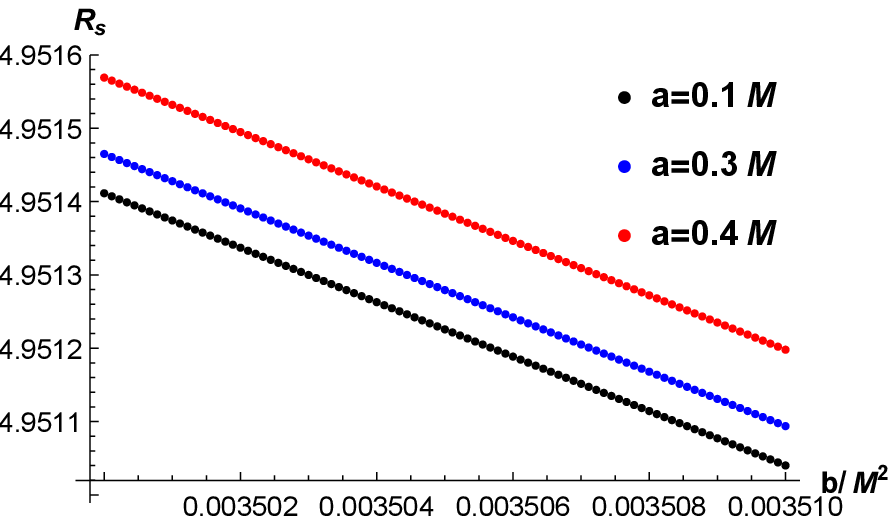}
\end{subfigure}%
\begin{subfigure}{.5\textwidth}
\centering
\includegraphics[width=.7\linewidth]{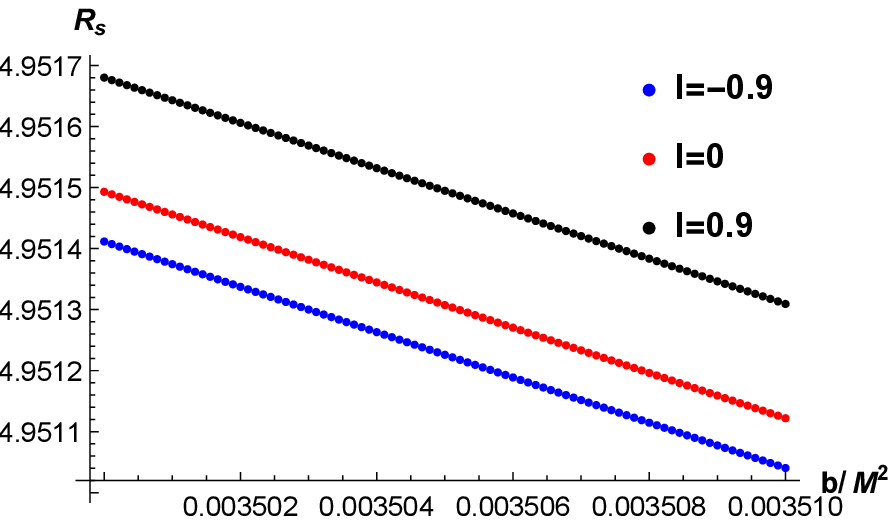}
\end{subfigure}
\caption{The left panel shows variation of $R_{s}$ for various
values of $a$ with $l=-0.2$ and $\theta=\pi/2$, and the right
panel shows the variation of $R_{s}$ for various values of $l$
with $a=0.3M$ and $\theta=\pi/2$. } \label{fig:test}
\end{figure}

\begin{figure}[H]
\centering
\begin{subfigure}{.5\textwidth}
\centering
\includegraphics[width=.7\linewidth]{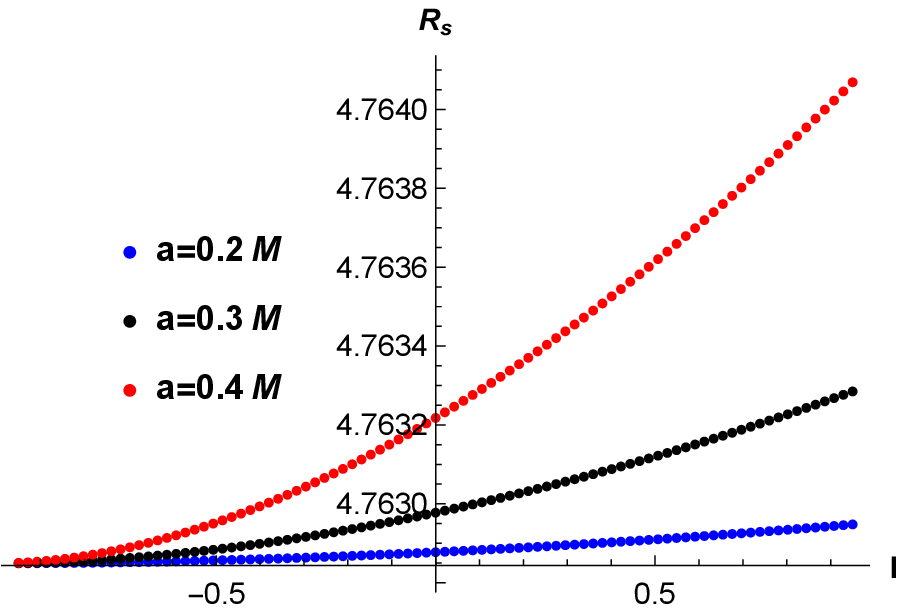}
\end{subfigure}%
\begin{subfigure}{.5\textwidth}
\centering
\includegraphics[width=.7\linewidth]{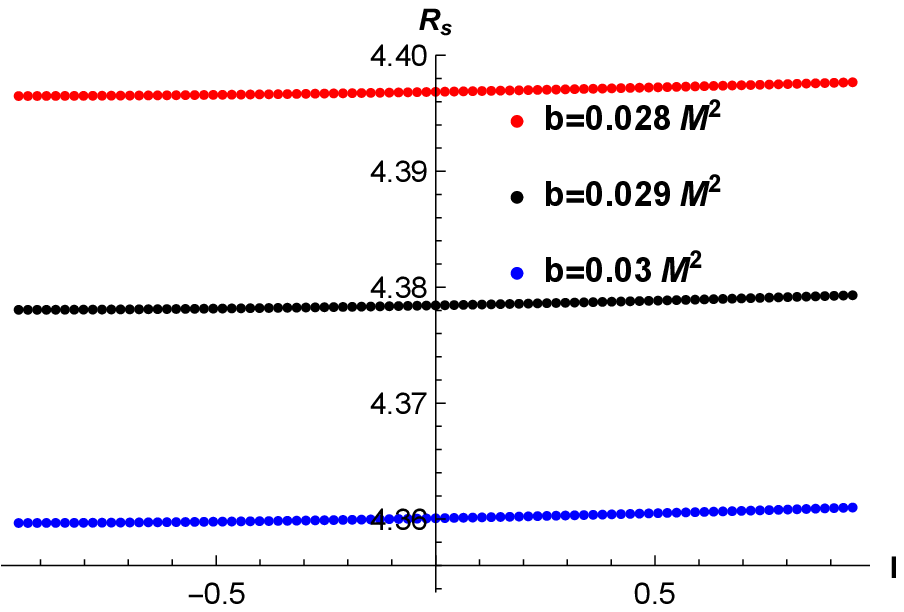}
\end{subfigure}
\caption{The left panel shows variation of $R_{s}$ for various
values of $a$ with $b=0.01M^{2}$ and $\theta=\pi/2$ and the right
panel shows the variation of $R_{s}$ for various values of $b$
with $a=0.3M$ and $\theta=\pi/2$. } \label{fig:test}
\end{figure}

\begin{figure}[H]
\centering
\begin{subfigure}{.5\textwidth}
\centering
\includegraphics[width=.7\linewidth]{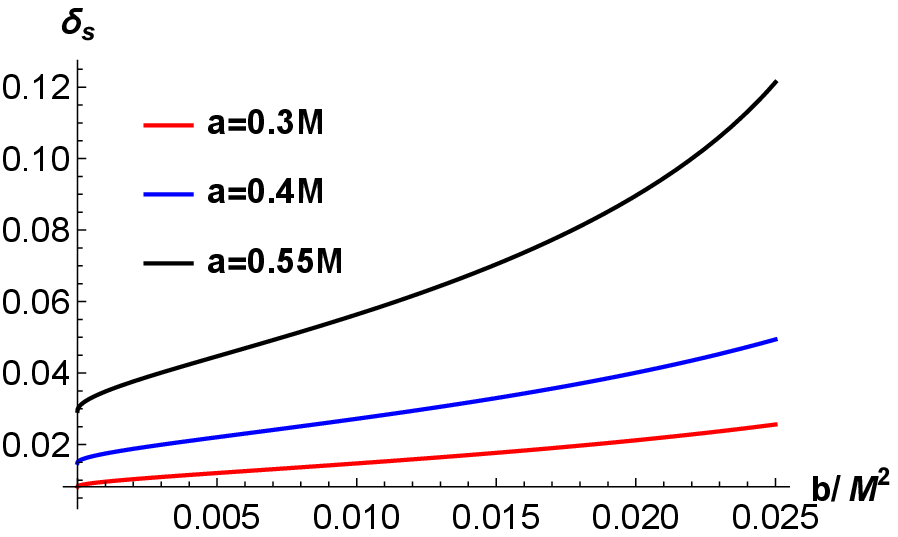}
\end{subfigure}%
\begin{subfigure}{.5\textwidth}
\centering
\includegraphics[width=.7\linewidth]{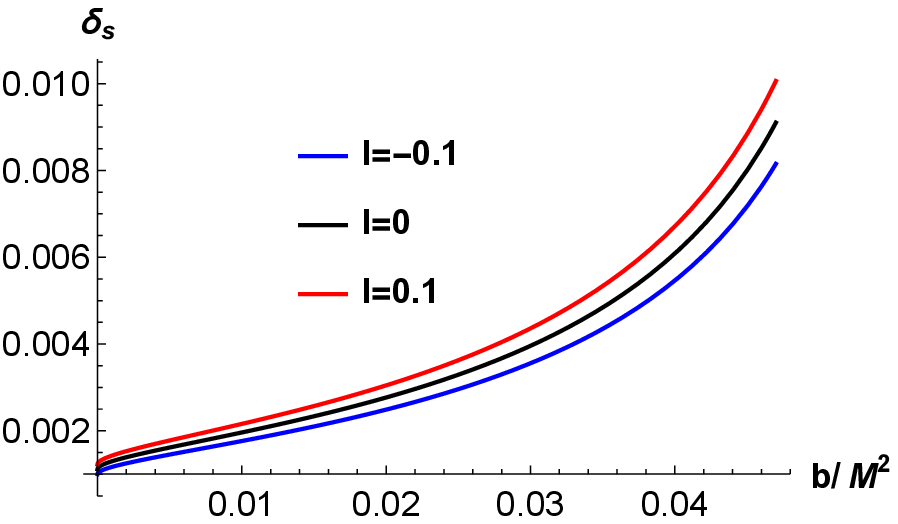}
\end{subfigure}
\caption{The left panel shows variation of $\delta_{s}$ for
various values of $a$ with $l=-0.2$ and $\theta=\pi/2$ and the
right panel shows the variation of $\delta_{s}$ for various values
of $l$ with $a=0.1M$ and $\theta=\pi/2$.} \label{fig:test}
\end{figure}

\begin{figure}[H]
\centering
\begin{subfigure}{.5\textwidth}
\centering
\includegraphics[width=.7\linewidth]{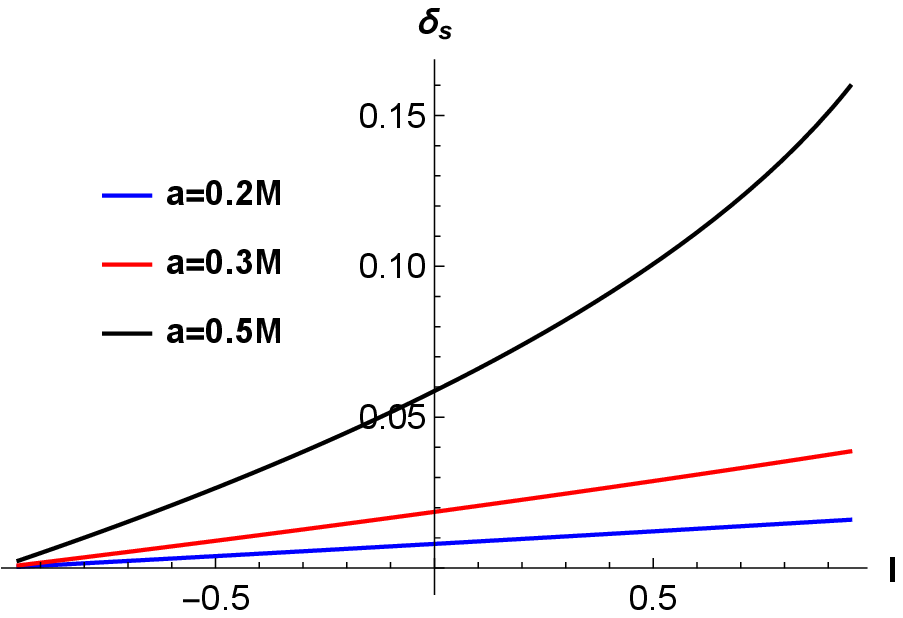}
\end{subfigure}%
\begin{subfigure}{.5\textwidth}
\centering
\includegraphics[width=.7\linewidth]{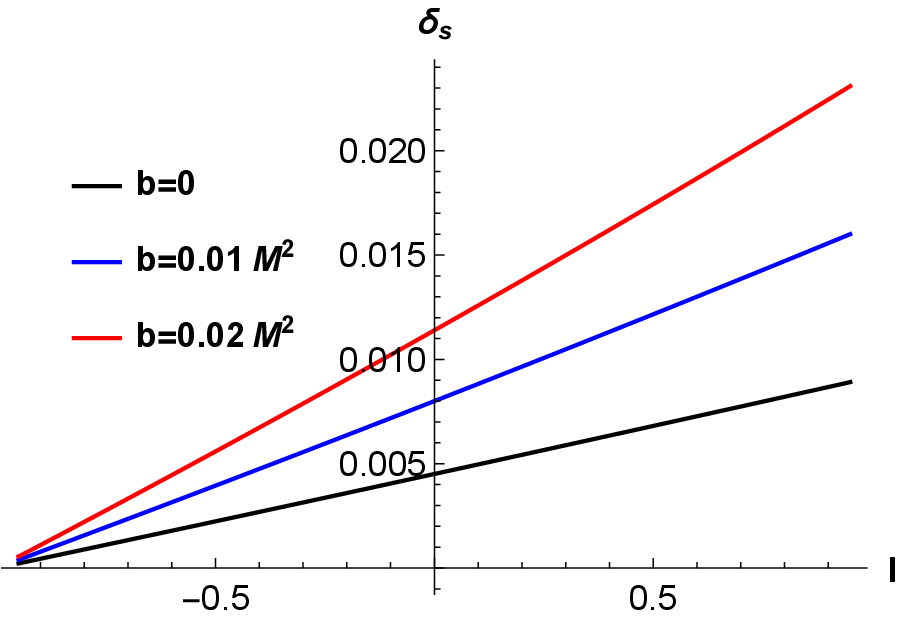}
\end{subfigure}
\caption{The left panel shows the variation of $\delta_{s}$ for
various values of $a$ with $b=.01M^{2}$ and $\theta=\pi/2$ and the
right panel one gives the variation of $\delta_{s}$ for various values
of $b$ with $a=0.2M$ and $\theta=\pi/2$.} \label{fig:test}
\end{figure}
From the above plots we observe that $R_{s }$ decreases with an
increase in $b$ for fixed values of $a$ and $l$, whereas
for fixed values of $a$ and $b$, it increases with an increase
in $l$. On the other hand, $\delta_{s}$ increases with an increase
in $l$ for fixed values of $a$ and $b$ as well as with an increase
in $b$ for fixed values of $a$ and $l$.

\section{Computation of energy emission rate}
In this part, we study the possible visibility of the non-commutative
Kerr-like black hole through shadow. In the vicinity of limiting
constant value,
the cross-section of the black
hole's absorption moderates lightly at high energy. We know that
a rotating black hole can
absorb electromagnetic waves, so the absorbing cross-section for
a spherically symmetric black hole is \cite{BM}
\begin{equation}
\sigma_{l i m}=\pi R_{s}^{2}.
\end{equation}
Using the above equation the energy emission rate is obtained
\cite{AA}:
\begin{equation}
\frac{d^{2} E}{d \omega d t}=\frac{2 \pi^{3}
R_{s}^{2}}{e^{\left(\frac{\omega}{T}\right)}-1} \omega^{3},
\end{equation}
where $T=\frac{\sqrt{1+\ell} \Delta^{\prime}\left(r_{+}\right)}{4
\pi\left[r_{+}^{2} +(1+\ell) a^{2}\right]}$ is the Hawking
temperature and $\omega$ the frequency.

\begin{figure}[H]
\centering
\begin{subfigure}{.5\textwidth}
\centering
\includegraphics[width=.7\linewidth]{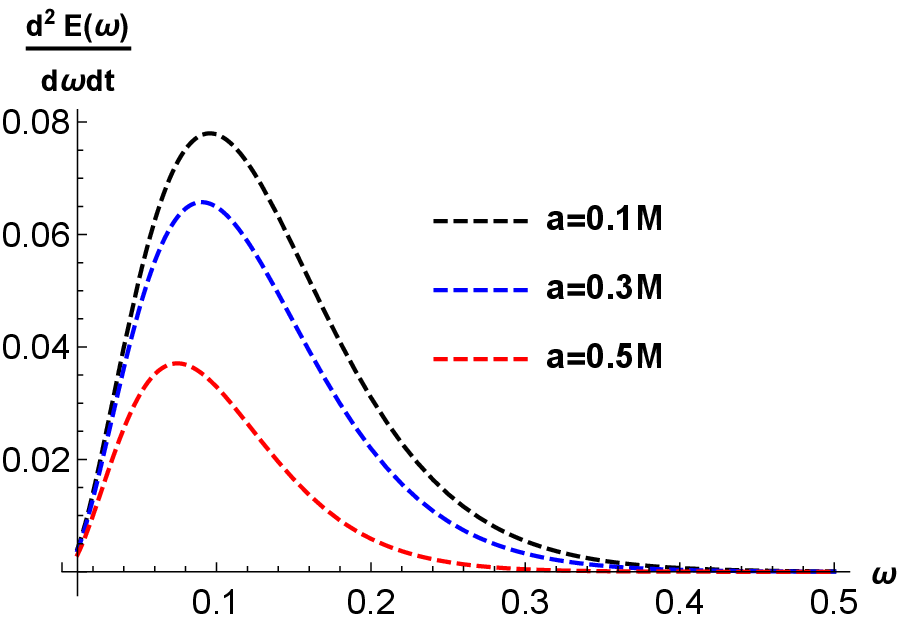}
\end{subfigure}%
\begin{subfigure}{.5\textwidth}
\centering
\includegraphics[width=.7\linewidth]{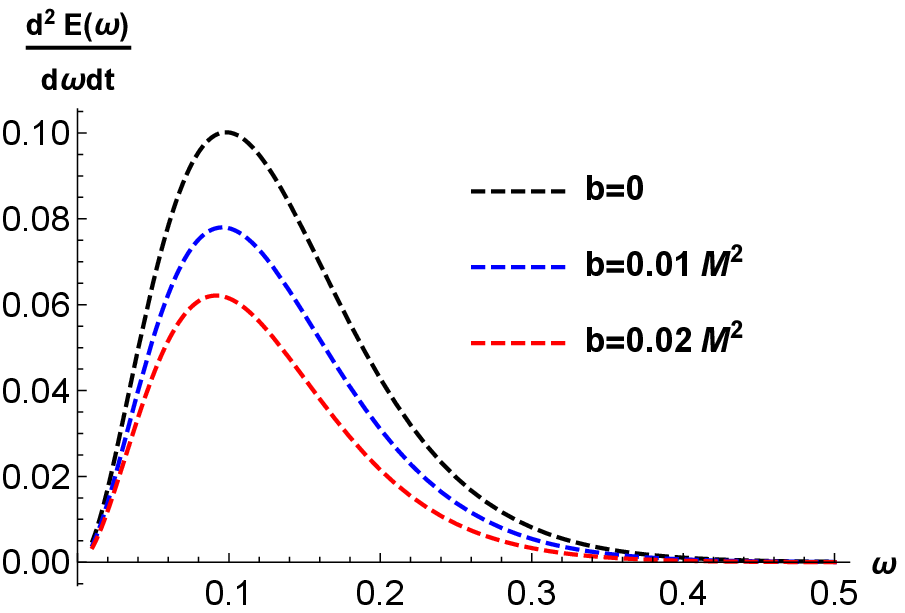}
\end{subfigure}
\caption{The left panel gives variation of emission rate against
$\omega$ for various values of $a$ with $b=.01M^{2}$ and $l=0.3$.
The right panel gives variation of emission rate against $\omega$
for various values of $b$ with $a=0.1M$ and $l=0.3$.}
\label{fig:test}
\end{figure}

\begin{figure}[H]
\centering
\begin{subfigure}{.5\textwidth}
\centering
\includegraphics[width=.7\linewidth]{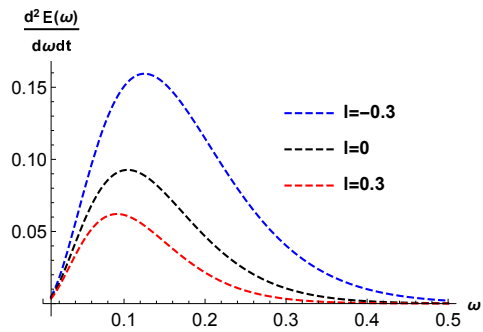}
\end{subfigure}
\caption{It gives variation of emission rate against $\omega$ for
various values of $l$ with $b=0.02M^{2}$ and $a=0.1M$.}
\label{fig:test}
\end{figure}
In Fig. 17 and 18, we have shown the plots of energy emission rate
versus $\omega$ for various cases. It is clear from the plots that
the emission rate decreases with an increase in the value of $b$
for any set of fixed values of $a$ and $l$. It also decreases with
an increase in $l$, for $a$ and $b$ being fixed, and with an
increase in $a$, when $l$ and $b$ remain fixed. Let us now
consider the case when some medium is present which is a more
natural one.

We now consider the situation when the black hole is veiled with a
dispersive medium like plasma and compute the emission rate in
this situation. In the presence of plasma, the celestial
coordinates are given by \cite{OUR}
\begin{eqnarray}\nonumber
\alpha(\xi, \eta ; \theta)&=& -\frac{\xi_{s} \csc
\theta}{n},\\\nonumber \beta(\xi, \eta ;
\theta)&=&\frac{\sqrt{\eta+(1+l) a^{2} \cos ^{2} \theta-\xi^{2}
\cot
^{2} \theta -\left(n^{2}-1 \right)a^{2}\left(1+l \right)sin^{2}\theta}}{n},\\
\end{eqnarray}
where $n=\sqrt{1-\frac{k}{r}}$ \cite{ROGERS} is the refractive
index of the plasma, k being the plasma constant. Using these
expressions combined with the expression of emission rate we
investigate the variation of rate of emission for various
situations.

\begin{figure}[H]
\centering
\begin{subfigure}{.5\textwidth}
\centering
\includegraphics[width=.7\linewidth]{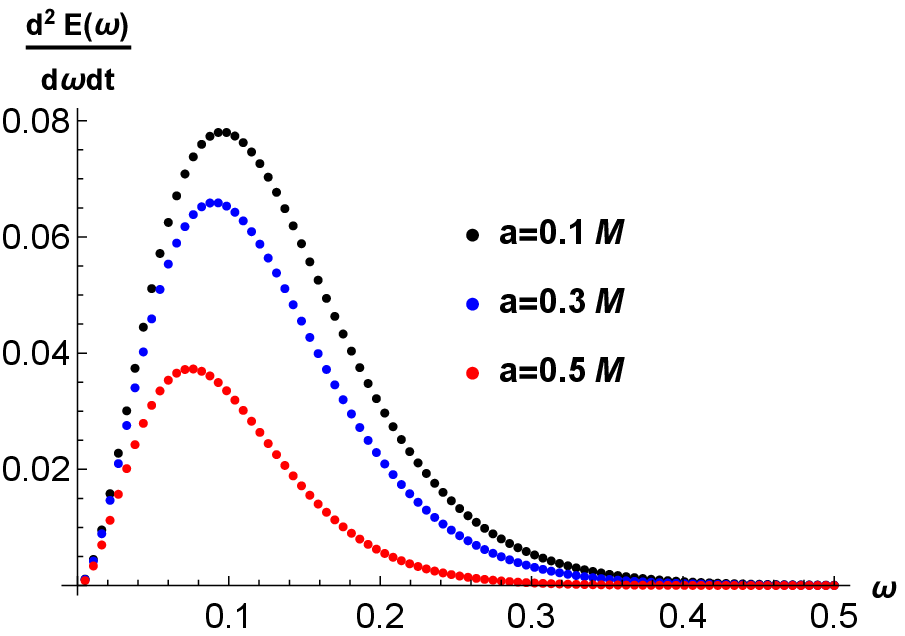}
\end{subfigure}%
\begin{subfigure}{.5\textwidth}
\centering
\includegraphics[width=.7\linewidth]{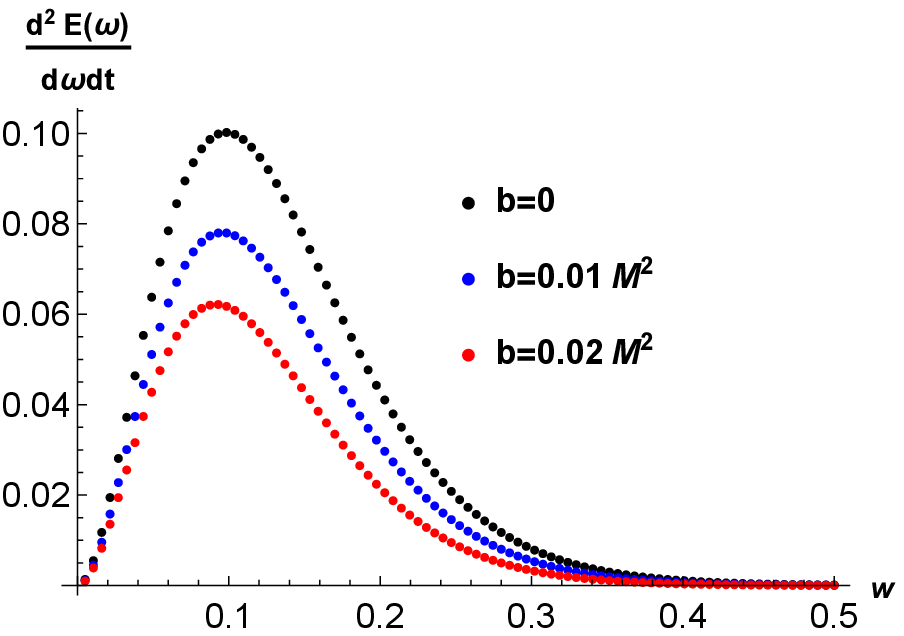}
\end{subfigure}
\caption{The left panel gives the variation of emission rate
against $\omega$ for various values of $a$ with
$b=0.01M^{2}$,$k=0.2M$, and $l=0.3$. The right panel gives
variation of emission rate against $\omega$ for various values of
$b$ with $a=0.1M$, $k=0.2M$, and $l=0.3$.} \label{fig:test}
\end{figure}

\begin{figure}[H]
\centering
\begin{subfigure}{.5\textwidth}
\centering
\includegraphics[width=.7\linewidth]{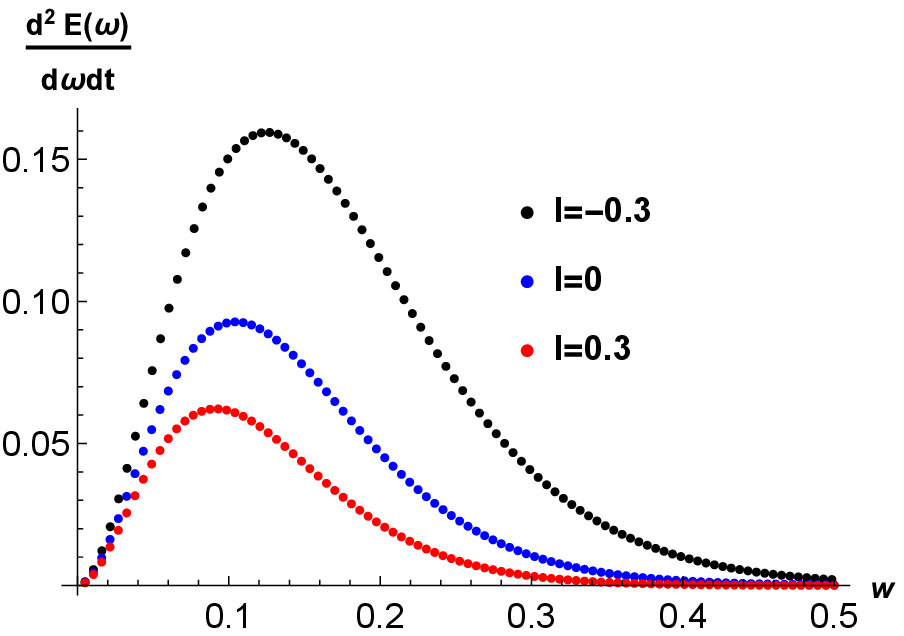}
\end{subfigure}%
\begin{subfigure}{.5\textwidth}
\centering
\includegraphics[width=.7\linewidth]{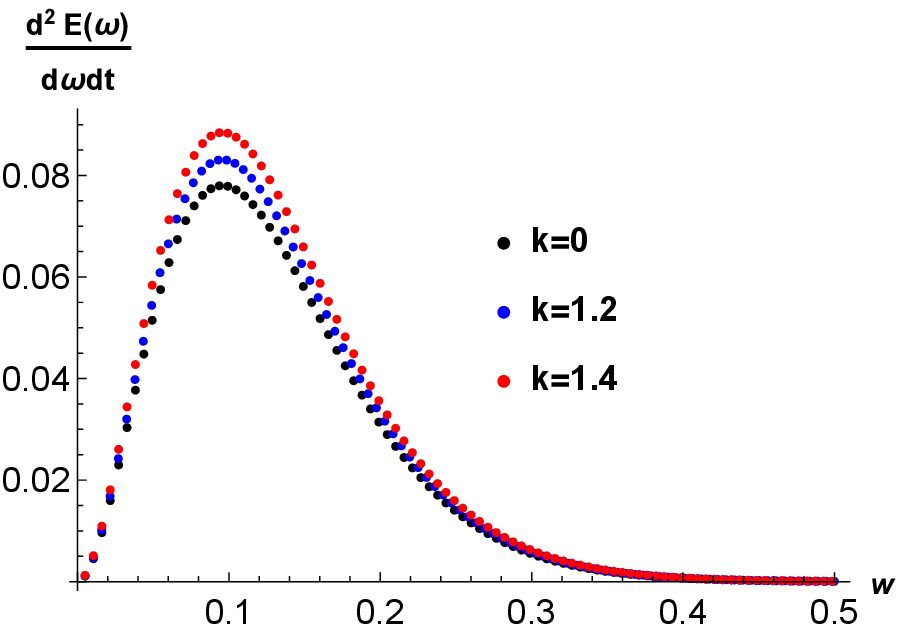}
\end{subfigure}
\caption{The left panel gives variation of emission rate against
$\omega$ for various values of $l$ with $b=.02M^{2}$, $k=0.2M$,
and $a=0.1$. The right one gives the variation of emission rate
against $\omega$ for various values of $k$ with $a=0.1M$
,$b=0.01M^{2}$, and $l=0.3$. } \label{fig:test}
\end{figure}
Here we observe that the rate of emission increases with the
plasma constant $k$ for fixed values of $a, b$, and $l$. The other
variations are similar to those which we have seen without plasma,
though with reduced values.

\section{Constraining from the observed data for $\mathrm{M}87^{*}$}
This section is devoted to the constraining of the parameter which
are involved in the modified theory. We compare the shadows
produced from the numerical calculation by the non-commutative
Kerr-like black holes with the observed one for the
$\mathrm{M}87^{*}$ black hole. For comparison, we consider the
experimentally obtained astronomical data for the deviation from
circularity $\Delta \leq 0.10$ and angular diameter
$\theta_{d}=42\pm 3 \mu as$. The boundary of the shadow is
described by the polar coordinate $(R(\phi),\phi)$ with the origin
at the center of the shadow $(\alpha_{C}, \beta_{C})$ where
$\alpha_{C}=\frac{|\alpha_{max}+\alpha_{min}|}{2}$, and
$\beta_{C}=0$.

If a point $(\alpha, \beta)$ over the boundary of the image
subtends an angle $\phi$ on the $\alpha$ axis at the geometric
center, $\left(\alpha_{C}, 0\right)$, and $R(\phi)$ be the
distance between the point $(\alpha, \beta)$ and
$\left(\alpha_{C}, 0\right)$, then the average radius
$R_{\text{avg}}$ of the image is given by \cite{CBK}
\begin{equation}
R_{\text {avg}}^{2} \equiv \frac{1}{2 \pi} \int_{0}^{2 \pi} d \phi R^{2}(\phi), \\
\end{equation}
where $R(\phi) \equiv
\sqrt{\left(\alpha(\phi)-\alpha_{C}\right)^{2}+\beta(\phi)^{2}}$,
and $\phi = tan^{-1}\frac{\beta(\phi)}{\alpha(\phi)-\alpha_{C}}$.

With the above inputs, the circularity deviation $\Delta C$ is
defined by \cite{TJDP}
\begin{equation}
\Delta C \equiv 2\sqrt{\frac{1}{2 \pi} \int_{0}^{2 \pi} d
\phi\left(R(\phi)-R_{\text {avg }}\right)^{2}}.
\end{equation}
We also consider the angular diameter of the shadow which is
define by
\begin{equation}
\theta_{d}=\frac{2}{d}\sqrt{\frac{A}{\pi}},
\end{equation}
where $A=2\int_{r_{-}}^{r_{+}} \beta d\alpha $ is the area of the
shadow and $d=16.8 Mpc$ is the distance of $M87^{*}$ from the
Earth. These relations will enable us to accomplish a comparison
between the theoretical predictions for non-commutative Kerr-like
black-hole shadows and the experimental findings of the EHT
collaboration. In the figures below the deviation from
circularity, $\Delta C$ is shown for non-commutative Kerr-like
black holes for inclination angles $\theta=90^{o}$ and
$\theta=17^{o}$ respectively.

\begin{figure}[H]
\centering
\begin{subfigure}{.25\textwidth}
\centering
\includegraphics[scale=.65]{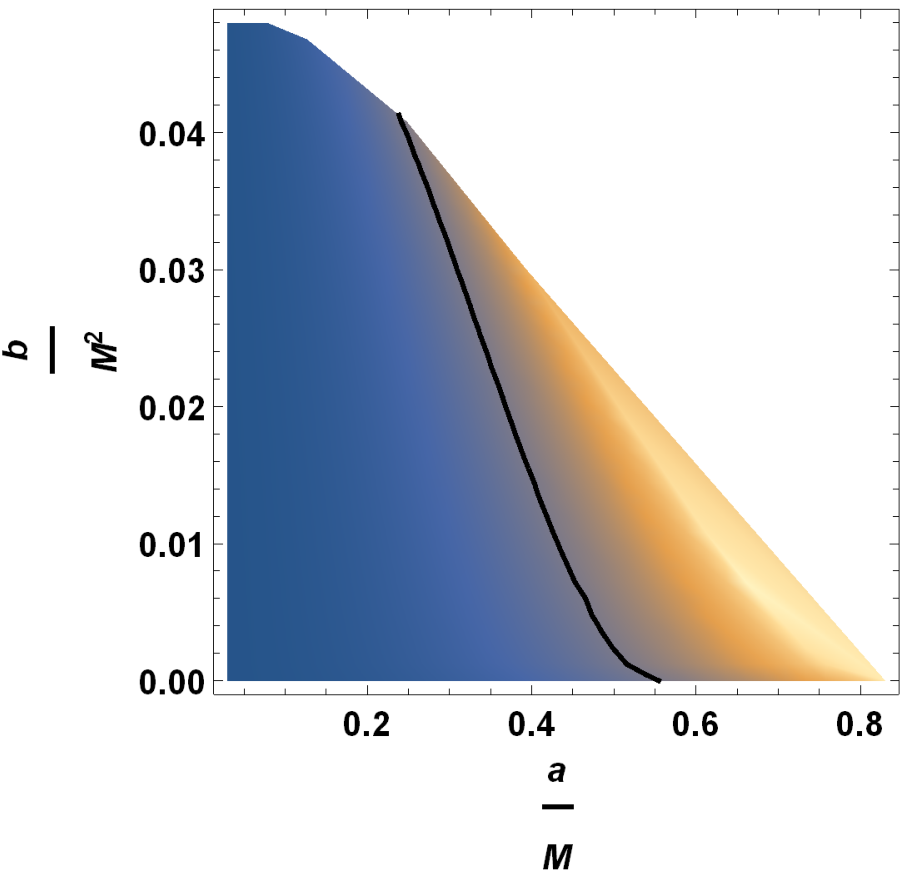}
%\caption{Critical radius for various values of $l$ with $k=.1,b=.1$ and $\theta=\pi/2$}
\end{subfigure}%
\begin{subfigure}{.28\textwidth}
\centering
\raisebox{.2\height}{\includegraphics[scale=.55]{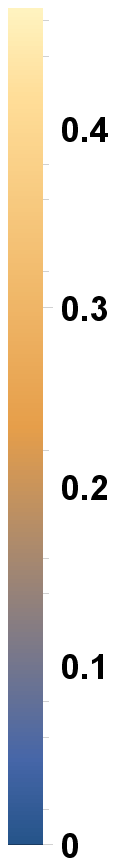}}\hspace{1.5em}%
%\caption{Critical radius for various values of $l$ with $k=.1,b=.1$ and $\theta=\pi/2$}
\end{subfigure}%
\begin{subfigure}{.25\textwidth}
\centering
\includegraphics[scale=.65]{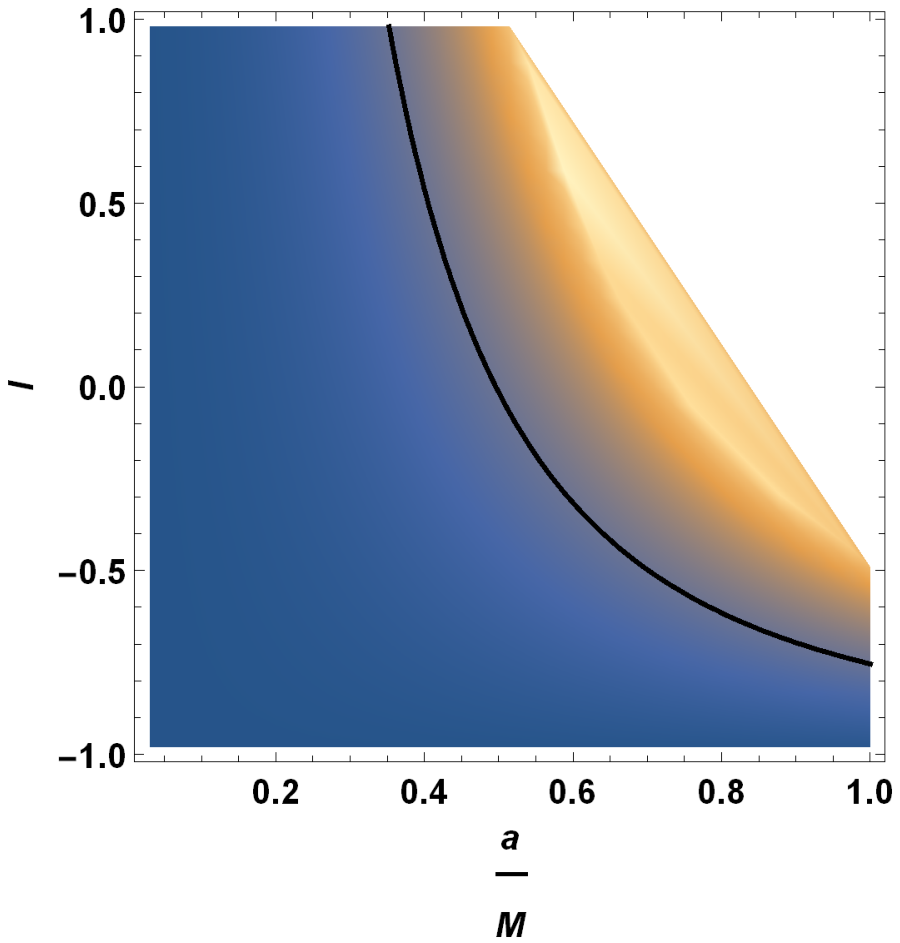}
%\caption{Critical radius for various values of $l$ with $k=.1,b=.1$ and $\theta=\pi/2$}
\end{subfigure}%
\begin{subfigure}{.28\textwidth}
\centering
\raisebox{.19\height}{\includegraphics[scale=.6]{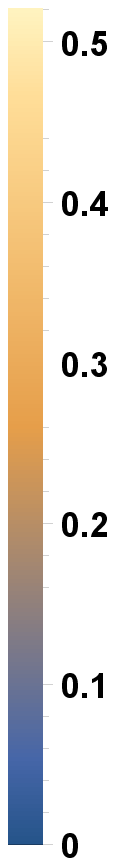}}
%\caption{Critical radius for various values of $l$ with $k=.1,b=.1$ and $\theta=\pi/2$}
\end{subfigure}
\caption{The left panel is for $l=0.4$, and the right panel is for
$b=0.01M^{2}$ where the inclination angle is $90^{o}$. The black
solid lines correspond to $\Delta C=0.1$. }
%\label{fig:test}
\end{figure}
\smallskip
\begin{figure}[H]
\centering
\begin{subfigure}{.25\textwidth}
\centering
\includegraphics[scale=.65]{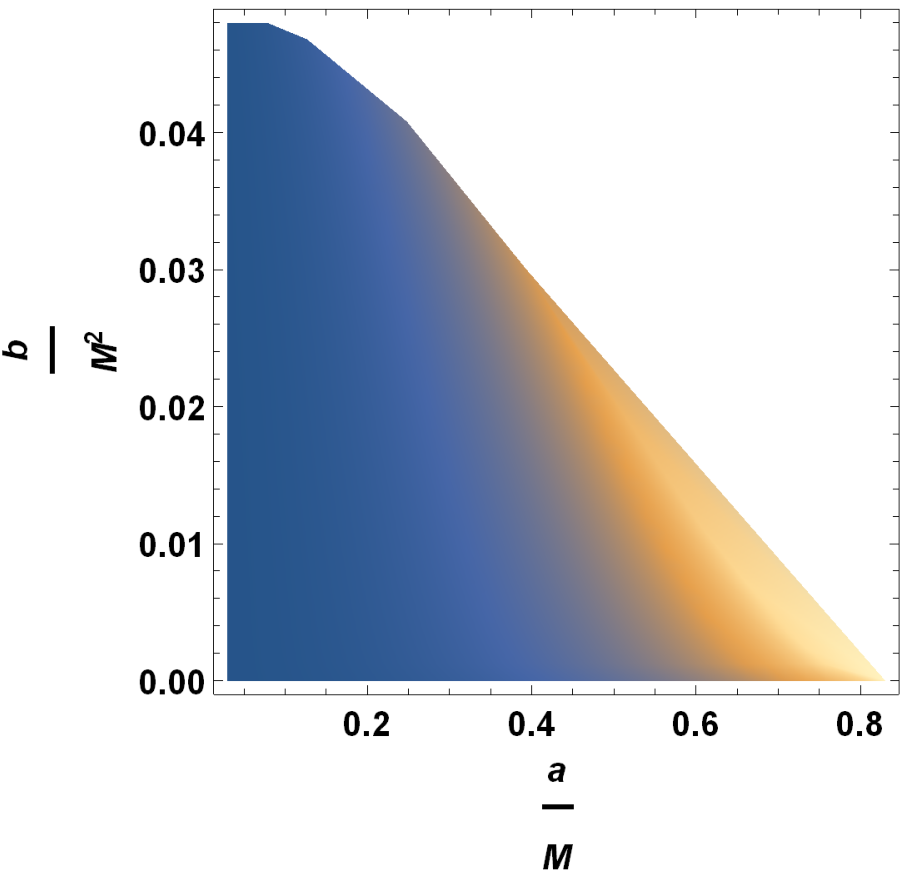}\hspace{1.5em}%
%\caption{Critical radius for various values of $l$ with $k=.1,b=.1$ and $\theta=\pi/2$}
\end{subfigure}%
\begin{subfigure}{.28\textwidth}
\centering
\raisebox{.2\height}{\includegraphics[scale=.55]{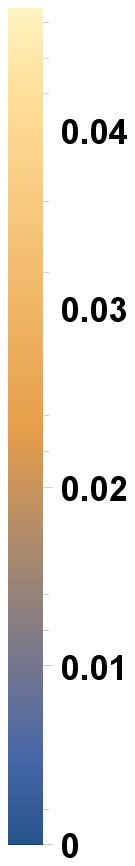}}\hspace{1.5em}%
%\caption{Critical radius for various values of $l$ with $k=.1,b=.1$ and $\theta=\pi/2$}
\end{subfigure}%
\begin{subfigure}{.25\textwidth}
\centering
\includegraphics[scale=.65]{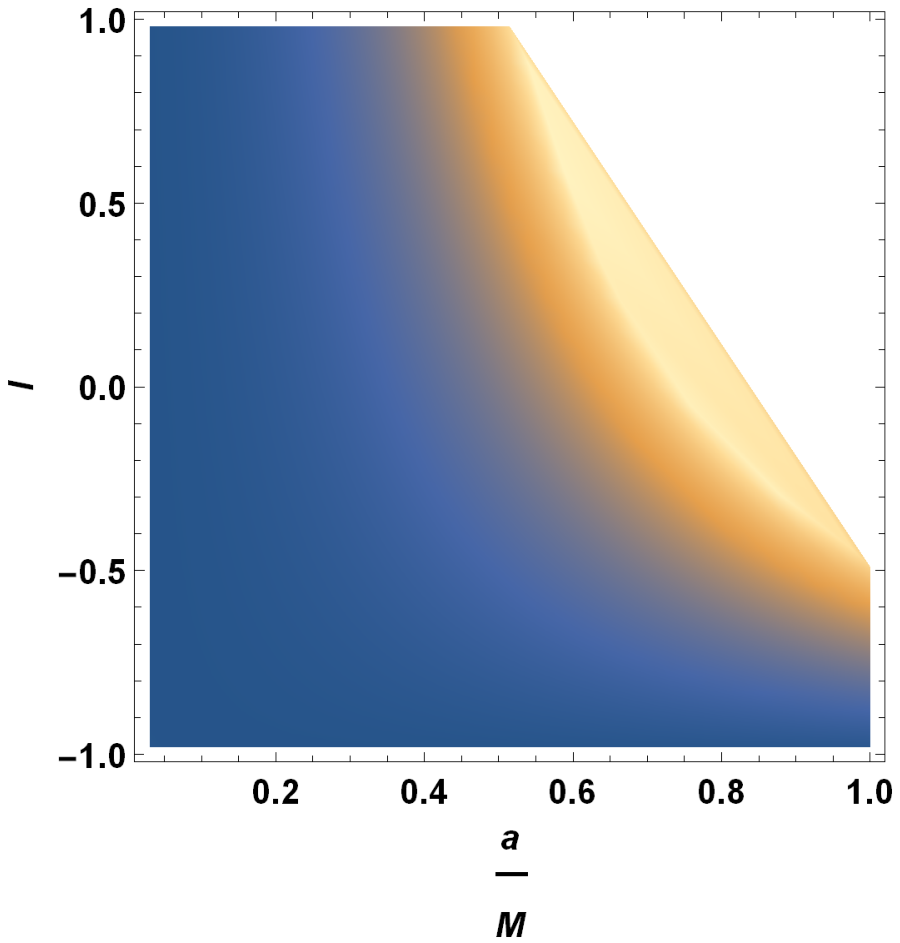}\hspace{1.5em}%
%\caption{Critical radius for various values of $l$ with $k=.1,b=.1$ and $\theta=\pi/2$}
\end{subfigure}%
\begin{subfigure}{.28\textwidth}
\centering
\raisebox{.19\height}{\includegraphics[scale=.6]{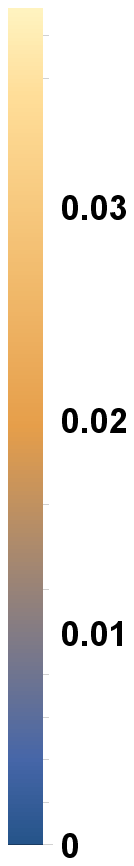}}
%\caption{Critical radius for various values of $l$ with $k=.1,b=.1$ and $\theta=\pi/2$}
\end{subfigure}
\caption{The left panel is for $l=0.4$, and the right one is for
$b=0.01M^{2}$ where the inclination angle is $17^{o}$. }
%\label{fig:test}
\end{figure}
In the figures depicted below the angular diameter $\theta_{d}$ is
shown for non-commutative Kerr-like black holes for inclination
angles $\theta=90^{o}$ and $\theta=17^{o}$ respectively.
%\medskip

\begin{figure}[H]
\centering
\begin{subfigure}{.25\textwidth}
\centering
\includegraphics[scale=.65]{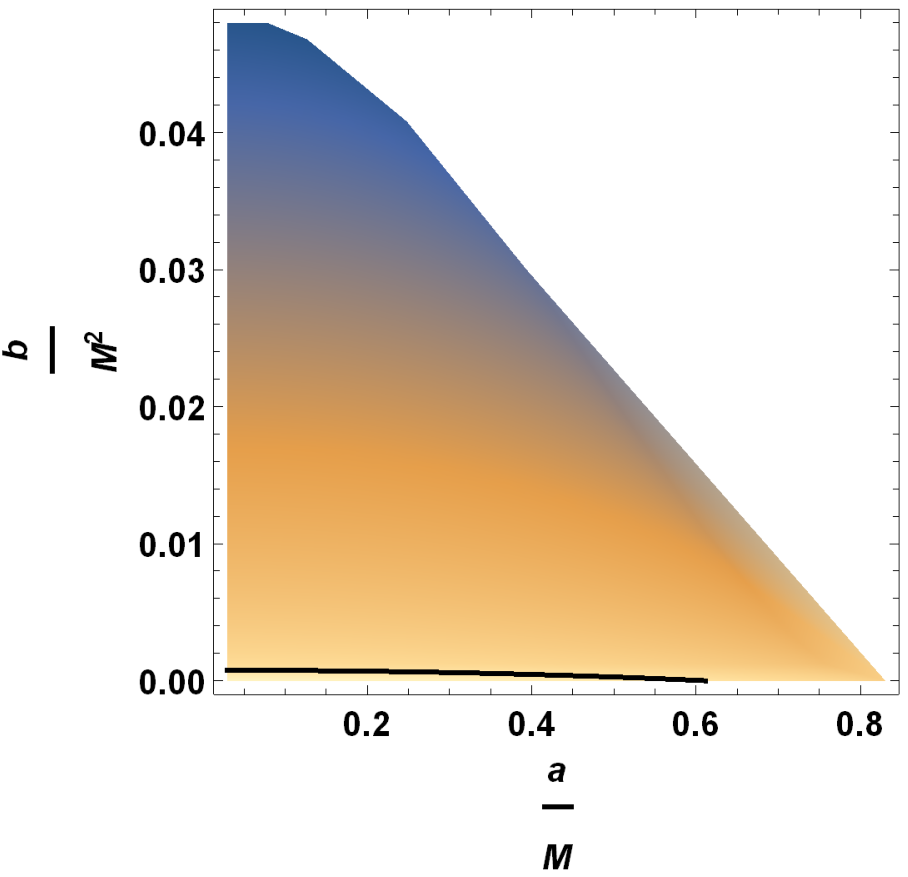}\hspace{1.5em}%
%\caption{Critical radius for various values of $l$ with $k=.1,b=.1$ and $\theta=\pi/2$}
\end{subfigure}%
\begin{subfigure}{.28\textwidth}
\centering
\raisebox{.2\height}{\includegraphics[scale=.55]{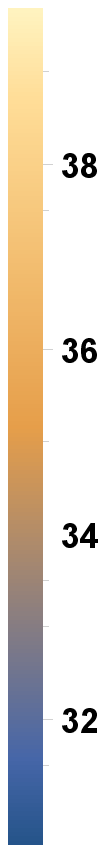}}\hspace{1.5em}%
%\caption{Critical radius for various values of $l$ with $k=.1,b=.1$ and $\theta=\pi/2$}
\end{subfigure}%
\begin{subfigure}{.25\textwidth}
\centering
\includegraphics[scale=.65]{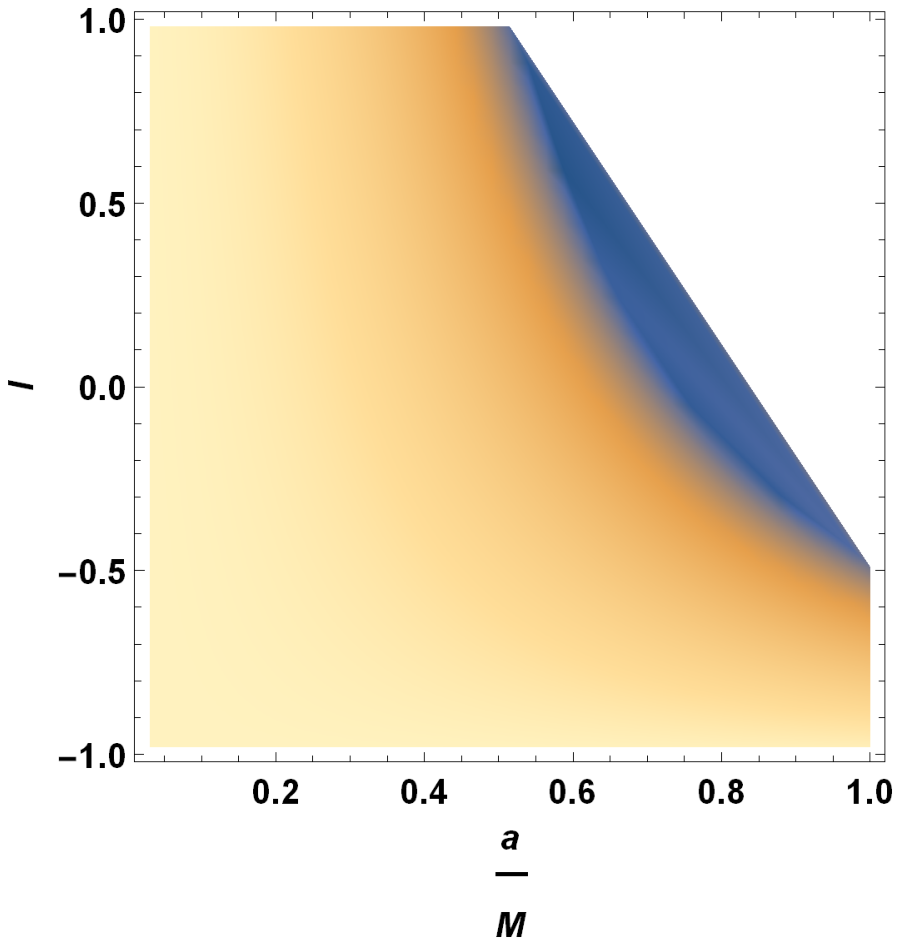}\hspace{1.5em}%
%\caption{Critical radius for various values of $l$ with $k=.1,b=.1$ and $\theta=\pi/2$}
\end{subfigure}%
\begin{subfigure}{.28\textwidth}
\centering
\raisebox{.19\height}{\includegraphics[scale=.6]{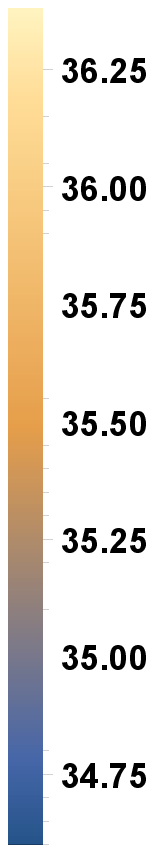}}
%\caption{Critical radius for various values of $l$ with $k=.1,b=.1$ and $\theta=\pi/2$}
\end{subfigure}
\caption{The left panel is for $l=0.4$, and the right panel is for
$b=0.01M^{2}$ where the inclination angle is $90^{o}$. The black
solid lines correspond to $\theta_{d}=39 \mu as$.}
%\label{fig:test}
\end{figure}
\smallskip
\begin{figure}[H]
\centering
\begin{subfigure}{.25\textwidth}
\centering
\includegraphics[scale=.65]{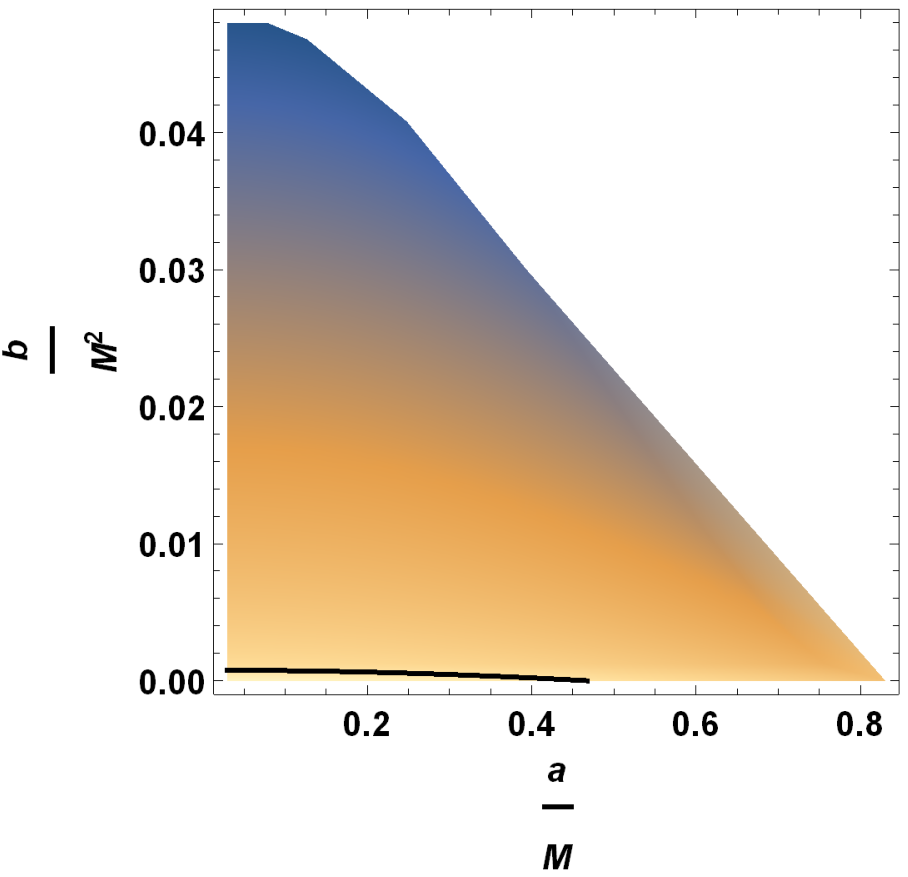}\hspace{1.5em}%
%\caption{Critical radius for various values of $l$ with $k=.1,b=.1$ and $\theta=\pi/2$}
\end{subfigure}%
\begin{subfigure}{.28\textwidth}
\centering
\raisebox{.2\height}{\includegraphics[scale=.55]{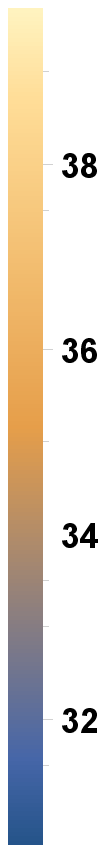}}\hspace{1.5em}%
%\caption{Critical radius for various values of $l$ with $k=.1,b=.1$ and $\theta=\pi/2$}
\end{subfigure}%
\begin{subfigure}{.25\textwidth}
\centering
\includegraphics[scale=.65]{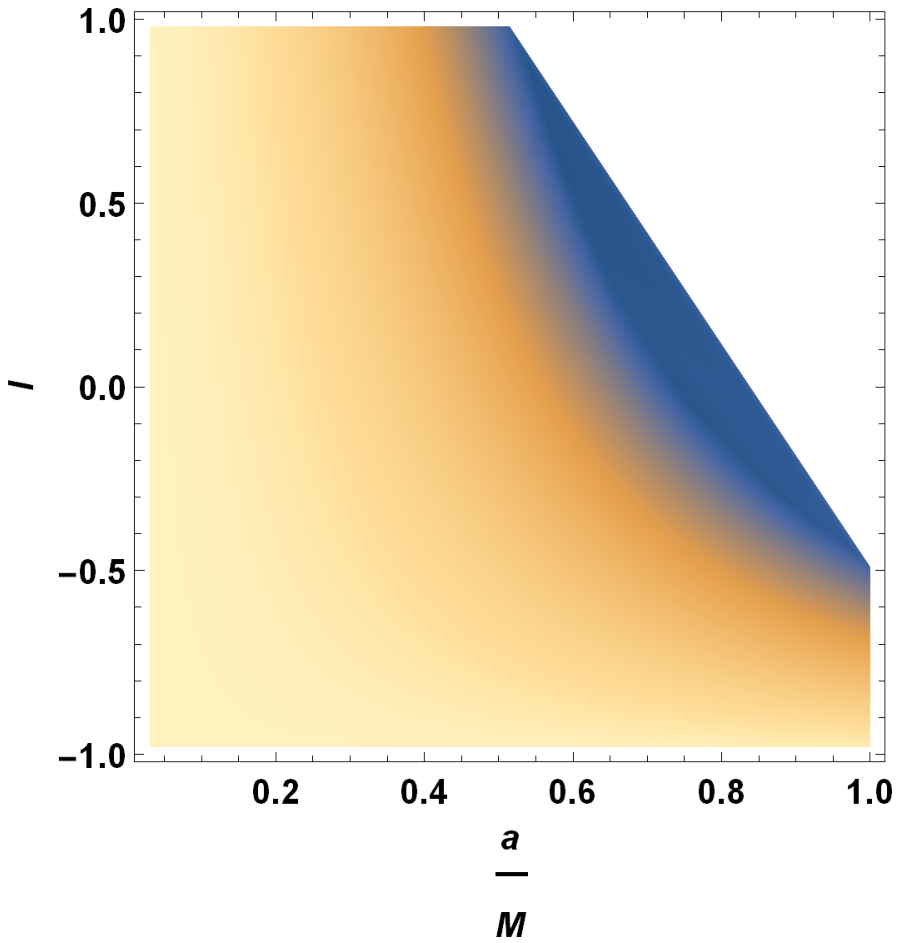}\hspace{1.5em}%
%\caption{Critical radius for various values of $l$ with $k=.1,b=.1$ and $\theta=\pi/2$}
\end{subfigure}%
\begin{subfigure}{.28\textwidth}
\centering
\raisebox{.19\height}{\includegraphics[scale=.6]{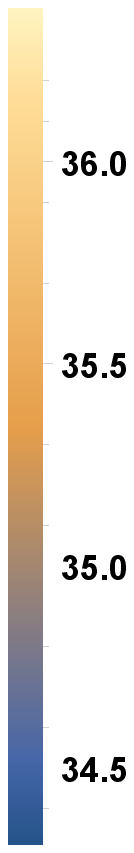}}
%\caption{Critical radius for various values of $l$ with $k=.1,b=.1$ and $\theta=\pi/2$}
\end{subfigure}
\caption{The left panel is for $l=0.4$, and the right panel is for
$b=0.01M^{2}$ where the inclination angle is $17^{o}$. The black
solid lines correspond to $\theta_{d}=39 \mu as$.}
%\label{fig:test}
\end{figure}

From the above plots, we can conclude that the constrain $\Delta
C\leq0.1$ is satisfied for finite parameter space when the
inclination angle is $90^{o}$, whereas, when the inclination angle
is $17^{o}$, the constrain is satisfied for the entire parameter
space. For inclination angles $\theta=90^{o}$ and $\theta=17^{o}$,
the constrain $\theta_{d}=42\pm3\mu$ within $1\sigma$ region is
satisfied for finite parameter space. The circular asymmetry in
the $M87^{*}$ shadow can also be defined in terms of the axial
ratio $D_{X}$ which is the ratio of the major to the minor
diameter of the shadow \cite{KA1}. It is defined by \cite{BCS}
\begin{equation}
D_{X}=\frac{\Delta Y}{\Delta X}=\frac{\beta_{t}-\beta_{b}}{\alpha_{r}-\alpha_{p}}.
\end{equation}
We should have $1 < D_{X}\lesssim4/3$ in accordance with the EHT
observations of $M87^{*}$ \cite{KA1}. Note that $D_{X}$ is another
way of defining $\Delta C$. Axial ratio of $4:3$ indeed
corresponds to a $\Delta C \leq 0.1$ \cite{KA1}. In the figures
below axial ratio, $D_{X}$ is shown for non-commutative Kerr-like
black holes for inclination angles $\theta=90^{o}$ and
$\theta=17^{o}$ respectively.
%\medskip

\begin{figure}[H]
\centering
\begin{subfigure}{.25\textwidth}
\centering
\includegraphics[scale=.65]{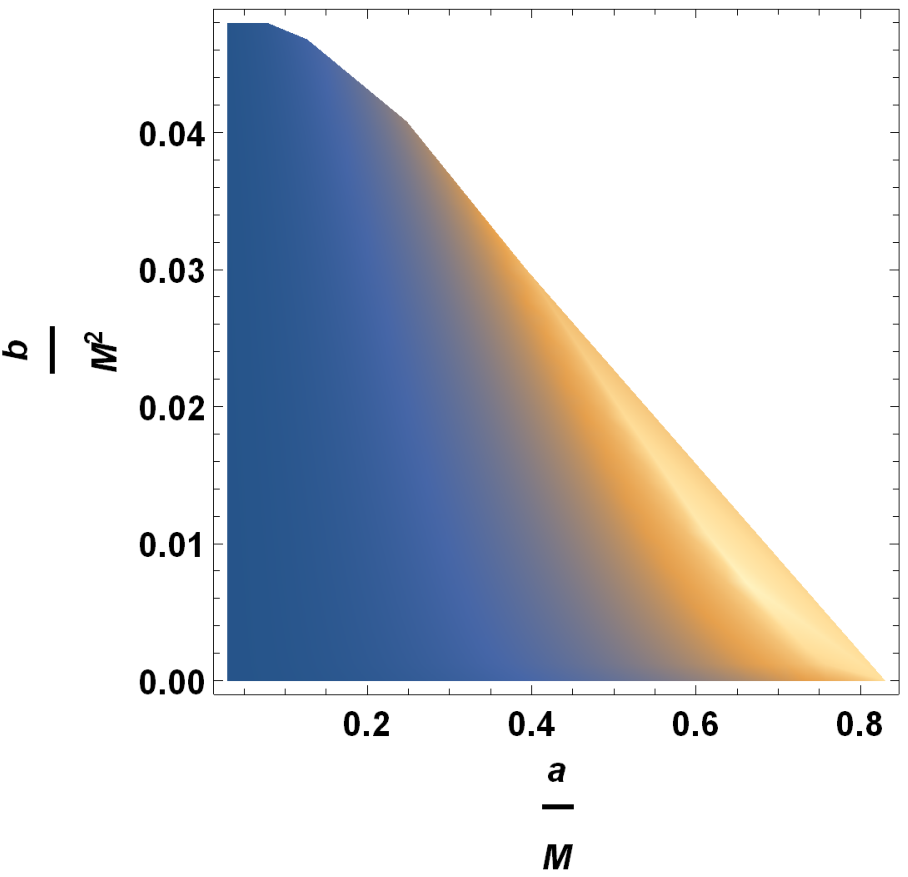}\hspace{1.5em}%
%\caption{Critical radius for various values of $l$ with $k=.1,b=.1$ and $\theta=\pi/2$}
\end{subfigure}%
\begin{subfigure}{.28\textwidth}
\centering
\raisebox{.2\height}{\includegraphics[scale=.55]{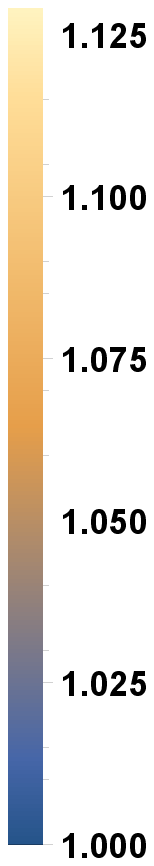}}\hspace{1.5em}%
%\caption{Critical radius for various values of $l$ with $k=.1,b=.1$ and $\theta=\pi/2$}
\end{subfigure}%
\begin{subfigure}{.25\textwidth}
\centering
\includegraphics[scale=.65]{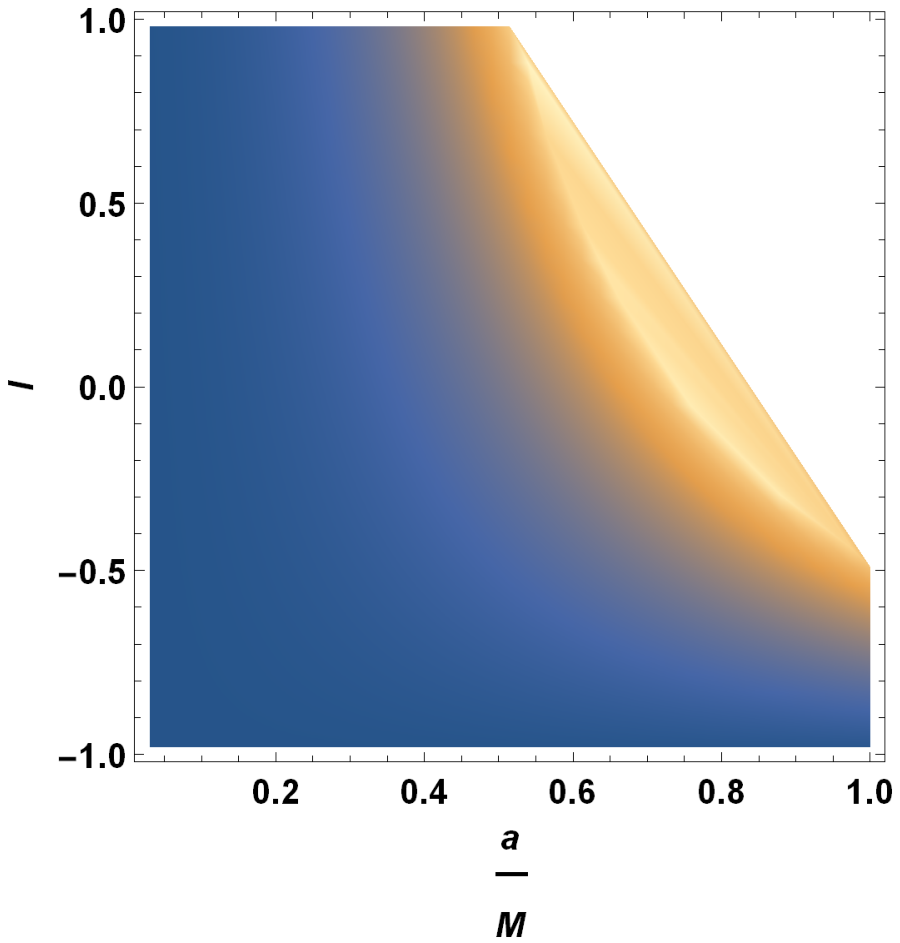}\hspace{1.5em}%
%\caption{Critical radius for various values of $l$ with $k=.1,b=.1$ and $\theta=\pi/2$}
\end{subfigure}%
\begin{subfigure}{.28\textwidth}
\centering
\raisebox{.19\height}{\includegraphics[scale=.6]{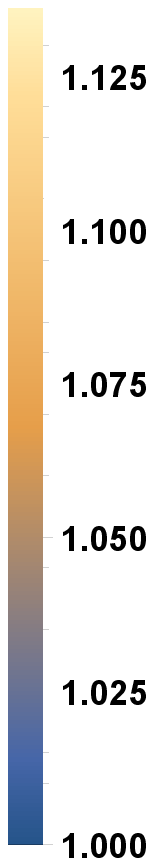}}
%\caption{Critical radius for various values of $l$ with $k=.1,b=.1$ and $\theta=\pi/2$}
\end{subfigure}
\caption{The left panel is for $l=0.4$ and the right panel is for
$b=0.01M^{2}$ where the inclination angle is $90^{o}$.}
%\label{fig:test}
\end{figure}
\smallskip
\begin{figure}[H]
\centering
\begin{subfigure}{.25\textwidth}
\centering
\includegraphics[scale=.65]{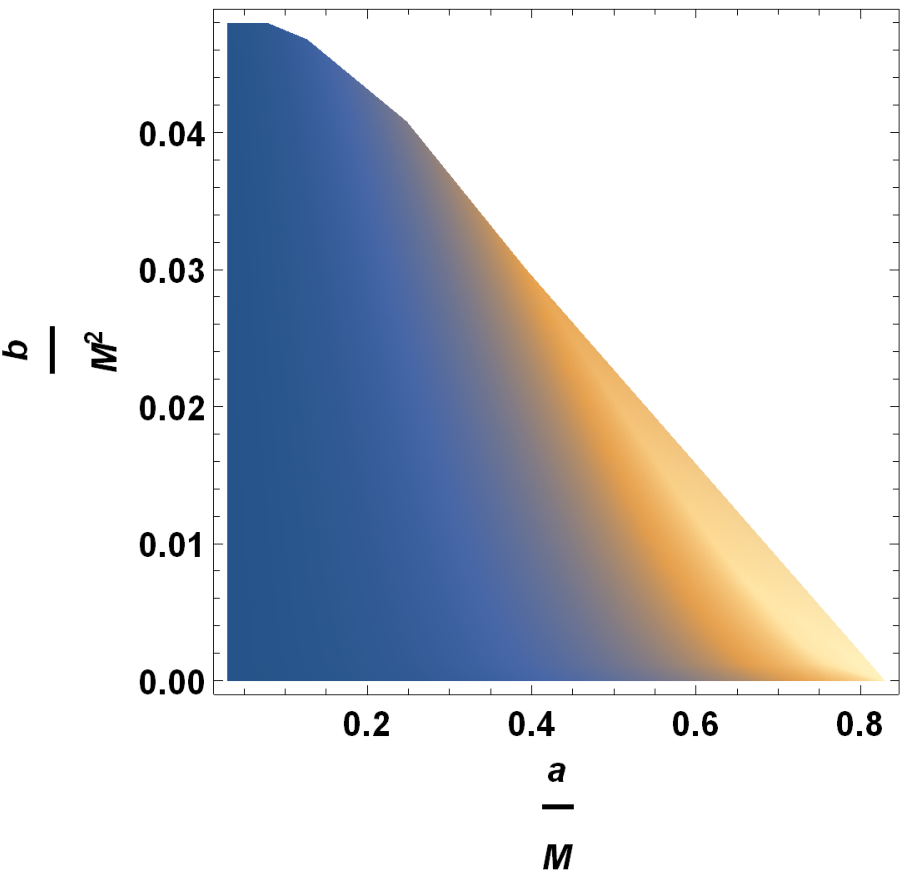}\hspace{1.5em}%
%\caption{Critical radius for various values of $l$ with $k=.1,b=.1$ and $\theta=\pi/2$}
\end{subfigure}%
\begin{subfigure}{.28\textwidth}
\centering
\raisebox{.2\height}{\includegraphics[scale=.55]{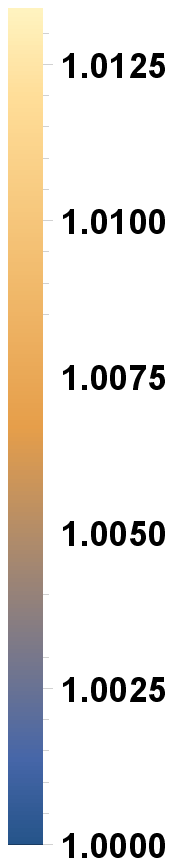}}\hspace{1.5em}%
%\caption{Critical radius for various values of $l$ with $k=.1,b=.1$ and $\theta=\pi/2$}
\end{subfigure}%
\begin{subfigure}{.25\textwidth}
\centering
\includegraphics[scale=.65]{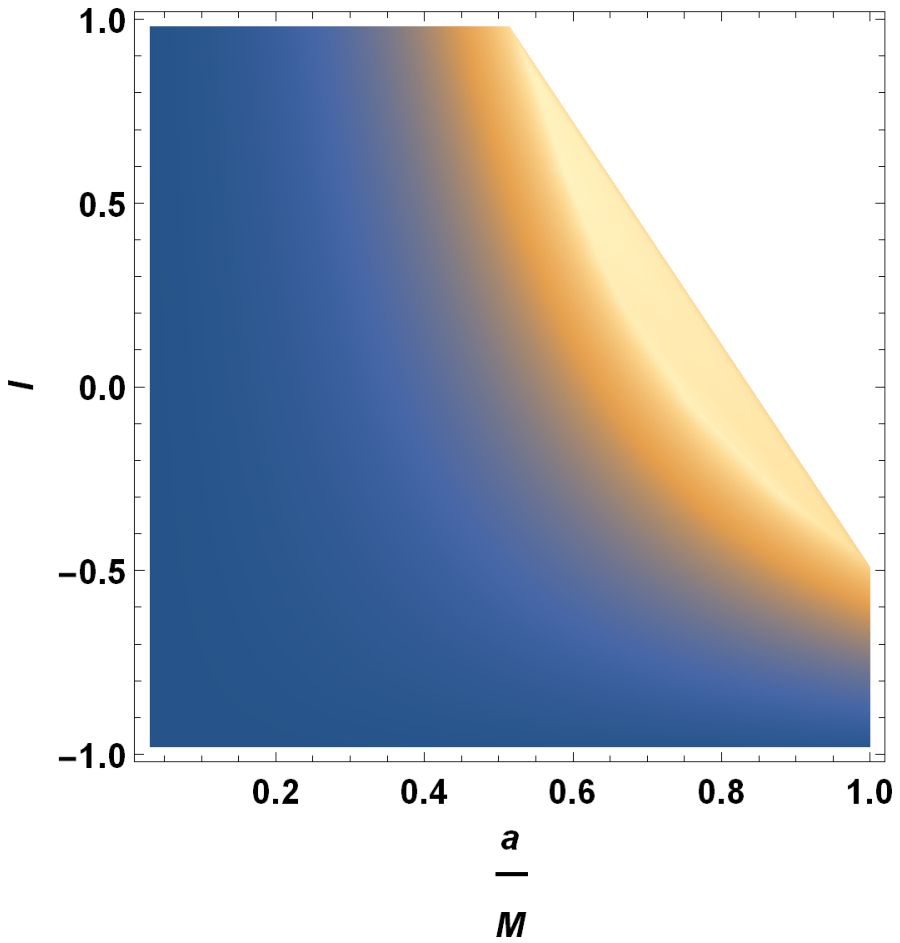}\hspace{1.5em}%
%\caption{Critical radius for various values of $l$ with $k=.1,b=.1$ and $\theta=\pi/2$}
\end{subfigure}%
\begin{subfigure}{.28\textwidth}
\centering
\raisebox{.19\height}{\includegraphics[scale=.6]{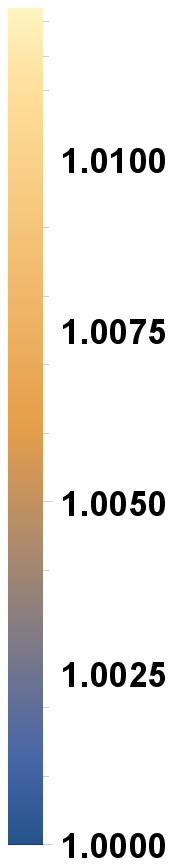}}
%\caption{Critical radius for various values of $l$ with $k=.1,b=.1$ and $\theta=\pi/2$}
\end{subfigure}
\caption{The left panel is for $l=0.4$ and the right panel is for
$b=0.01M^{2}$ where the inclination angle is $17^{o}$.}
%\label{fig:test}
\end{figure}
From the plots above we see that the condition $1 <D_{X} \lesssim
4/3$ is satisfied for the entire parameter space of
non-commutative Kerr-like black holes. Thus non-commutative
Kerr-like black holes are remarkably consistent with EHT images of
$M87^{*}$. Therefore, we can not rule out non-commutative
Kerr-like black holes from the observational data of $M87^{*}$
black hole shadow.

We can have the bound of the parameter $b$ associated with the
non-commutativity of the spacetime in a similar way we determined
the bound of the parameter $l$ in \cite{OUR}. By modeling
M$87^{*}$ black hole as Kerr black hole, the author of the article
\cite{RODRIGO} obtained a lower limit of $a$ for the M$87^{*}$
black hole. Bringing this result under consideration in \cite{OUR}
we put the interval of interest for $a$ as $[0.50M,0.99M]$, and
using the experimental constraints $\Delta C \leq 0.10$ and
$\theta_{d}=42\pm3 \mu as$ with the information
$a\in[0.50M,0.99M]$, we observed that $l\in(-1,0.621031]$. In a
similar way taking into account the bounds $a\in[0.50M,0.99M]$ and
$l\in(-1,0.621031]$ and the experimental constraints $\Delta
C\leq0.1$ and $\theta_{d}=39\pm3\mu as$, we get a bound on the
parameter $b$ which is linked with the non-commutativity of
spacetime. We find that the parameter $b\in[0,0.000505973M^{2}]$.
It is intriguing to have an upper bound of $b$ which is found out
to be $0.000505973M^{2}$. To the best of our knowledge, the bound
of the parameter $b$ from the shadow of the astronomical black hol
has not yet been reported so far.

\section{Summary and Conclusion}
In this work, we have developed a framework where quantum
correction due to the Lorentz violation and noncommutativity of
spacetime have been taken into account on the same footing. The
spacetime background renders a non-commutative Kerr-like LV black
hole. We have extensively studied different aspects of the
non-commutative and LV Kerr-like black hole. The spin, the mass,
the LV parameter, and the non-commutative parameter involved in it
determine the gravitational field of this black hole. First of
all, we study geometry in detail concerning its horizon structure
and ergosphere.

The study of the two important optical phenomena in the vicinity
of this black hole was our main objective in this article. In this
respect, we first have consider the superradiance phenomena and
find that it crucially depends on the parameter $l$ and $b$ apart
from itsdependence on $a$ which is licked with the spin of the
black hole. The superradiance process enhances with the decrease
in the value of the LV parameters and it diminishes when the
increase in value of the LV parameter. We also observe that with
the increase in the value of the parameter $b$ the superradiance
process gets diminished. However, with the increase in the value
of $a$ the superradiance process increases.

Next, we have brought into our investigation the effect of Lorentz
violating parameter $l$ and non-commutative parameter $b$ on the
size of the black hole shadow. We have observed that the size of
the black hole shadow increases with an increase in the value of
the parameter $l$, and it decreases with an increase in the value
of the parameter $b$. Thus, it can be safely concluded that
Lorentz violation and non-commutativity, both, have significant
impacts on black hole shadow.

We have also studied emission rates in the presence of plasma and
also in the absence of it to make a comparison between these two
cases. Our study shows that although the emission rate changes
with the variation of the plasma parameter, the nature of the
emission curves remains unaltered in both cases: in the presence
and in the absence of plasma. These results have clearly
established the influence of the LV parameter and non-commutative
parameter on emission rate. Besides, it being a generic study, we
can also obtain the results for Kerr and Kerr-like black holes
with suitable limits

We have made an attempt to constrain parameters in our modified
theories using the observations of EHT collaboration. For
inclination angle $\theta=90^{o}$, the deviation from circularity
$\Delta C\leq0.1$ and angular diameter $\theta_{d}=42\pm3\mu$as
within $1\sigma$ region are satisfied for finite parameter space
$(\frac{b}{M^{2}} - \frac{a}{M})$ and $(l- \frac{a}{M})$. For
inclination angle $\theta=17^{o}$, the circularity deviation
$\Delta C\leq0.1$ is satisfied for entire parameter space
$(\frac{b}{M^{2}} - \frac{a}{M})$ and $(l - \frac{a}{M})$. The
angular diameter $\theta_{d}=42\pm3\mu$ as within $1\sigma$ is
satisfied for finite parameter space
$(\frac{b}{M^{2}}-\frac{a}{M})$ and $(l-\frac{a}{M})$. The axis
ratio $D_{X}$ satisfies the constraint $1<D_{X}\lesssim4/3$ for
the entire parameter space at both the inclination angles
$\theta=90^{o}$ as well as $\theta=17^{o}$. Therefore our study
enables to establish the fact that non-commutative Kerr-like black
holes are remarkably consistent with EHT images of $M87^{*}$. It
demands that ruling out non-commutative Kerr-like black holes from
the observational data of black hole shadow would be illogical.
Thus non-commutative Kerr-like black hole may be considered as a
suitable candidate for the astrophysical black hole. It has also
been shown that the possible upper bound of $b$ which is
associated with the non-commutativity is $0.000505973M^{2}$. It is
intriguing and indeed a novel way to constrain the parameter
associated with the non-commutativity from the shadow of an
astronomical black hole.

Till now we do not have any available data to constrain the
parameter $l$ and $b$ from the superradiance effect and from the
energy emission process. So the parameter con not be constrained
by knowledge of the superradiance phenomena and the energy
emission process. We have made all the plots for the superradiance
and the energy emission process maintaining the constraint
obtained from the EHT data concerning the shadow of the $M87^*$.

\end{document}